\documentclass[aip,twocolumn,reprint,superscriptaddress]{revtex4-1}
\usepackage{amsmath}
\usepackage{amssymb}
\usepackage[latin9]{inputenc}
\usepackage{graphicx}
\usepackage{hyperref}
\usepackage{physics}

\usepackage{graphicx}
\usepackage[color= blue!20]{todonotes}
\usepackage[normalem]{ulem}
\usepackage{color}
\usepackage{ulem}
\usepackage[symbol]{footmisc}
\newcommand\DB{\textcolor{blue}}

\usepackage{xcolor}
\hypersetup{
	colorlinks,
	linkcolor={blue!50!black},
	citecolor={blue!50!black}
}
\DeclareMathOperator{\atan2}{atan2}

\begin{document}
	
\addtocontents{toc}{\protect\setcounter{tocdepth}{0}}

\onecolumngrid

\noindent\textbf{\textsf{\Large Four-wave-cooling to the single phonon level in Kerr optomechanics}}

\normalsize
\vspace{.3cm}
\noindent\textsf{ D.~Bothner$^{\dagger,1,2}$, I.~C.~Rodrigues$^{\dagger,1}$ and G.~A.~Steele$^{1}$}

\vspace{.2cm}
\noindent\textit{$^1$Kavli Institute of Nanoscience, Delft University of Technology, PO Box 5046, 2600 GA Delft, The Netherlands\\$^2$Physikalisches Institut, Center for Quantum Science (CQ) and LISA$^+$, Universit\"at T\"ubingen, Auf der Morgenstelle 14, 72076 T\"ubingen, Germany\\$^\dagger$\normalfont{these authors contributed equally}}

\vspace{.5cm}

\date{\today}

{\addtolength{\leftskip}{10 mm}
\addtolength{\rightskip}{10 mm}

The field of cavity optomechanics has achieved groundbreaking photonic control and detection of mechanical oscillators, based on their coupling to linear electromagnetic modes.
Lately, however, there is an uprising interest in exploring cavity nonlinearities as a powerful new resource in radiation-pressure interacting systems.
Here, we present a flux-mediated optomechanical device combining a nonlinear Josephson-based superconducting quantum interference cavity with a mechanical nanobeam.
We demonstrate how the intrinsic Kerr nonlinearity of the microwave circuit can be used for a counter-intuitive blue-detuned sideband-cooling scheme based on multi-tone cavity driving and intracavity four-wave-mixing.
Based on the large single-photon coupling rate of the system of up to $g_0 = 2\pi\cdot 3.6\,$kHz and a high mechanical quality factor $Q_\mathrm{m} \approx 4\cdot 10^{5}$, we achieve an effective four-wave cooperativity of $\mathcal{C}_\mathrm{fw} > 100$ and demonstrate four-wave cooling of the mechanical oscillator close to its quantum groundstate, achieving a final occupancy of $n_\mathrm{m} \sim 1.6$.
Our results significantly advance the recently developed platform of flux-mediated optomechanics and demonstrate how cavity Kerr nonlinearities can be utilized for novel control schemes in cavity optomechanics.
}
\vspace{.5cm}

\twocolumngrid
\noindent\textbf{\textsf{\small INTRODUCTION}}
\vspace{2mm}

Cavity optomechanical systems are the leading platform for the detection and manipulation of mechanical oscillators with electromagnetic fields from the nano- to the macro-scale \cite{Aspelmeyer14}.
Displacement detection with an imprecision below the standard quantum limit \cite{Teufel09, Mason19}, sideband-cooling to the motional quantum groundstate \cite{Teufel11, Chan11}, the preparation of nonclassical states of motion \cite{Wollman15, Riedinger16, Reed17, Ma20}, quantum entanglement of distinct mechanical oscillators \cite{Riedinger18, OckeloenKorppi18}, topological energy transfer using exceptional points \cite{Xu16} and microwave-to-optical frequency transducers \cite{Andrews14, Forsch20} are just some of the highlights that have been reported during the last decade.
Essentially all of these impressing results have been achieved with linear cavities and linear mechanical oscillators, but the exploration of intrinsic cavity nonlinearities, often considered undesired and parasitic in optomechanics as they impose limitations on the maximally achievable multi-photon coupling rate \cite{Peterson19}, has attracted increasing interest lately \cite{Nation08, Kumar10, Mikkelsen17, Asjad19, Gan19, Qiu19, Lau20}.
An exciting new scheme to couple a mechanical oscillator to microwave photons in a superconducting LC circuit has very recently been realized: Flux-mediated optomechanical coupling \cite{Rodrigues19, Zoepfl20, Schmidt20, Bera20}.
In this approach, the displacement of a mechanical oscillator is transduced to magnetic flux threading a superconducting quantum interference device (SQUID) embedded in a microwave LC circuit as flux-dependent inductance \cite{Xue07, Nation08, Shevchuk17}.
Due to the scaling of the optomechanical single-photon coupling rate $g_0$ with the external magnetic transduction field in flux-mediated optomechanics \cite{Shevchuk17, Rodrigues19}, record single-photon coupling rates for the microwave domain have been reported \cite{Zoepfl20, Schmidt20, Bera20}.
In future devices, the optomechanical single-photon regime \cite{Nunnenkamp11, Rabl11} or even the ultrastrong coupling to superconducting qubits \cite{Kounalakis20} seem feasible.
In addition to being a flux-tunable inductor, a SQUID simultaneously constitutes a flexible and highly controllable Kerr nonlinearity, which is widely utilized in superconducting qubits \cite{Krantz19}, Josephson parametric amplifiers \cite{CastellanosBeltran08} and four-wave-mixing based bosonic code quantum information processing \cite{Leghtas15}.
Therefore, flux-mediated optomechanics is also an ideal platform for realizing and studying Kerr optomechanics and for the development of new detection and control schemes of mechanical motion.
\begin{figure*}
	\centerline{\includegraphics[trim = {0cm, 0cm, 0cm, 0cm}, clip=True, width=0.99\textwidth]{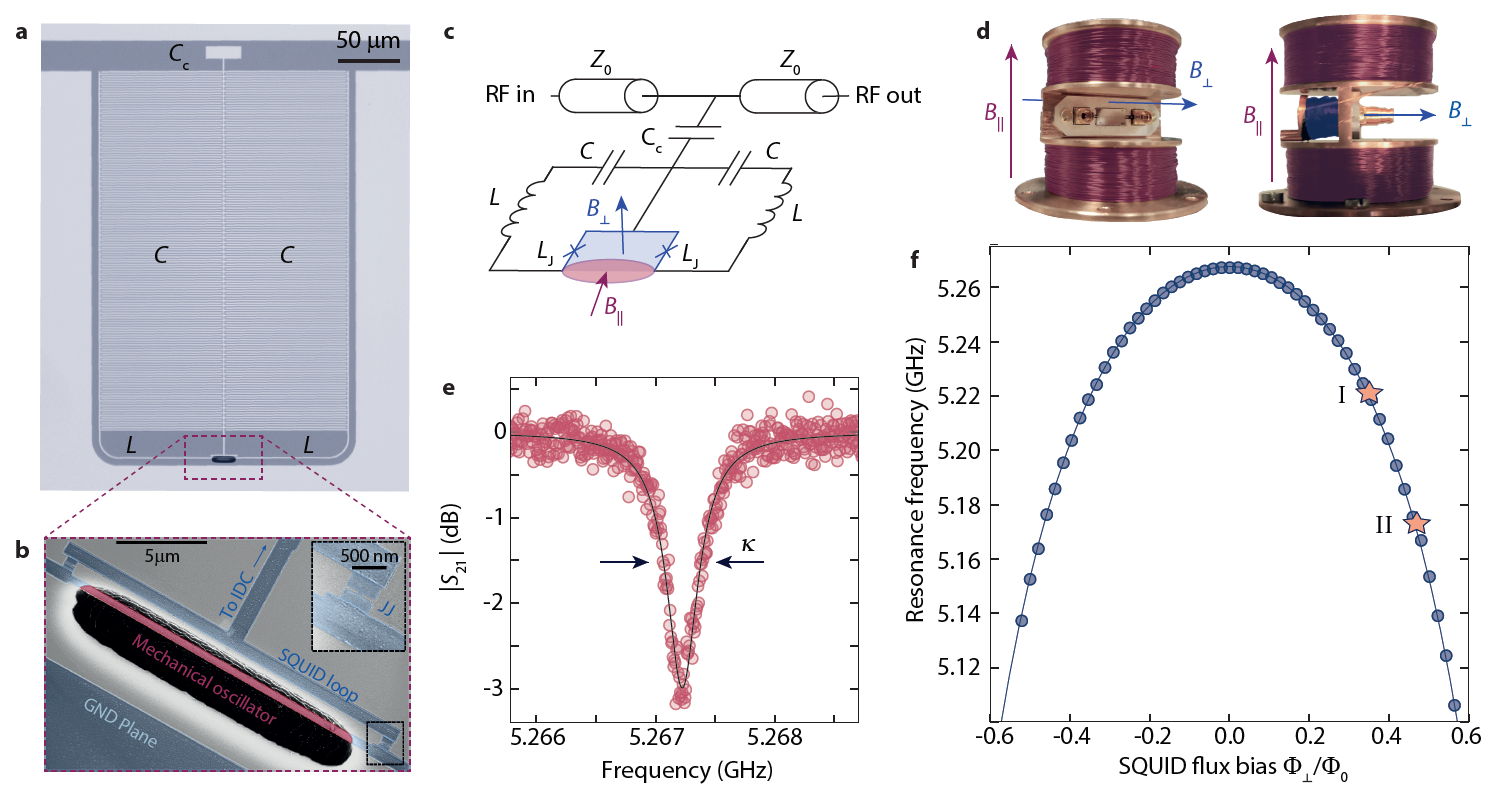}}
	\caption{\textsf{\textbf{A superconducting quantum interference cavity parametrically coupled to a mechanical nanobeam.} \textbf{a} Optical micrograph of the circuit. Bright parts are Aluminum, dark parts are Silicon substrate. The LC circuit combines two interdigitated capacitors $C$ with two linear inductors $L$, connected through a superconducting quantum interference device (SQUID) with total Josephson inductance $L_\mathrm{S} = L_\mathrm{J}/2$. The circuit is capacitively coupled to a $Z_0 = 50\,\Omega$ coplanar waveguide feedline (top of image) with a coupling capacitor $C_\mathrm{c}$ and surrounded by ground-plane. Scale bar corresponds to $50\,\mu$m. \textbf{b} Scanning electron micrograph of the constriction-type SQUID, showing the two Josephson junctions and the mechanical oscillator as part of the loop released from the substrate. Inset shows a zoom-in to one of the nano-bridge Josephson junctions. \textbf{c} Circuit equivalent of the device. For the experiments, two magnetic fields can be applied. The field $B_\perp$ is oriented perpendicular to the chip plane and is used to set the flux bias working point of the SQUID $\Phi_\perp$. The parallel field $B_\parallel$ transduces mechanical displacement of the out-of-plane mode to additional flux $\Delta\Phi_\parallel = B_\parallel l \Delta x$ threading the SQUID loop. \textbf{d} shows the sample integrated into a printed circuit board with two microwave connectors and mounted into a 2D vector magnet. The large split coil is used to generate $B_\parallel$, a small single coil behind the chip generates $B_\perp$. \textbf{e} Transmission response $|S_{21}|$ of the cavity at $B_\parallel = 25\,$mT and $B_\perp = 0$. From a fit to the data, we extract the resonance frequency $\omega_0 = 2\pi\cdot 5.2673\,$GHz, the total linewidth $\kappa = 2\pi\cdot 380\,$kHz, and the external linewidth $\kappa_\mathrm{e} = 2\pi\cdot 110\,$kHz. Data are shown as circles, fit as black line. \textbf{e} Resonance frequency $\omega_0$ vs magnetic flux $\Phi_\perp$, normalized to one flux quantum $\Phi_0$ at $B_\parallel = 25\,$mT. Circles are data, line is a fit. The two operation points for this paper are marked with stars and denoted "I" for $\omega_0 \approx 2\pi\cdot5.22\,$GHz and "II" for $\omega_0 \approx 2\pi\cdot5.17\,$GHz. Details on measurements and fits can be found in Supplementary Note~4.}}
	\label{fig:Device}
\end{figure*}
Here, we implement a flux-mediated optomechanical device with a large single-photon couling rate of up to $g_0 \approx 2\pi\cdot3.6\,$kHz and demonstrate sideband cooling of the mechanical oscillator close to its quantum groundstate by intracavity four-wave mixing (FWM).
By using a strong parametric cavity drive, we activate the emergence of two Kerr quasi-modes in the SQUID circuit and realize an optomechanical coupling of these quasi-modes to the mechanical oscillator by an additional optomechanical sideband pump field.
The drive-activated Kerr-modes show enhanced properties such as a reduced effective linewidth compared to the undriven circuit and we achieve effective single-photon cooperativities $\mathcal{C}_\mathrm{0} \gtrsim 10$.
Strikingly, we find that blue-detuned optomechanical sideband-pumping on one of the Kerr-modes leads to dynamical backaction with the characteristics of red-sideband pumping in a standard optomechanical system, in particular to positive optical damping.
We use this FWM based blue-detuned optical damping to cool the mechanical oscillator extremely close to its quantum groundstate with a residual occupation of $n_\mathrm{m} \sim 1.6$.
Our results demonstrate how cavity Kerr nonlinearities can be used in optomechanics to achieve both, enhanced device performance and new control schemes for mechanical oscillators.
At the same time they reveal the potential of flux-mediated optomechanics regarding low-power groundstate-cooling of mechanical oscillators and the future preparation of quantum states of motion.
\vspace{10mm}%

\noindent\textbf{\textsf{\small RESULTS}}
\vspace{2mm}

\noindent\textbf{\textsf{\small The device}}
\vspace{3mm}

Our device combines a superconducting quantum interference LC circuit with a mechanical nanobeam oscillator embedded into the loop of the SQUID, cf. Fig.~\ref{fig:Device}.
Details on device fabrication are given in Supplementary Note~1.
At the core of the circuit, the SQUID acts as a magnetic-flux-dependent inductance $L_\mathrm{S}(\Phi)$, where $\Phi$ is the total magnetic flux threading the $21\times 3\,\mu$m$^2$ large loop.
For the tunable optomechanical coupling between the displacement of the mechanical nanobeam and the microwave circuit, two distinct external magnetic fields are required.
First, a magnetic field perpendicular to the chip surface $B_\perp$ is used to change the magnetic flux bias $\Phi_\perp$ through the SQUID loop, allowing to tune the circuit resonance frequency $\omega_0$ and flux responsivity $\mathcal{F} = \partial\omega_0/\partial\Phi$.
Secondly, a magnetic in-plane field $B_\parallel$ is used to transduce the out-of-plane displacement $\Delta x$ of the mechanical oscillator to additional flux $\Delta\Phi_\parallel = B_\parallel l_\mathrm{m} \Delta x$, where $l_\mathrm{m} = 18\,\mu$m is the length of the mechanical beam.
To apply these two fields, the chip is mounted into a home-made 2D vector magnet, consisting of a large split coil for $B_\parallel$ and an additional small coil mounted below the chip for the generation of $B_\perp$, cf. Fig.~\ref{fig:Device}\textbf{d}.
The whole configuration is placed in a cryoperm magnetic shielding and attached to the mK plate of a dilution refrigerator with a base temperature $T_\mathrm{b} \approx 15\,$mK.
More details on the measurement setup are given in Supplementary Note~2.
We perform the experiments presented here at in-plane fields of $B_\parallel = 21\,$mT and $B_\parallel = 25\,$mT.
Figure~\ref{fig:Device}\textbf{e} shows the transmission response of the cavity for $B_\parallel = 25\,$mT and at the bias-flux sweetspot.
It has a resonance frequency $\omega_0 = 2\pi\cdot 5.2673\,$GHz, a total linewidth $\kappa = 2\pi\cdot 380\,$kHz and an external linewidth $\kappa_\mathrm{e} = 2\pi\cdot 110\,$kHz.
Figure~\ref{fig:Device}\textbf{f} shows how the cavity resonance frequency can be tuned by $\sim 150\,$MHz by changing the applied flux bias $\Phi_\perp$ threading the SQUID loop.
The curves and cavity parameters at $B_\parallel = 21\,$mT only deviate slightly from the ones given here, the corresponding additional data can be found in Supplementary Note~4.
Due to an improved SQUID design and fabrication, the cavity flux responsivity $\mathcal{F}$ is increased by one order of magnitude compared to our previous results \cite{Rodrigues19}, which leads to a significantly enhanced single-photon coupling rate
\begin{equation}
g_0 = \mathcal{F}B_\parallel l_\mathrm{m} x_\mathrm{zpf}
\end{equation}
where $x_\mathrm{zpf}$ is the mechanical zero-point fluctuation amplitude.
The mechanical nanobeam, visible in Fig.~\ref{fig:Device}\textbf{b} and released from the substrate in an isotropic reactive ion etching process using SF$_6$ plasma \cite{Norte18}, is $500\,$nm wide and $70\,$nm thick.
From its total mass of $m \approx 1.9\,$pg and the resonance frequency of the out-of-plane mode $\Omega_\mathrm{m}\approx 2\pi\cdot 5.32\,$MHz, we get $x_\mathrm{zpf} = \sqrt{\frac{\hbar}{2m\Omega_\mathrm{m}}} \approx 30\,$fm.
For an in-plane field of $B_\parallel = 25\,$mT, and the two flux-bias points $\Phi_\mathrm{I}$ and $\Phi_\mathrm{II}$, cf. Fig.~\ref{fig:Device}\textbf{f}, we obtain single-photon coupling rates $g_{0, \mathrm{I}} = 2\pi\cdot{2.2}\,$kHz and $g_{0, \mathrm{II}} = 2\pi\cdot{3.6}\,$kHz with $\mathcal{F}_\mathrm{I} = 2\pi\cdot 300\,$MHz$/\Phi_0$ and $\mathcal{F}_\mathrm{II} = 2\pi\cdot\,520\,$MHz$/\Phi_0$.
For the smaller in-plane field of $B_\parallel = 21\,$mT, the $g_0$-values are scaled accordingly, cf. Supplementary Note~5.
The final important parameter of the device is its Kerr nonlinearity, which at the flux sweetspot is $\mathcal{K}/2\pi = - 30\,$kHz.
For the two flux bias operation points I and II we obtain $\mathcal{K}_\mathrm{I}/2\pi = - 40\,$kHz and $\mathcal{K}_\mathrm{II}/2\pi = - 55\,$kHz, respectively.
More details on the determination of the circuit parameters and their flux dependence can be found in Supplementary Note~4. \vspace{10mm}

\noindent\textbf{\textsf{\small Driven Kerr-modes and dynamical Kerr backaction}}
\vspace{3mm}

\begin{figure*}
	\centerline{\includegraphics[trim = {0cm, 0cm, 0cm, 0cm}, clip=True, width=0.98\textwidth]{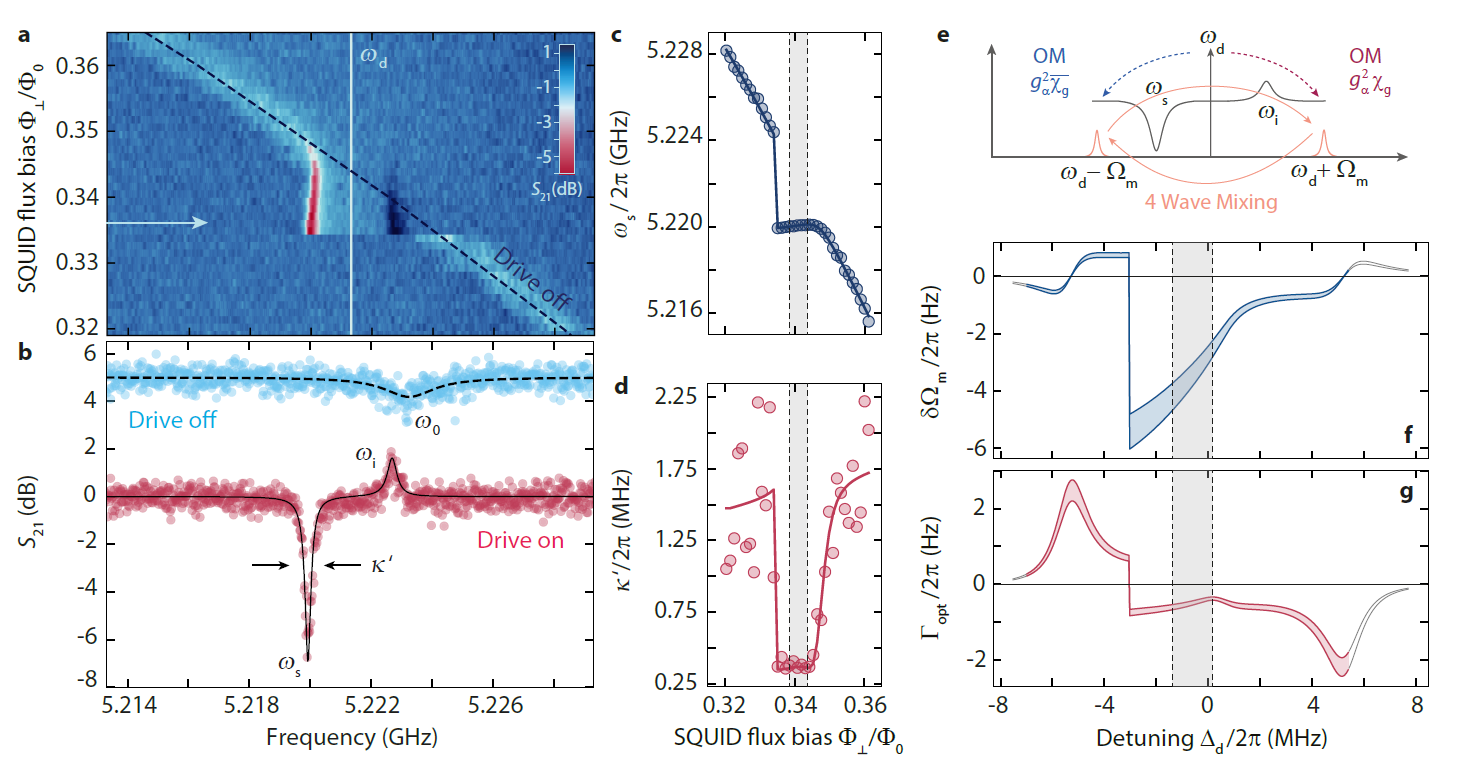}}
	\caption{\textsf{\textbf{Activating the driven Kerr quasi-mode state and single-tone dynamical Kerr backaction.} \textbf{a} displays color-coded the magnitude of the SQUID cavity response $S_{21}$ for varying SQUID flux bias $\Phi_\perp/\Phi_0$ in the presence of a strong drive placed at $\omega_\mathrm{d}$. The flux bias range corresponds to a small variation of $\Phi_\perp$ around operation point I and the in-plane field is $B_\parallel = 21\,$mT. When the flux-tunable resonance frequency $\omega_0$, indicated as dashed line and labelled "Drive off", is far detuned from the drive tone, the cavity response exhibits a single broad absorption resonance. As the detuning between cavity and drive $\Delta_\mathrm{d} = \omega_\mathrm{d}-\omega_0$ is reduced, the cavity response is significantly modified and the original resonance is developing into a double-mode structure. The appearance of these driven Kerr quasi-modes indicates the onset of parametric amplification and degenerate FWM in the SQUID circuit. We denote the two modes as signal and idler resonance with the resonance frequencies $\omega_\mathrm{s}$ and $\omega_\mathrm{i}$, respectively. Arrow on the left indicates the position of the linescan shown in panel \textbf{b}. In addition to the linescan from \textbf{a} (red circles) and the result of the analytical response calculation (solid black line), we show the equivalent linescan without parametric drive (blue circles) and its corresponding theoretical response (dashed black line). The curves without parametric drive are offset by $+5\,$dB for clarity. Panels \textbf{c} and \textbf{d} show the extracted resonance frequency $\omega_\mathrm{s}$ and effective linewidth $\kappa'$ of the signal resonance vs flux bias. Lines show the result of modelling the effective quantities with the driven Kerr cavity equations and taking into account flux-noise broadening and two-level systems. The regime of operation for the experiments reported below is indicated by dashed lines and shaded areas. In this regime, the linewidth is nearly constant with $\kappa'/2\pi \approx 340\,$kHz. The width of the operation range corresponds to the flux noise standard deviation, which we estimate to be $\sigma_\Phi \sim 5\,$m$\Phi_0$. Panel \textbf{e} illustrates the contributions to the dynamical Kerr backaction of the intracavity drive fields to the nanobeam. Optomechanical (OM) up- and downscattering induces cooling and heating/amplification to the mechanical mode, respectively, where $g_\alpha$ is the multiphoton coupling rate and $\chi_\mathrm{g}$ is the probe susceptibility of the driven Kerr oscillator. In addition, interference between up- and downscattered fields due to degenerate FWM has to be taken into account. \textbf{f} and \textbf{g} show the calculated optical spring and optical damping due to dynamical Kerr backaction. The two blue/red lines and shaded area correspond to $g_0/2\pi = (1.78 \pm 0.1)\,$kHz. The detuning range $\Delta_\mathrm{d}$ is slightly increased compared to \textbf{a}-\textbf{d}. In the additional range, the backaction is plotted in gray. The device operation range is indicated by the shaded area in between the vertical dashed lines.}}
	\label{fig:Idler}
\end{figure*}
Owing to the Kerr anharmonicity $\mathcal{K}$, the application of a strong microwave drive tone close to the cavity resonance frequency $\omega_0$ significantly modifies the cavity response to an additional probe field.
In Fig.~\ref{fig:Idler}, we discuss this modified response in the presence of a parametric drive tone with a fixed frequency $\omega_\mathrm{d}$, when the cavity is tuned to cross this drive tone by means of the bias field $B_\perp$.
For large detunings between cavity and drive, the circuit response $S_{21}$ exhibits a standard single-mode resonance lineshape. 
However, as the detuning $\Delta_\mathrm{d} = \omega_\mathrm{d} - \omega_0$ is reduced, the driven cavity susceptibility 
\begin{equation}
\chi_\mathrm{g}(\Omega) = \frac{\tilde{\chi}_\mathrm{p}(\Omega)}{1 - \mathcal{K}^2n_\mathrm{d}^2\tilde{\chi}_\mathrm{p}(\Omega)\tilde{\chi}_\mathrm{p}^*(-\Omega)}
\end{equation}
deviates considerably from a single linear cavity, leading to the regime of parametric amplification and degenerate four-wave mixing, which is experimentally identified by the appearance of a second mode.
Here, $\Omega = \omega - \omega_\mathrm{d}$ denotes the detuning between the probe field at $\omega$ and the parametric drive and
\begin{equation}
\tilde{\chi}_\mathrm{p}(\Omega) = \frac{1}{\frac{\kappa}{2} + i\left(\Delta_\mathrm{d}  - 2\mathcal{K}n_\mathrm{d} + \Omega \right)}.
\end{equation}
The two Kerr quasi-modes, which we denote as signal and idler resonance, appear symmetrically around the drive with complex resonance frequencies
\begin{equation}
\omega_{i/s} = \omega_\mathrm{d} + i\frac{\kappa}{2} \pm \sqrt{\left( \Delta_\mathrm{d} - \mathcal{K}n_\mathrm{d} \right) \left( \Delta_\mathrm{d} - 3\mathcal{K}n_\mathrm{d} \right)}
\end{equation}
where $n_\mathrm{d}$ is the parametric drive intracavity photon number.
These Kerr-modes have been observed and discussed also in the context of optical cavities and mechanical oscillators \cite{Drummond80, Khandekar15, Huber20}.
The signal mode can be identified by the shifted and significantly deepened cavity absorption dip and the idler mode by the resonance peak, indicating net transmission gain by Josephson parametric amplification.
With the activation of the quasi-mode state, we also obtain a highly stabilized effective resonance frequency and linewidth, while the bare cavity suffers from considerable frequency fluctuations due to flux noise. 
Due to the reduction of frequency fluctuations in combination with a saturation of two-level system losses by the parametric drive (cf. Supplementary Note 6), the effective cavity linewidth is reduced from the flux-noise broadened $\kappa'_\mathrm{off} \sim 2\pi\cdot 1.5\,$MHz to the driven $\kappa'_\mathrm{on} \approx 2\pi\cdot 340\,$kHz.
An analysis of the signal mode resonance frequency and linewidth in the presence of the parametric drive is provided in Figs.~\ref{fig:Idler}\textbf{c} and \textbf{d}.
Within a small region of flux bias values, the drive-tone induced Kerr shift compensates for the flux-noise induced frequency shifts by means of an internal feedback loop.
Strikingly, this mechanism yields a stabilization of the driven resonance, which thereby becomes the natural choice of operation regime during the following experiments.
In an optomechanical system, any intracavity field also acts back on the mechanical oscillator by altering its resonance frequency and decay rate, an effect known as dynamical backaction \cite{Schliesser06, Teufel09}.
Therefore, the effect of the parametric drive to the mechanical oscillator also requires some careful consideration.
From the linearized equations of motion for the mechanical amplitude field $\hat{b}$ and the intracavity fluctuation field $\hat{a}$ in a single-tone driven Kerr cavity
\begin{eqnarray}
\dot{\hat{b}} & = & \left(i\Omega_\mathrm{m} - \frac{\Gamma_\mathrm{m}}{2}\right)\hat{b} - i g_\alpha\left( \hat{a} + \hat{a}^\dagger\right) + \sqrt{\Gamma_\mathrm{m}}\hat{\zeta}\\
\dot{\hat{a}} & = & \left[ -i\left(\Delta_\mathrm{d} - 2\mathcal{K}n_\mathrm{d} \right) -\frac{\kappa}{2}\right]\hat{a} + i\mathcal{K}n_\mathrm{d}\hat{a}^\dagger \nonumber \\
& & -i g_\alpha\left(\hat{b} + \hat{b}^\dagger \right) + \sqrt{\kappa_\mathrm{i}}\hat{\xi}_\mathrm{i} + \sqrt{\kappa_\mathrm{e}}\hat{\xi}_\mathrm{e}
\end{eqnarray}
with multi-photon coupling rate $g_\alpha = \sqrt{n_\mathrm{d}}g_0$ and input fields $\hat{\zeta}$, $\hat{\xi}_\mathrm{i}$ and $\hat{\xi}_\mathrm{e}$, the effective mechanical susceptibility can be derived as
\begin{equation}
\chi_0^\mathrm{eff}(\Omega) = \frac{1}{\frac{\Gamma_\mathrm{m}}{2} + i\left(\Omega - \Omega_\mathrm{m}\right) + \Sigma_\mathrm{k}(\Omega_\mathrm{m})}
\end{equation}
for the weak-coupling and high-$Q_\mathrm{m}$ limit, which is safely fulfilled for our mechanical oscillator with a linewidth of $\Gamma_\mathrm{m} \approx 2\pi\cdot 13\,$Hz.
The single-tone dynamical Kerr backaction 
\begin{equation}
\Sigma_\mathrm{k}(\Omega_\mathrm{m}) = g_\alpha^2\big[\chi_\mathrm{g}\left(1 -\overline{\mathcal{A}}\right) - \overline{\chi}_\mathrm{g}\left(1 - \mathcal{A}\right) \big]   
\end{equation}
with $\chi_\mathrm{g} = \chi_\mathrm{g}(\Omega_\mathrm{m})$ and $\overline{\chi}_\mathrm{g} = \chi_\mathrm{g}^*(-\Omega_\mathrm{m})$ has almost the same form as in linear optomechanics, but with a modified cavity susceptibility $\chi_\mathrm{g}$.
A striking difference, however, is found in the terms $\mathcal{A} = -i\mathcal{K}n_\mathrm{d}\tilde{\chi}_\mathrm{p}(\Omega_\mathrm{m})$ and $\overline{\mathcal{A}} = i\mathcal{K}n_\mathrm{d}\tilde{\chi}_\mathrm{p}^*(-\Omega_\mathrm{m})$.
These terms correspond to an interference of the red and blue mechanical sideband fields, which occurs due to intracavity four-wave mixing in a driven Kerr cavity.
By this FWM, the two standard mechanical sidebands become idler fields of each other.
A schematic of the dynamical backaction and the sideband interference is shown in Fig.~\ref{fig:Idler}\textbf{e}.
The optical spring $\delta\Omega_\mathrm{m} = -\mathrm{Im}\left[\Sigma_\mathrm{k}(\Omega_\mathrm{m}) \right]$ and optical damping $\Gamma_\mathrm{opt} = 2\mathrm{Re}\left[ \Sigma_\mathrm{k}(\Omega_\mathrm{m})\right]$ caused by the dynamical Kerr backaction are displayed in Figs~\ref{fig:Idler}\textbf{f} and \textbf{g}.
When the drive is located around one mechanical frequency detuned from the cavity $|\Delta_\mathrm{d}| \approx \Omega_\mathrm{m} = 2\pi\cdot 5.32\,$MHz, the backaction looks very similar to that of a linear cavity.
However, when the drive and the cavity are near-resonant, the backaction is strongly dominated by the intracavity photon number and a Duffing-like behaviour can be observed with a sudden transition from high- to low-amplitude state at $\Delta_\mathrm{d} \approx - 2\pi\cdot 3\,$MHz.
In the operation regime for the experiments described here, the drive-induced backaction for operation point I is small with $\Gamma_\mathrm{opt}/2\pi \sim - 1\,
$Hz and $\delta\Omega_\mathrm{m}/2\pi \sim -5\,$Hz.
Using the bare mechanical linewidth $\Gamma_\mathrm{m} \sim 2\pi\cdot 13\,$Hz, the corresponding phonon occupation is therefore increased by about $10\%$, a detailed calculation and discussion of the resulting mechanical mode occupation is given in Supplementary Note~7.
Due to the considerable cavity flux noise outside of the driven quasi-mode regime, we unfortunateley cannot experimentally access the dynamical Kerr backaction for the detuning range shown in Fig.~\ref{fig:Idler}.
Nevertheless, with a larger single-photon coupling rate $g_0$ at operation point II and a stronger drive tone, we observe regimes of mechanical instability induced by the dynamical Kerr backaction, which are in excellent agreement with the prediction from the theory. 
The corresponding data and analysis are explained in detail in Supplementary Note~7.
The presented formalism for the dynamical Kerr backaction can also directly be applied to the sideband-unresolved regime and explain the experimental findings of a recent experiment with a similar SQUID cavity optomechanical device \cite{Zoepfl20}, cf. also Supplementary Note~7.\vspace{10mm}
\begin{figure*}
	\centerline{\includegraphics[trim = {0.0cm, 0.0cm, 0.0cm, 0.0cm}, clip=True, width=0.95\textwidth]{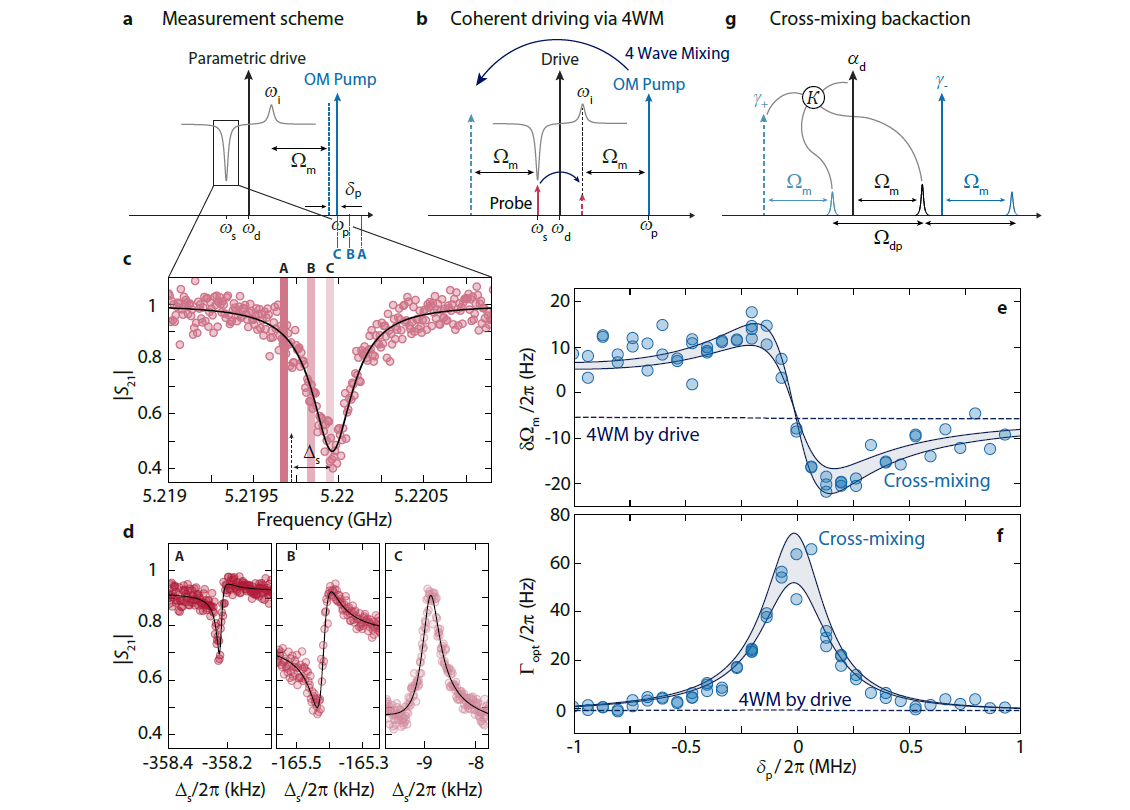}}
	\caption{\textsf{\textbf{Four-wave-OMIT and four-wave-backaction for optomechanical blue-sideband pumping of the idler quasi-mode.} \textbf{a} shows the experimental protocol. The SQUID cavity is prepared in the quasi-mode state by a strong parametric drive (PD). In addition, we apply an optomechanical (OM) pump tone on the blue sideband of the idler resonance (IR) $\omega_\mathrm{p} = \omega_\mathrm{i} + \Omega_\mathrm{m} + \delta_\mathrm{p}$. Finally, we use a weak probe tone around the signal resonance (SR) to detect optomechanically induced transparency. We repeat this scheme for varying detunings $\delta_\mathrm{p}$. \textbf{b} explains how this protocol to first order leads to coherent driving of the mechanical oscillator. By PD-induced intracavity 4WM, the OM pump (probe tone) gets an idler field on the opposite side of the drive, which has the right detuning to the probe tone (pump) $\sim \Omega_\mathrm{m}$ to coherently drive the mechanical oscillator. \textbf{c} shows the signal resonance transmission $S_{21}$ measured with the weak probe field (OM pump off). Circles are data, line is a fit. Vertical bars labelled with A, B, and C indicate zoom regions for the corresponding panels shown in \textbf{c} and $\Delta_\mathrm{s} = \omega - \omega_\mathrm{s}$ denotes the detuning between probe field and SR. \textbf{d} probe tone response (OM pump on) in three narrow frequency windows around $\omega \approx 2\omega_\mathrm{d} - \omega_\mathrm{p} + \Omega_\mathrm{m}$ for three different pump detunings $\delta_\mathrm{p}$, cf. panel \textbf{a}. Note that the frequency difference between OM pump and probe field is $\Omega \approx \Omega_\mathrm{m} - 2\Omega_\mathrm{dp}$, which implies that when the pump field frequency is reduced, the probe field frequency is increasing. Each probe tone response displays a narrow-band resonance, indicating optomechanically induced transparency (OMIT) via excitation of the mechanical oscillator. For each $\delta_\mathrm{p}$, we fit the OMIT response (lines in \textbf{c}) and extract the effective mechanical resonance frequency $\Omega_\mathrm{eff} = \Omega_\mathrm{m} + \delta\Omega_\mathrm{m}$ and the effective mechanical linewidth $\Gamma_\mathrm{eff} = \Gamma_\mathrm{m} + \Gamma_\mathrm{opt}$. The contributions $\delta\Omega_\mathrm{m}$ and $\Gamma_\mathrm{opt}$, induced by dynamical backaction of all intracavity fields, are plotted in panels \textbf{e} and \textbf{f} as circles vs $\delta_\mathrm{p}$. The result of analytical calculations is shown as two solid lines with shaded area, where the range described by the lines captures uncertainties in the device parameters, cf. Supplementary Note~11. The dashed line shows the result of equivalent calculations without cross-mixing (non-degenerate 4WM) terms. \textbf{f} illustrates schematically one four-wave cross-mixing term that leads to the observed dynamical backaction. Hereby, two mechanical sidebands with frequency difference $\Omega_\mathrm{dp} = \omega_\mathrm{d} - \omega_\mathrm{p}$ and both, the PD and the OM pump, contribute to the interaction.}}
	\label{fig:FWOMIT}
\end{figure*}

\noindent\textbf{\textsf{\small Multi-tone dynamical four-wave backaction}}
\vspace{3mm}

An interesting question arising now is how the Kerr quasi-modes couple to the mechanical nanobeam, when an additional optomechanical pump tone is applied to one of the Kerr-mode sidebands.
One might expect that the coupling to the mechanical oscillator is suppressed in this state, similar to the reduced impact of flux noise, as the Kerr-mode frequencies $\omega_\mathrm{s}$ and $\omega_\mathrm{i}$ display only a very weak dependence on flux through the SQUID.
Fluctuations of the bare resonance frequency, however, lead to modulations of $\alpha_\mathrm{d}$ and parametric gain, and therefore will impact the mechanical oscillator by inducing changes in the radiation-pressure force.
A straightforward way to investigate this setting experimentally is to apply an additional optomechanical pump tone on the red sideband of the signal resonance, i.e., with a pump frequency $\omega_\mathrm{p} \approx \omega_\mathrm{s} - \Omega_\mathrm{m}$.
Once in this configuration, a weak probe signal around $\omega \approx \omega_\mathrm{p} + \Omega_\mathrm{m}$ can be used to detect optomechanically induced transparency (OMIT) \cite{Weis10} and thereby characterize the optomechanical interaction.
A detailed theoretical description as well as a discussion of the experimental findings for this red-sideband pumping setup is given in Supplementary Notes~12-15.
A conceptually less straightforward and more exciting possibility is to pump the idler resonance on its blue sideband $\omega_\mathrm{p} \approx \omega_\mathrm{i} + \Omega_\mathrm{m}$, cf. Fig.~\ref{fig:FWOMIT}\textbf{a}.
A blue-detuned pump is commonly associated with amplification/heating due to the favoured Stokes-scattering to lower energy photons.
The Kerr-mode susceptibility $\chi_\mathrm{g}$ close to the idler resonance, however, resembles that of an "inverted" mode.
Any small intracavity field in the driven Kerr cavity experiences in addition a mirroring effect due to degenerate four-wave mixing with the parametric drive tone.
The presence of the blue-sideband pump field enriches this situation even further.
Then the Kerr cavity is effectively oscillating with $\Omega_\mathrm{dp} = \omega_\mathrm{d} - \omega_\mathrm{p}$ due to the presence of two strong fields, and effects arising from non-degenerate four-wave mixing can impact probe fields and mechanical sideband fields and finally also the OMIT response and the backaction to the mechanical oscillator.
A clear signature of the parametric state and four-wave mixing is the appearance of optomechanically induced transparency in the probe response of the signal resonance, when the idler Kerr-mode is pumped on its blue sideband.
Corresponding data are shown in Fig.~\ref{fig:FWOMIT}\textbf{b} and \textbf{c}.
Here and in stark contrast to the usual OMIT protocol, the frequency detuning between the idler blue-sideband pump and the probe tone is not even close to the mechanical resonance frequency but given by $\Omega = \omega - \omega_\mathrm{p} \approx 2\Omega_\mathrm{dp} - \Omega_\mathrm{m}$.
To first order, the observation of this transparency can be understood by considering the intracavity generated tones in addition to the ones that are sent externally.
The parametric drive generates an intracavity field with amplitude $\alpha_\mathrm{d}$ at $\omega_\mathrm{d}$, and the optomechanical pump at $\omega_\mathrm{p}$ generates an intracavity field with amplitude $\gamma_-$.
Just by this doubly-driven configuration, a third intracavity "pump" field is generated by degenerate FWM at $\omega_+ = \omega_\mathrm{p} + 2\Omega_\mathrm{dp}$ and we denote its amplitude as $\gamma_+$.
Therefore, when $\omega_\mathrm{p} = \omega_\mathrm{i} + \Omega_\mathrm{m}$, the $\gamma_+$-field is located at the red sideband of the signal resonance $\omega_+ = \omega_\mathrm{s} - \Omega_\mathrm{m}$.
The beating between a probe field at $\omega \approx \omega_\mathrm{s}$ and the $\gamma_+$-field is then near-resonant with the mechanical oscillator and will drive it into coherent motion.
A second beating component, which is driving the mechanical oscillator, originates from the beating of the $\gamma_-$-field and the idler field of the weak probe itself, cf. Fig.~\ref{fig:FWOMIT}\textbf{a}. 
These two are also near-resonant with the mechanical oscillator.
Once in coherent motion, the mechanical oscillator generates sidebands to all intracavity field Fourier components, some of which interfere with the original probe tone causing the observed appearance of four-wave OMIT.
To characterize the dynamical backaction imprinted by the intracavity fields on the mechanical oscillator in the presence of the $\alpha_\mathrm{d}, \gamma_-$ and $\gamma_+$ fields, we measure the optomechanical transparency response for varying detuning $\delta_\mathrm{p}$ between the $\gamma_-$-field and the idler-mode blue sideband, cf. Fig.~\ref{fig:FWOMIT}.
For each detuning, we determine the effective mechanical resonance frequency $\Omega_\mathrm{eff}$ and effective mechanical linewidth $\Gamma_\mathrm{eff}$ from a fit to the transparency signal and subtract the intrinsic values $\Omega_\mathrm{m}$ and $\Gamma_\mathrm{m}$. 
The remaining contributions to the resonance frequency and linewidth $\delta\Omega_\mathrm{m}$ and $\Gamma_\mathrm{opt}$, respectively, correspond to the optical spring and optical damping by the microwave fields. 
The result, shown in Fig.~\ref{fig:FWOMIT}\textbf{d} and \textbf{e}, is quite surprising.
Even though the optomechanical pump field is blue-detuned to all cavity resonances $\omega_0, \omega_\mathrm{s}$ and $\omega_\mathrm{i}$, we observe dynamical backaction with characteristics resembling red-sideband pumping in linear optomechanical systems.
Most strikingly, we find a positive optical damping, which is usually a clear signature for red-sideband physics and the basis for sideband-cooling of the mechanical mode\cite{Teufel11}.
We use a linearized, optomechanical multi-tone Kerr cavity model, and implement the hierarchy from the experiment $\alpha_\mathrm{d} \gg \gamma_\mp \gg \langle \hat{a} \rangle$ to reveal which interactions are responsible for the observed behaviour, cf. Supplementary Notes 8-10.
The resulting effective mechanical susceptibility
\begin{equation}
\chi_0^\mathrm{eff}(\Omega) = \frac{1}{\frac{\Gamma_\mathrm{m}}{2} + i\left(\Omega - \Omega_\mathrm{m}\right) + \Sigma_\mathrm{fw}(\Omega_\mathrm{m})}
\end{equation}
has still the same form as for a standard optomechanical system, and all the FWM contributions can be captured in $\mathcal{J}$-factors in the dynamical four-wave backaction
\begin{equation}
\Sigma_\mathrm{fw}(\Omega_m) = \sum_{j = -, \alpha, +}^{} |g_j|^2\left[\chi_{\mathrm{g}, j}\mathcal{J}_j - \overline{\chi}_{\mathrm{g}, j}\overline{\mathcal{J}}_j \right]
\end{equation} 
with $g_- = \gamma_- g_0$, $g_+ = \gamma_+ g_0$, $\chi_\mathrm{g, -} = \chi_\mathrm{g}(\Omega_\mathrm{m})$, $\chi_\mathrm{g, \alpha} = \chi_\mathrm{g}(\Omega_\mathrm{m} + \Omega_\mathrm{dp})$ and $\chi_\mathrm{g, +} = \chi_\mathrm{g}(\Omega_\mathrm{m} + 2\Omega_\mathrm{dp})$.
Closed-form expressions for the $\mathcal{J}$ are given in Supplementary Note~9.
We identify non-degenerate four-wave mixing terms in the $\mathcal{J}$-factors as the dominant origin of the observed backaction.
These terms have contributions from the drive field $\alpha_\mathrm{d}$, from one of the $\gamma_\pm$ fields and couple any two distinct mechanical sidebands which have the frequency difference $\pm\Omega_\mathrm{dp}$, cf. Fig.~\ref{fig:FWOMIT}\textbf{f} for a schematic of one of these terms.
Hence, these terms correspond to intracavity cross-mixing based on $\alpha_\mathrm{d}$ and $\gamma_\pm$ fields.
Using independently determined system parameters, we find excellent agreement between the experimental data and the analytical model when we take these cross-mixing terms into account, cf. solid lines in Fig.~\ref{fig:FWOMIT}\textbf{d} and \textbf{e}.
If we take only the degenerate FWM terms into account, which are induced by the presence of $\alpha_\mathrm{d}$, we find a small and nearly constant backaction for all $\delta_\mathrm{p}$, cf. dashed lines.\vspace{10mm}
%

%
%
%
%
%
%
%

\noindent\textbf{\textsf{\small Blue-detuned four-wave cooling close to the groundstate}}
\vspace{3mm}

Positive optical damping is commonly related to cooling of the mechanical mode.
Therefore, the blue-detuned pumping scheme described in Fig.~\ref{fig:FWOMIT} seems feasible to be utilized as a counter-intuitive, yet innovative, method to eliminate the residual thermal excitations in the mechanical resonator.

\begin{figure*}
	\centerline{\includegraphics[trim = {0cm, 0cm, 0cm, 0cm}, clip=True, width=0.96\textwidth]{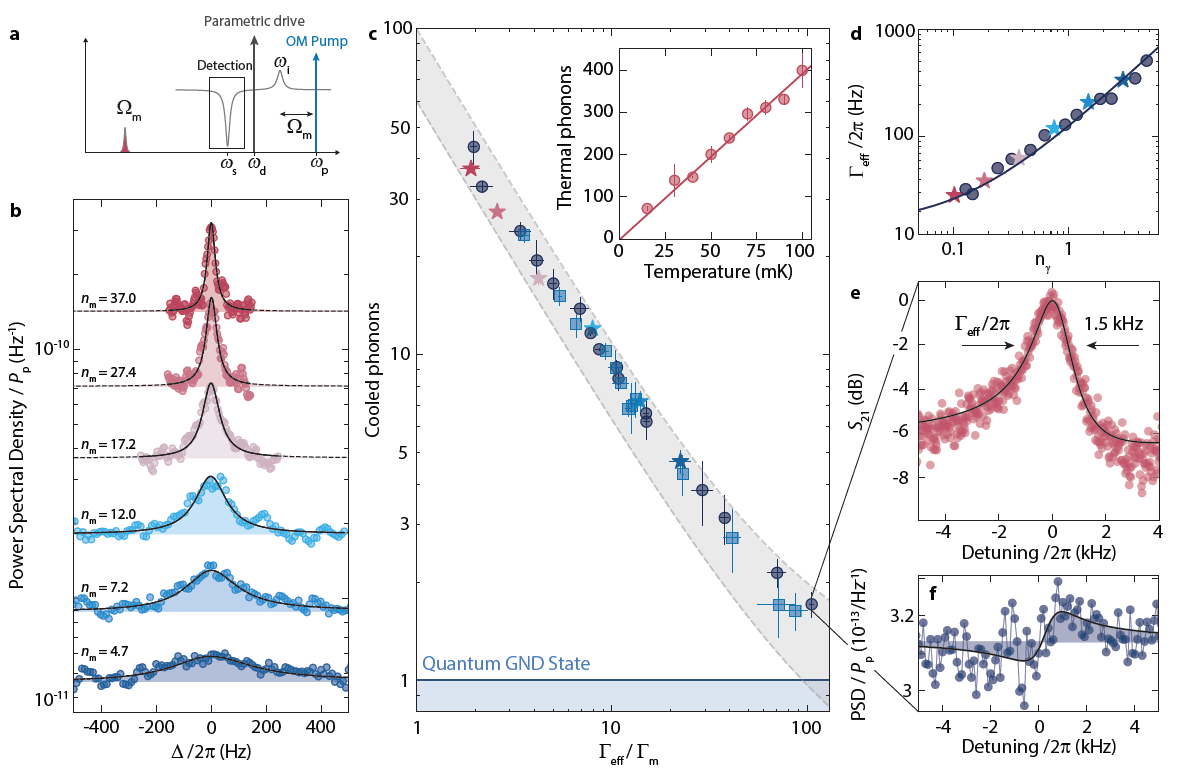}}
	\caption{\textsf{\textbf{Blue-detuned four-wave-cooling of a mechanical oscillator close to its quantum groundstate.} \textbf{a} Schematic representation of the experiment. A parametric drive is used to activate the quasi-mode state and an OM pump is sent to the blue sideband of the idler resonance $\omega_\mathrm{p} \approx \omega_\mathrm{i} + \Omega_\mathrm{m}$. The signal resonance output power spectral density is measured using a spectrum analyzer around $\omega = \omega_\mathrm{p} + 2\Omega_\mathrm{dp} + \Omega_\mathrm{m} \approx \omega_\mathrm{s}$. \textbf{b} Power spectral densities normalized to the optomechanical pump input power $P_\mathrm{p}$ for various pump powers. Frequency axis is given with respect to $\omega = \omega_\mathrm{p} + 2\Omega_\mathrm{dp} + \Omega_\mathrm{m}$. With increasing pump power, the linewidth of the upconverted mechanical noise spectrum is increasing, indicating four-wave dynamical backaction damping. Simultaneously, the area of the normalized signal decreases, indicating cooling of the mode. From fits (lines and shaded areas) to the data (points), we determine the resulting phonon occupation $n_\mathrm{m}$. In \textbf{c} we show the cooled phonon number vs $\Gamma_\mathrm{eff}/\Gamma_\mathrm{m}$ in a collection of several different datasets. Intracavity drive photon numbers vary between different points in the range $40 < n_\mathrm{d} < 100$. Circles correspond to data from measurements at operation point I and squares to data from operation point II. Stars show the points that correspond to the data shown in \textbf{b}, taken at operation point I. All measurements have been taken at $B_\parallel = 25\,$mT. Inset shows the result of a thermal calibration measurement, indicating that the mechanical oscillator mode equilibrates with the fridge base temperature and the residual thermal occupation at $T_\mathrm{b} = 15\,$mK is $n_\mathrm{m}^\mathrm{th} \approx 70-90$. Dashed lines and shaded area display the theoretically calculated range of four-wave-cooled phonon occupation, taking into account a possible range of $60\leq n_\mathrm{m}^\mathrm{th} \leq 100$ and $45 \leq n_\mathrm{d} \leq 90$. Parametric amplification of cavity quantum noise limits the minimally achievable phonon occupation in our parameter regime to $n_\mathrm{m}^\mathrm{lim} \sim 0.6$. For the highest powers, we exceed this theoretical limit by only a factor $\sim 3$. \textbf{d} shows the effective effective mechanical linewidth vs intracavity sideband photon number $n_\gamma = |\gamma_-|^2 + |\gamma_+|^2$ for points from \textbf{c}, which have nearly constant $n_\mathrm{d} \approx 60 \pm 10$, demonstrating that we achieve significant cooling with a small number of photons. Line corresponds to theory with $\Gamma_\mathrm{m} = 2\pi\cdot 15\,$Hz \textbf{e} shows an OMIT scan at the point of largest cooling with an effective linewidth $\Gamma_\mathrm{eff} \approx 2\pi\cdot 1.5\,$kHz, which corresponds to an effective four-wave cooperativity of $\mathcal{C}_\mathrm{fw} \gtrsim 100$. \textbf{f} shows the corresponding power spectral density in units of quanta with noise squashing due to a small, but finite effective temperature of the cavity by amplified quantum noise. Error bars in \textbf{c} consider uncertainties in the fitting procedure and in the bare mechanical linewidth, for details see Supplementary Note~14.}}
	\label{fig:FWCooling}
\end{figure*}

To characterize the mechanical mode temperature, we detect the upconverted thermal displacement fluctuations in the signal resonance output field with a spectrum analyzer.
For this measurement, the SQUID cavity in the quasi-mode state is pumped with an optomechanical tone on the blue sideband of the idler mode.
Using a probe tone, we then measure the signal mode response $S_{21}$ in a wide frequency range and the OMIT response in a narrow range.
Finally, we detect the output spectrum in the same frequency window where the OMIT is observed.
A collection of spectra for varying optomechanical pump power $P_ \mathrm{p}$ is presented in Fig.~\ref{fig:FWCooling}\textbf{b}.
From a careful analysis of the combined data sets, cf. Supplementary Notes 8-13, the equilibrium phonon occupation of the mechanical oscillator as well as the phonon occupation resulting from four-wave-cooling can be inferred.
The mechanical oscillator is well thermalized to the mixing chamber base temperature and its residual phonon occupation at the lowest operation temperature $T_\mathrm{b} = 15\,$mK is about $n_\mathrm{m}^\mathrm{th} \approx 70-90$ phonons.
With increasing optical damping caused by the blue-detuned pump tone, we observe a corresponding reduction of the initial thermal occupation and the cooling factor is determined by $\Gamma_\mathrm{opt}$, very similar to usual optomechanical sideband-cooling.
The observed four-wave cooling is also very robust with respect to pump and drive strengths and we achieve at both flux bias operation points a final four-wave-cooled occupation extremely close to the quantum groundstate $n_\mathrm{m} \sim 1.6$.
Due to the high single-photon coupling rates, it requires only a small amount of effective sideband photons $n_\gamma = |\gamma_-|^2 + |\gamma_+|^2 \lesssim 10$ to achieve these low occupations.
The fact that we use strongly driven Kerr quasi-modes as cold bath, however, modifies the minimally achievable occupation.
Due to Josephson parametric amplification of quantum noise in the quasi-mode state, the cavity will acquire an effective temperature, even if the bare cavity is in the quantum groundstate.
This drive-induced cavity heating defines the cooling limit for the mechanical resonator.
In the state we are operating here, the Josephson gain is small and the effective thermal occupation of the cavity is still considerably below $1$.
We estimate the current cooling limit due to amplified quantum noise to be $\sim 0.6$, where the exact value depends on the drive strength $n_\mathrm{d}$ and on the bias-flux operation point.
With higher bias flux stability the cavity could be stabilized at a point where the Josephson gain is small enough to enable $n_\mathrm{m}^\mathrm{lim}<0.3$.
Achieving the lowest occupation in the current device requires a careful balancing of drive and pump strength and for the highest pump powers, we observe the onset of additional cavity shifts and line broadening, possibly related to drive depletion or higher-order nonlinear effects.
With slightly optimized device parameters regarding $\mathcal{K}$ and $g_0$, we should therefore be able to cool to $n_\mathrm{m} < 1$.
We emphasize though, that the blue-detuned cooling scheme allowed to achieve a significantly lower phonon occupation than signal-mode red-sideband pumping.
With a pump on the red signal-mode sideband, a second cavity bifurcation instability occurs at moderately high pump powers, as the red sideband pump is attracting the cavity, while the blue-detuned pump is repelling it.
The related jump to a high-amplitude state with a different signal resonance frequency, prevents us from cooling below $n_\mathrm{m}^\mathrm{red} \sim 4$.
The corresponding red-sideband cooling data and analysis can be found in Supplementary Note~16.\vspace{10mm}

\noindent\textbf{\textsf{\small DISCUSSION}}
\vspace{3mm}

The results we presented here demonstrate clearly that the young field of flux-mediated optomechanics is quickly advancing towards an exciting and competitive optomechanical platform, which intrinsically allows for novel ways of manipulating mechanical motion.
Our device provides a large single-photon coupling rate of up to $g_0 = 2\pi\cdot 3.6\,$kHz and achieves large cooperativities of up to $\mathcal{C}_\mathrm{fw} > 100$ for small numbers of intracavity photons.
By using strong parametric driving, we show how the intrinsic Josephson-based Kerr nonlinearity can be utilized as a resource for improved sideband-resolution and frequency-stability and for the implementation of a novel four-wave-mixing-based phonon control scheme.
In combination, these properties enabled us to use four-wave-cooling in a Kerr cavity to prepare a MHz mechanical nanobeam resonator close to its quantum groundstate.
Future device improvements can be achieved by reducing the SQUID loop inductance further in order to increase the flux responsivity and the single-photon coupling rate.
One order of magnitude is a feasible goal in this direction, as related platforms have already demonstrated such high responsivities \cite{Zoepfl20, Schmidt20}.
This improvement alone would bring the device to a cooperativity of $10^4$ and to the onset of the strong-coupling regime with $g \sim 2\pi\cdot 150\,$kHz $\sim \kappa/2$.
With increased in-plane fields, up to $\sim1\,$T with e.g. Niobium or granular Aluminum, those numbers could be improved by another order of magnitude.
In the current device, however, the main limiting factor to achieve higher coupling rates and cooling the mechanical oscillator into the groundstate was external flux noise coupling into the SQUID in large in-plane fields.
We suspect that the origin of this flux noise is in the vector magnet leads and the used current sources, respectively, or in parasitic out-of-plane components that lead to flux instabilities, vortex avalanches and microwave-triggered vortex motion in proximity to the SQUID.
Flux noise in the leads and current sources could potentially be reduced by using a superconducting magnet in persistent current mode.
And although our current setup can locally cancel parasitic out-of-plane fields, it cannot do so over the complete chip simultaneously due to the geometry of the small coil.
A global compensation might be necessary, however, to completely avoid any flux instabilities arising from the out-of-plane fields, which can cause flux fluctuations also in large distances from their occurrence.
Using intrinsic Kerr nonlinearities as a resource in optomechanical systems has just begun.
Further interesting directions in Kerr optomechanics might involve intracavity squeezing, intracavity Josephson parametric amplification, intracavity cat-state generation, groundstate cooling in the sideband unresolved regime or enhanced quantum transduction.
Significantly larger Kerr nonlinearites than the ones presented here, implemented in superconducting transmon qubits, have also been discussed recently for mechanical quantum state preparation \cite{Khosla18, Kounalakis19, Kounalakis20}.
Similar schemes investigating and exploiting the Kerr nonlinearity of SQUID circuits could furthermore be implemented naturally in the platform of photon-pressure coupled circuits \cite{Eichler18, Bothner20, Rodrigues20}.\vspace{10mm}

\noindent\textbf{\textsf{\small References}}
 \vspace{10mm}

\noindent\textbf{\textsf{\small Acknowledgements}}
\vspace{3mm}

This research was supported by the Netherlands Organisation for Scientific Research (NWO) in the Innovational Research Incentives Scheme -- VIDI, project 680-47-526.
This project has received funding from the European Research Council (ERC) under the European Union's Horizon 2020 research and innovation programme (grant agreement No 681476 - QOMD) and from the European Union's Horizon 2020 research and innovation programme under grant agreement No 732894 - HOT.
The authors thank Ronald Bode for the construction of the 2D vector magnet, Olaf Benningshof and Raymond Schouten for useful discussions and D. Koelle and R. Kleiner for providing the High-Finesse current source used to power the in-plane magnet.\vspace{10mm}

\noindent\textbf{\textsf{\small Author contributions}}
\vspace{3mm}

All authors conceived the experiment.
DB and ICR designed the device, performed the experiments, and analyzed the data.
ICR fabricated the device.
DB developed the theoretical treatment.
DB and ICR edited the manuscript with input from GAS.
All authors discussed the results and the manuscript.
GAS supervised the project.\vspace{10mm}

\noindent\textbf{\textsf{\small Competing interest}}
\vspace{3mm}

The authors declare no competing interests.

\clearpage

\widetext

\noindent\textbf{\textsf{\Large  Supplementary Material for: Four-wave-cooling to the single phonon level in Kerr optomechanics}}

\normalsize
\vspace{.3cm}

\noindent\textsf{D.~Bothner$^\dagger$, I.C.~Rodrigues$^\dagger$ and G.~A.~Steele}

\vspace{.2cm}
\noindent{$^\dagger$these authors contributed equally}

\renewcommand{\thefigure}{S\arabic{figure}}
\renewcommand{\theequation}{S\arabic{equation}}

\renewcommand{\thesection}{S\arabic{section}}
\renewcommand{\bibnumfmt}[1]{[S#1]}
\renewcommand{\citenumfont}[1]{S#1}

\setcounter{figure}{0}
\setcounter{equation}{0}

\addtocontents{toc}{\protect\setcounter{tocdepth}{1}}

\tableofcontents

\section*{Supplementary Note 1: Device fabrication}
\label{Section:Fab}
Here we present a step-by-step description of the device fabrication.
The individual steps are schematically shown in Supplementary Fig.~\ref{fig:Fab}, where we omitted step 0, the patterning of the electron beam lithography (EBL) alignment markers, as well as the wafer dicing steps and the final device mounting.

\begin{figure}[h]
	\centerline {\includegraphics[trim={3cm 1cm 3cm 1cm},clip=True,scale=0.62]{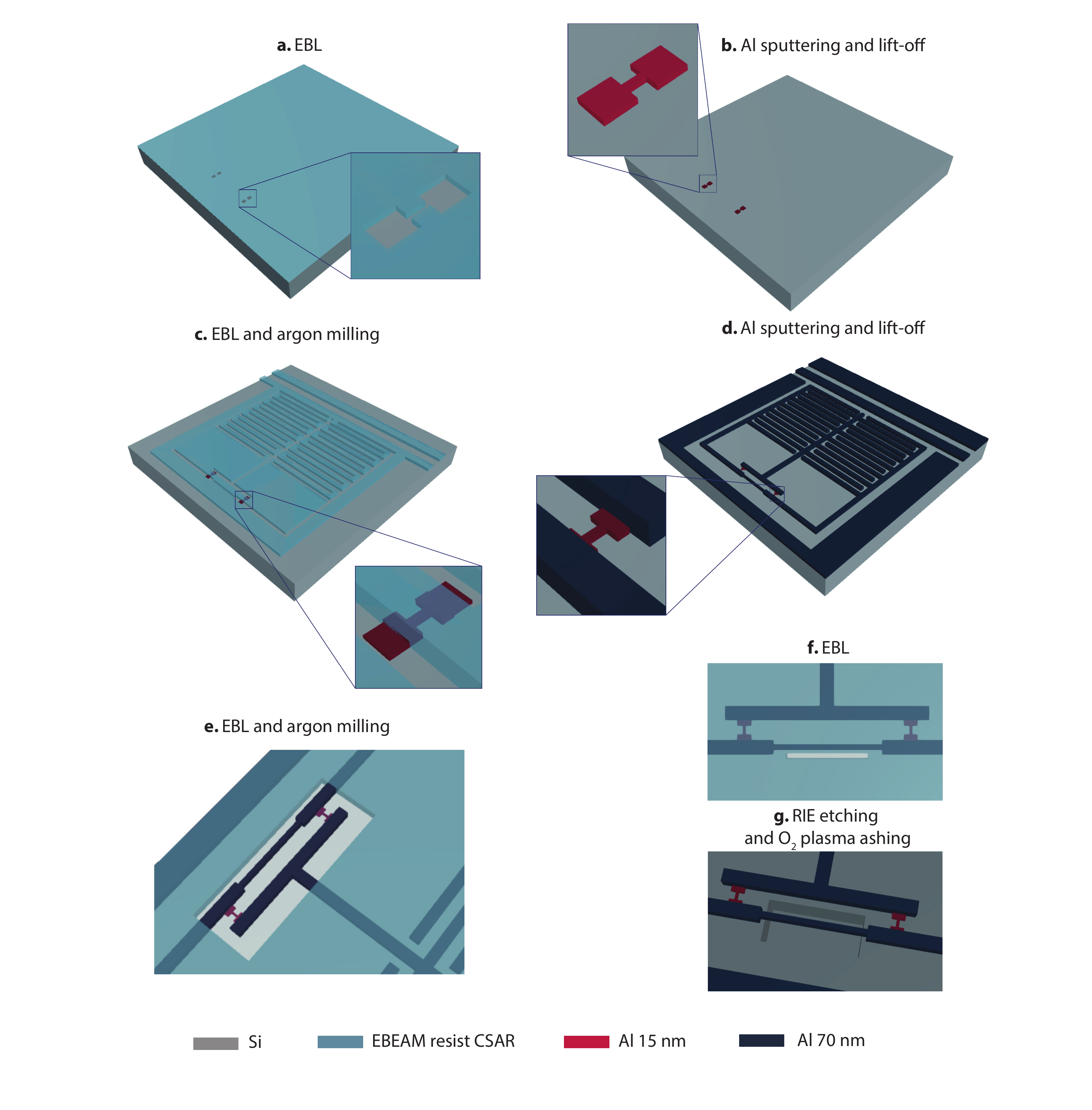}}
	\caption{\textsf{\textbf{Schematic device fabrication.} \textbf{a}, \textbf{b} show the deposition and patterning of the nanobridge junctions and contact pads (Step 1). \textbf{c}, \textbf{d} show the patterning and deposition of the remaining superconducting structures (Step 2). \textbf{e} shows the nanobridge thinning by Argon ion milling on the SQUID (Step 3).  \textbf{f}, \textbf{g} show the window patterning and nanobeam release (Step 5). Dimensions are not to scale. A detailed description of the individual steps is given in the text.}}
	\label{fig:Fab}
\end{figure}

\begin{itemize}

	\item \textbf{Step 0: Marker patterning}
	The fabrication of the device starts by the patterning of alignment markers on top of a 2 inch silicon wafer using electron beam lithography (EBL).
	The marker structures are patterned using a CSAR62.13 resist mask and a sputter deposition of $50\,$nm Molybdenum-Rhenium alloy.
	After undergoing a lift-off process, the only remaining structures on the wafer are the markers.
	The complete 2 inch wafer is then diced into individual $14\times14\,$mm$^2$ chips, which are used individually for the subsequent fabrication steps.
	On each of these fabrication chips, we structure 2 device chips with dimensions of $5\times10\,$mm$^2$, each of which contain one coplanar waveguide microwave feedline and seven quantum interference LC circuits.
	\item \textbf{Step 1: Junctions patterning}
	As first real step of the device fabrication we pattern two nanobridges (the later Josephson junctions) for each LC circuit using CSAR62.09, cf. Supplementary Fig. \ref{fig:Fab}\textbf{a}.
	The two bridges of each pair of nanobridges forming one superconducting quantum interference device (SQUID) are hereby always identical.
	All bridges have a length of $\sim 100\,$nm but vary in width between $30$ and $60\,$nm for different SQUIDs in order to compensate for small variations and uncertainties in final structure size and select the most suitable device during the experiment.
	The nanobridges also have two $700\times1150\,$nm$^2$ large pads for achieving good galvanic contact to the rest of the circuit, which is patterned in fabrication step 2. 
	After the EBL exposure, the sample is developed in Pentylacetate for $60\,$seconds followed by a 1:1 solution of MIBK:IPA (Methyl IsoButyl Ketone:IsoPropyl Alcohol) for another $60\,$seconds and finally rinsed in IPA. 
	Once the resist is developed, the chip is loaded into a sputtering machine where a $15\,$nm think layer of Aluminum ($1\%$ Silicon) is deposited.
	After the deposition, the sample is placed horizontally at the bottom of a glass beaker containing a small amount of room-temperature Anisole and left in an ultrasonic bath for a few minutes.
	During this time, the remaining resist is dissolved and the Aluminum layer sitting on top is lifted off, the result is schematically shown in Supplementary Fig. \ref{fig:Fab}\textbf{b}. 
	\item \textbf{Step 2: Microwave cavity patterning}
	After the junctions are patterned, we once again spin-coat the sample with CSAR62.13 and pattern the SQUID arms together with all the remaining superconducting structures. 
	After the EBL exposure, the sample is developed as for the previous fabrication step and afterwards loaded into a sputtering machine.
	Hereby, the nanobridges themselves are covered and protected by resist, cf. Supplementary Fig.~\ref{fig:Fab}\textbf{c}.
	At this point and prior to the deposition of the second Aluminum layer, an Argon milling process is perfomed in-situ in order eliminate any oxide present on top of the contact pads.
	This measure is necessary to generate good electrical contact between the two layers. 
	After the sputtering process of the second, $70\,$nm thick Aluminum ($1\%$ Silicon) layer, the sample undergoes an ultrasonic lift-off process similar to the one in Step 1, the result is shown schematically in Supplementary Fig.~\ref{fig:Fab}\textbf{d}.
	\item \textbf{Step 3: Nanobridge thinning by Ar ion milling}
	In order to reduce the cross-section and the critical current of the nanobridges even further, we apply a short ion milling step to the SQUID at this point.
	To do so, we pattern and develop another layer of CSAR62.13 on top of the device as described in Steps 1 and 2, which protects the whole chip except for rectangular windows around the SQUIDs themselves, cf. Supplementary Fig.~\ref{fig:Fab}\textbf{e}.
	From test measurements, we observe that if we do not protect the rest of the circuit from the milling in this step, we obtain a significant reduction of the circuit quality factor, which we think might be due to ion implanation into the substrate.
	Note that with the milling parameters we use for this step, we do not get a directional milling, but mainly a narrowing of the nanobridges from the sides.
	This is also the reason why we need the contact pads in the first place.
	If we work with bare nanobridges in Step 1, they are milled away completely during the essential in-situ native oxide removal in Step 2.

	\item \textbf{Step 4: Dicing}
	Right before the final release of the mechanical oscillator, the sample is once again diced to two smaller $5\times10\,$mm$^2$ sized chips in order to fit into the sample mountings and the microwave PCB (Printed Circuit Board).
	The remaining $2\,$mm at each edge of the original $14\times14\,$mm$^2$ large chip is only a margin for the fabrication and is disposed of.
	\item \textbf{Step 5: Mechanical beam release}
	For the final EBL step, a CSAR62.13 resist was once again used as mask and the development of the pattern was done in a similar way as for the first two layers.
	Once the etch mask, consisting of a small window close to the outer side of the SQUID loop (cf. Supplementary Fig.~\ref{fig:Fab}\textbf{f}), is patterned, the sample undergoes an isotropic, reactive ion etching process in SF$_6$ at a sample temperature of $\sim-10\,^{\circ}{\rm C}$ for two minutes \cite{SNorte18}.
	During this time the Silicon substrate under the SQUID arm/the mechanical beam is etched without attacking the aluminum layer forming the cavity and the mechanical beam.
	Once the beam is released, we proceeded with an O$_2$ plasma ashing step in order to remove the remaining resist from the sample.
	At this point the fabrication is completed, the result is shown schematically in Supplementary Fig.~\ref{fig:Fab}\textbf{g}.
	\item \textbf{Step 6: Device mounting}
	After the fabrication, the sample is glued into a microwave printed circuit board (PCB) using GE varnish and wirebonded both to ground and to $50\,\Omega$ connector lines. 
	An optical image of the chip and the PCB, both mounted into a magnet, is shown in Fig.~1 of the main paper.
\end{itemize}

\section*{Supplementary Note 2: Measurement setup}

\begin{figure}[h]
	\centerline {\includegraphics[trim={0cm 4.5cm 0.5cm 1.7cm},clip=True,scale=0.8]{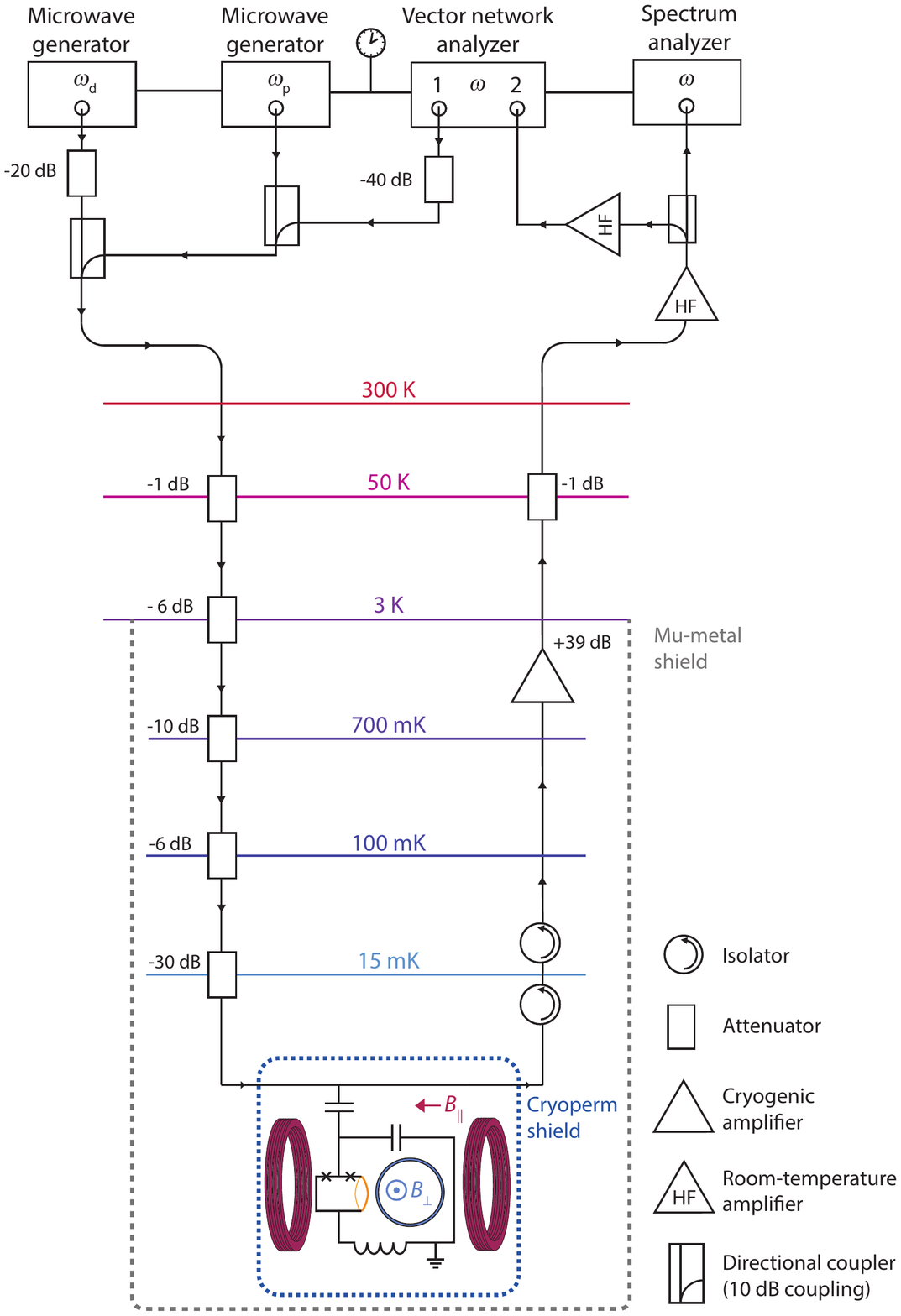}}
	\caption{\textsf{\textbf{Schematic of the measurement setup.} Detailed information is provided in text.}}
	\label{fig:Setup}
\end{figure}

\subsection*{Setup configuration}

The experiments reported in this paper were performed in a dilution refrigerator with a base temperature $T_\mathrm{b} \approx 15\,$mK.
Within the outer vacuum can of the system, a mu-metal shield is installed to provide basic magnetic shielding for the whole sample space from the $3\,$K plate to the mK plate.
A schematic diagram of the experimental setup and of the external measurement configuration used in the reported experiments can be seen in Supplementary Fig.~\ref{fig:Setup}.
The PCB, onto which the fabricated sample was glued and wirebonded, is mounted into a 2D vector magnet casing and connected to two coaxial lines.
The complete configuration including the vector magnet is placed in a magnetic cryoperm shield. 
The vector magnet combines two distinct superconducting magnets, a small one for the generation of an out-of-plane field and a larger split coil for the in-plane field.
The coils are used to independently generate a magnetic field in the two different directions by providing a DC current to the corresponding coil.
A more detailed information about the design and setup of the vector magnet is provided in the following subsection.
Since the optomechanical circuit that we present in this paper was designed in a side-coupled geometry, the input and output signals were sent/received through separate coaxial lines in order to measure the transmission spectrum of the feedline to which the system is coupled.
The input line is heavily attenuated in order to balance the thermal radiation from the line to the base temperature of the fridge and the output line contains a cryogenic HEMT (High-Electron-Mobility Transistor) amplifier working in a range from 4 to 8 GHz and two isolators to block the thermal radiation from the HEMT to reach the sample.
Outside of the refrigerator, we used a single measurement scheme for all the different experiments.
The VNA was used to measure the response spectrum $S_\textrm{21}$ of the electromechanical system, one microwave generator sends a coherent signal at $\omega_\textrm{d}$ as parametric drive for the SQUID cavity and the second microwave generator sends a tone at $\omega_\textrm{p}$ as optomechanical pump for the parametrically driven cavity.
Finally, a spectrum analyzer was used to record the output power spectrum around the cavity resonance. 
For all experiments, the microwave sources and vector network analyzers (VNA) as well as the spectrum analyzer used a single reference clock of one of the devices.

\subsection*{Vector magnet design}

Figure~1 of the main paper shows photographs of the sample mounted on the PCB and fixed in the vector magnet bobbin.
The two large parallel coils on each side of the sample are wound from a single wire (niobium-titanium in copper-nickel matrix) and in the same orientation and therefore form a Helmholtz-like split coil (the distance between the coils is slightly larger than their effective radius), which creates a nearly homogeneous in-plane magnetic field at the location of the device.
At room temperature the coil has a resistance of $R_\parallel \approx 6\,\textrm{k}\Omega$, which approximately corresponds to 2000 windings of superconducting wire on each side.
From the coil geometry and the number of windings, we estimate the current-to-field conversion factor to be $70\,$mT/A.
On the backside of the sample/PCB platform within the magnet bobbin is a second small coil mounted for providing the out-of-plane magnetic field used to tune the SQUID flux bias point, cf. main paper Fig.~1.
This out-of-plane coil can also be used to compensate for a parasitic out-of-plane component of the in-plane field due to misalignments of the sample/PCB with respect to the in-plane field axis (estimated to be around $2^\circ-3^\circ$ from the SQUID flux response).
For in-plane fields $B_\parallel \lesssim 25\,$mT, however, the compensation is not yet critical. 
For larger in-plane fields, vortices start to penetrate the film and there is a dramatic reduction in the cavity quality factor observable.
The room-temperature resistance of the out-of-plane coil is $R_\perp \approx 120\,\Omega$ which corresponds to approximately 400 turns of superconducting wire and to a conversion factor of $1\,$mT/A.
The superconducting wires leading to each of the coils from the $3\,$K plate are twisted in pairs, in order to reduced the amount of captured flux noise.
Furthermore, since the critical temperature of the wire is about $\sim 12\,$K, the wires can go unbroken until the $3\,$K stage.
Above this plate, the wires are no longer superconducting and therefore a transition to normal conducting wires is required.
For this, we connected each of the superconducting in-plane coil wires to 9 wires of a 24-line copper loom provided by Bluefors and each of the out-of-plane coil wires to 3 wires of the loom.
From the $3\,$K stage until room temperature the current flows in parallel through the respective loom wires, decreasing the additional heat load on the plate.
With this approach we are able to send $I_\parallel \sim 0.5\,$A through the in-plane coil without any considerable heat added to any of the plates and maintaining the fridge base temperature.
At room temperature we are left with 4 cables, two for each coil, which are used with individual directed current (DC) sources to independently generate the magnetic fields.
\section*{Supplementary Note 3: Power calibration}
In order to estimate the input power on the on-chip feedline of the device, we use the thermal noise of the HEMT (High-Electron-Mobility Transistor) amplifier as calibration method.
The cryogenic HEMT amplifier thermal noise power is given by
\begin{equation}
P_\mathrm{HEMT} = 10\,\log\left(\frac{k_\mathrm{B} T_{\mathrm{HEMT}}}{1\,\textrm{mW}}\right) + 10\,\log\left(\frac{\Delta f}{\textrm{Hz}}\right)
\label{eq:HEMT}
\end{equation}
where $k_\mathrm{B}$ is the Boltzmann constant, $T_{\mathrm{HEMT}}$ is the noise temperature of the amplifier, which, according to the specification datasheet, is approximately $2\,$K, and $\Delta f = 2000\,$Hz is the measurement IF bandwidth.
The calculated noise power is $P_\mathrm{HEMT} = -162.6\,$dBm, or as noise RMS voltage $\Delta V = 1.66\,$nV.
Taking into account the room temperature attenuators of $60\,$dB as well as additional $3\,$dB of room-temperature cable losses between the VNA output and the directional couplers for the pump tones and assuming an attenuation between the sample and the HEMT of $2\,$dB we extract a frequency-dependent input attenuation for the pump tones as shown in Supplementary Fig.~\ref{fig:NoiseCal}.
In addition and for confirmation, we perform a fixed-frequency measurement of the signal-to-noise ratio using the pump signal generator itself and a spectrum analyzer for selected frequency points around 5.22 and 5.17 GHz.
We observe agreement between the two methods better than $0.5\,$dB.

\begin{figure}[h]
	\centerline {\includegraphics[trim={0cm 0cm 0cm 0cm},clip=True,scale=0.6]{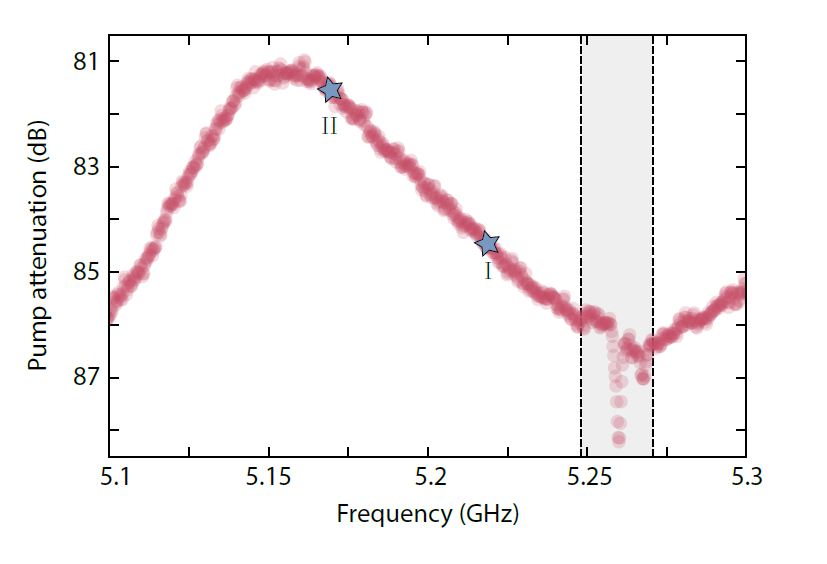}}
	\caption{\textsf{\textbf{Estimation of the frequency-dependent input line attenuation for the pump tone.} The shown data are obtained by measuring $501$ traces in the shown frequency range using the vector network analyzer shwon in Supplementary Fig.~\ref{fig:Setup}. For each frequency point, we determine from the $501$ traces the signal-to-noise ratio and with the assumption of a frequency-independent HEMT noise temperature and $2\,$dB losses between the sample and the HEMT, we get the input line attenuation as plotted. The gray area shows where the cavity was during the calibration. Due to its presence, the attenuation in this range can not be considered a reliable value. Our experiments, however, mainly take place around $5.22\,$GHz and $5.17\,$GHz (labelled with I and II, respectively) and therefore the presence of the cavity at around $5.26\,$GHz does not lead to any calibration problems. We also note, that we observe almost identical amplitude oscillations in the transmitted signal, indicating that we are indeed dealing with strong cable resonances.}}
	\label{fig:NoiseCal}
\end{figure}

\section*{Supplementary Note 4: The SQUID Cavity}

\subsection*{Circuit model}

\begin{figure}[h]
	\centerline {\includegraphics[trim={0cm 11.5cm 0cm 0.5cm},clip=True,scale=0.8]{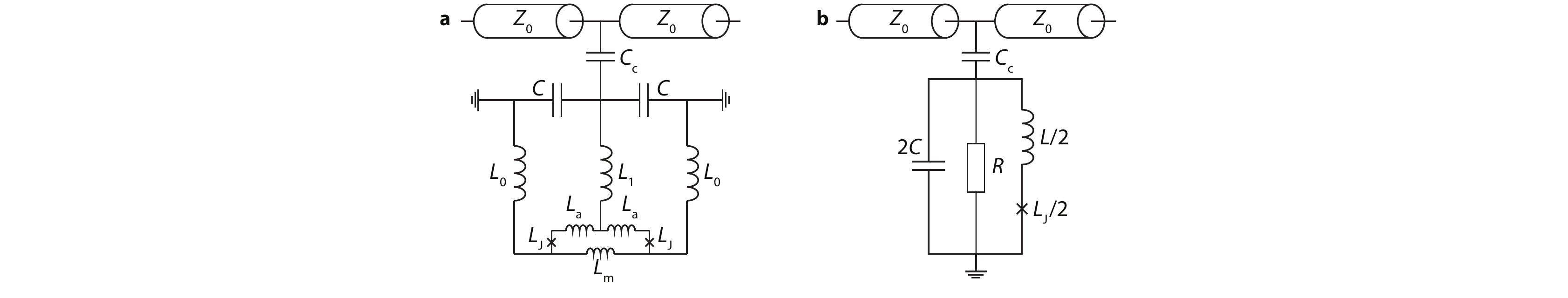}}
	\caption{\textsf{\textbf{The circuit model.} \textbf{a} Circuit equivalent of the SQUID cavity shwon in main paper Fig.~1. Each $C$ corresponds to one interdigitated capacitor (IDC) and $C_\mathrm{c}$ to the coupling capacitance to the feedline with characteristic impedance $Z_0$. The SQUID loop inductance $L_\mathrm{loop} = 2L_\mathrm{a} + L_\mathrm{m}$ has contributions from the non-released arms $L_\mathrm{a}$ and from the loop part that acts as mechanical oscillator $L_\mathrm{m}$. The remaining linear inductances $L_1$ and $L_0$ correspond to the inductances of the circuit wires and IDCs and each nanobridge Josephson junction is described by a Josephson inductance $L_\mathrm{J}$. \textbf{b} shows a simplified circuit model, where all linear contributions to the inductance are expressed through $L/2$, the nonlinear Josephson inductance is in good approximation given by $L_\mathrm{J}/2$ and the two IDCs are contained in the single capacitance $2C$. All internal losses of the circuit are captured by the resistor $R$. Another possible version for the circuit equivalent is shown in main paper Fig.~1 where all linear contributions to the inductance are split symmetrically between the two inductors $L$.}}
	\label{fig:Circuit}
\end{figure}

A simplified circuit equivalent of the SQUID cavity used in this experiment is shown in Supplementary Fig.~\ref{fig:Circuit}\textbf{a}.
We model it as a simple parallel $RLC$ circuit capacitively coupled by a coupling capacitance $C_\mathrm{c}$ to a microwave feedline with characteristic impedance $Z_0$ as shown in \textbf{b}, cf. also Ref.~\cite{SRodrigues19}.
The resistance in this model captures all intracavity losses.
The resonance frequency, external and internal linewidth of the circuit shown in $\mathrm{b}$ are given by
\begin{equation}
\omega_0 = \frac{1}{\sqrt{\left(2C + C_\mathrm{c}\right)\left(\frac{L}{2} + \frac{L_\mathrm{J}}{2}\right)}}, ~~~ \kappa_\mathrm{i} = \frac{1}{R(2C+C_\mathrm{c})}, ~~~ \kappa_\mathrm{e} = \frac{\omega_0^2 C_\mathrm{c}^2 Z_0}{2(2C+C_\mathrm{c})}
\end{equation}
respectively.
Each of the two physical capacitors in the main circuit, cf. main paper Fig. 1, is an interdigitated capacitor (IDC) with $N = 148$ fingers, each $100\,\mu$m long and $1\,\mu$m wide.
With the gap between two fingers of also $1\,\mu$m and the relative permittivity of the Silicon substrate $\epsilon_\mathrm{r} = 11.7$, we obtain for each of the IDCs $C \approx 824\,$fF using the analytical expressions provided in Ref.~\cite{Igreja04}.
The total capacitance is then approximately given by $C_\mathrm{tot} = 2C+C_\mathrm{c} \approx 1.65\,$pF, where we included also the (mostly negligible) coupling capacitance $C_\mathrm{c} \approx 6.5\,$fF.
The value for $C_\mathrm{c}$ was obtained via the external cavity linewidth of $\kappa_\mathrm{e} \approx 2\pi\cdot110\,$kHz, the feedline characteristic impedance $Z_0 = 50\,\Omega$ and the resonance frequency $\omega_0 = 2\pi\cdot5.267\,$GHz.
Using the resonance frequency, we can also estimate the total inductance as $L_\mathrm{tot} = \frac{1}{\omega_0^2 C_\mathrm{tot}} \approx 552\,$pH.

\subsection*{Response function and fitting routine}

\begin{figure}[h]
	\centerline {\includegraphics[trim={0cm 0cm 0.0cm 0cm},clip=True,scale=0.7]{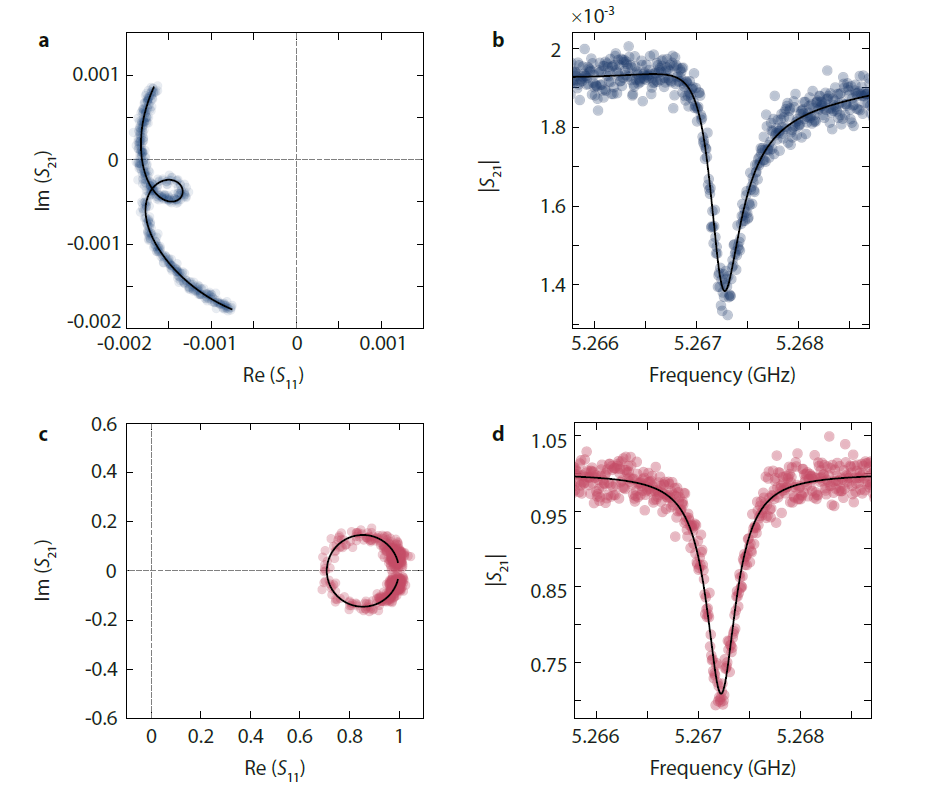}}
	\caption{\textsf{\textbf{Cavity response fitting and background-correction.} Raw data for the cavity response $S_{21}$ in the complex plane and in linear magnitude are shown in $\textbf{a}$ and $\textbf{b}$ as circles. Black line is a full fit including a phase rotation factor and a complex, frequency-dependent background. In $\textbf{c}$ and \textbf{d} the corresponding data after a background correction and a corresponding phase factor rotation are shown as circles, the corresponding background-corrected fit curves are shown as lines. Data correspond to an in-plane field $B_\parallel = 25\,$mT and SQUID bias flux $\Phi_\mathrm{b} = 0$. Dashed lines in \textbf{a} and \textbf{c} show the real and imaginary axes, respectively.}}
	\label{fig:CircleFits}
\end{figure}

In the linear regime, a capacitively side-coupled LC circuit is described by the $S_{21}$ response function
\begin{equation}
S_{21}^\mathrm{ideal} = 1- \frac{\kappa_\mathrm{e}}{\kappa_\mathrm{i}+\kappa_\mathrm{e}+2i\Delta}
\label{eq:Responsefunc}
\end{equation}
with detuning of the probe tone from the resonance frequency
\begin{eqnarray}
\Delta = \omega - \omega_0
\end{eqnarray}
and the internal and external linewidths $\kappa_\mathrm{i}$ and $\kappa_\mathrm{e}$, respectively.
Implicitly, we assume symmetric coupling to the left and right feedline part in this relation.
Due to considerable cable resonances in our setup, however, this assumption might be not strictly valid.
We also observe, that for a consistent modeling of all our datasets, small adjustments to $\kappa_\mathrm{e}$ in different experimental situations are leading to higher agreement between data and theory.
The different microwave components in the setup (cables, attenuators, directional couplers, isolators etc) affect the ideal cavity transmission spectrum by amplitude and phase modulations, and we consider a modification in the response function by introducing a frequency-dependent complex-valued microwave background.
The modified cavity response is written as
\begin{eqnarray}
S_{21} = (\alpha_0 + \alpha_1\omega)\left(1-\frac{\kappa_ee^{i\theta}}{\kappa_i+\kappa_e+2i\Delta}\right)e^{i(\beta_1\omega + \beta_0)}
\label{eqn:fitS21}
\end{eqnarray}
where we consider a frequency-dependent complex background
\begin{eqnarray}
S_{21}^\mathrm{bg} = (\alpha_0 + \alpha_1\omega)e^{i(\beta_1\omega + \beta_0)}
\label{eqn:Fit_BG}
\end{eqnarray}
and an additional, possible interference rotation of the resonance circle around its anchor point with the phase factor $e^{i\theta}$.
In our fitting routine the background is extracted by first excluding the cavity resonance from the response and fitting the remaining data with Eq.~(\ref{eqn:Fit_BG}).
After complex division of the data with the background model, the remaining cavity response is fitted independently.
As final step the original data are fitted with the full function for $S_{21}$ including the background again using the obtained fit values from the first two independent fits as starting values for the full fit.
From the final fit, we remove the background of the full dataset by complex division for the resonance data shown this paper.
Also, we correct for the additional rotation factor $e^{i\theta}$.
In Supplementary Fig.~\ref{fig:CircleFits}, we show an exemplary fit of the cavity response around resonance as raw data and as background-corrected data in both, the complex plane and in the magnitude of $S_{21}$.
From the fit to the data, taken at $B_\parallel = 25\,$mT and $B_\perp = 0$ (the sweetspot), we obtain $\omega_0 = 2\pi\cdot 5.2672\,$GHz, $\kappa_\mathrm{i} = 2\pi\cdot 269\,$kHz and $\kappa_\mathrm{e} \approx 2\pi\cdot 111\,$kHz.

\subsection*{The SQUID Josephson inductance}

The total flux $\Phi$ in a superconducting quantum interference device (SQUID) with non-negligible loop inductance $L_\mathrm{loop}$ is given by
\begin{equation}
\frac{\Phi}{\Phi_0} = \frac{\Phi_\mathrm{b}}{\Phi_0} + L_\mathrm{loop}J
\end{equation}
where $\Phi_\mathrm{b}$ is the bias flux by external magnetic fields, $J$ is the screening current circulating in the SQUID loop and $\Phi_0 = \frac{h}{2e} = 2.07\cdot10^{-15}\,$Tm$^2$ is the flux quantum.
Note that $L_\mathrm{loop}$ contains both, the geometric and the kinetic inductance contribution to the inductance of the SQUID loop.
In the absence of a bias current and for identical Josephson junctions with a sinusoidal current-phase relation, the circulating current is given by
\begin{equation}
J = -I_\mathrm{c}\sin{\left(\pi\frac{\Phi}{\Phi_0}\right)}
\end{equation}
with the zero-flux-bias of a single junction $I_\mathrm{c}$.
Using the screening parameter $\beta_L = \frac{2L_\mathrm{loop}I_\mathrm{c}}{\Phi_0}$, the relation for the total flux can be written as
\begin{equation}
\frac{\Phi}{\Phi_0} = \frac{\Phi_\mathrm{b}}{\Phi_0} - \frac{\beta_L}{2}\sin{\left(\pi\frac{\Phi}{\Phi_0}\right)}.
\label{eqn:conversion}
\end{equation}
We use this equation to numerically calculate the total flux in the SQUID for a given external flux.
With the total flux in the SQUID known, the Josephson inductance of a single junction
\begin{equation}
L_\mathrm{J}(\Phi) = \frac{\Phi_0}{2\pi I_\mathrm{c}\cos{\left(\pi\frac{\Phi}{\Phi_0}\right)}}
\end{equation}
and the total Josephson inductance of the SQUID
\begin{equation}
L_\mathrm{S}(\Phi) = \frac{L_\mathrm{J}(\Phi)}{2}
\end{equation}
can be determined.
\subsection*{Cavity field dependence}
Using the flux-dependence of the SQUID Josephson inductance and our simplified circuit model, the resonance frequency of the cavity as function of the perpendicular bias flux $\Phi_\perp$ can be written as
\begin{equation}
\omega_0(\Phi_\perp) = \frac{\omega_0(0)}{\sqrt{\Lambda + \frac{1-\Lambda}{\cos{\left(\pi\frac{\tilde{\Phi}}{\Phi_0}\right)}}}}
\label{eqn:arch}
\end{equation}
with the linear inductance participation ratio
\begin{equation}
\Lambda = \frac{L}{L+L_\mathrm{J0}}
\end{equation}
and the total flux in the SQUID
\begin{equation}
\tilde{\Phi} = \Phi\big|_{\Phi_\mathrm{b} = \Phi_\perp}.
\end{equation}
The zero-bias junction inductance is hereby given as $L_\mathrm{J0} = L_\mathrm{J}(\Phi = 0)$.

\begin{figure}[h]
	\centerline {\includegraphics[trim={0cm 12.5cm 0cm 3cm},clip=True,scale=0.65]{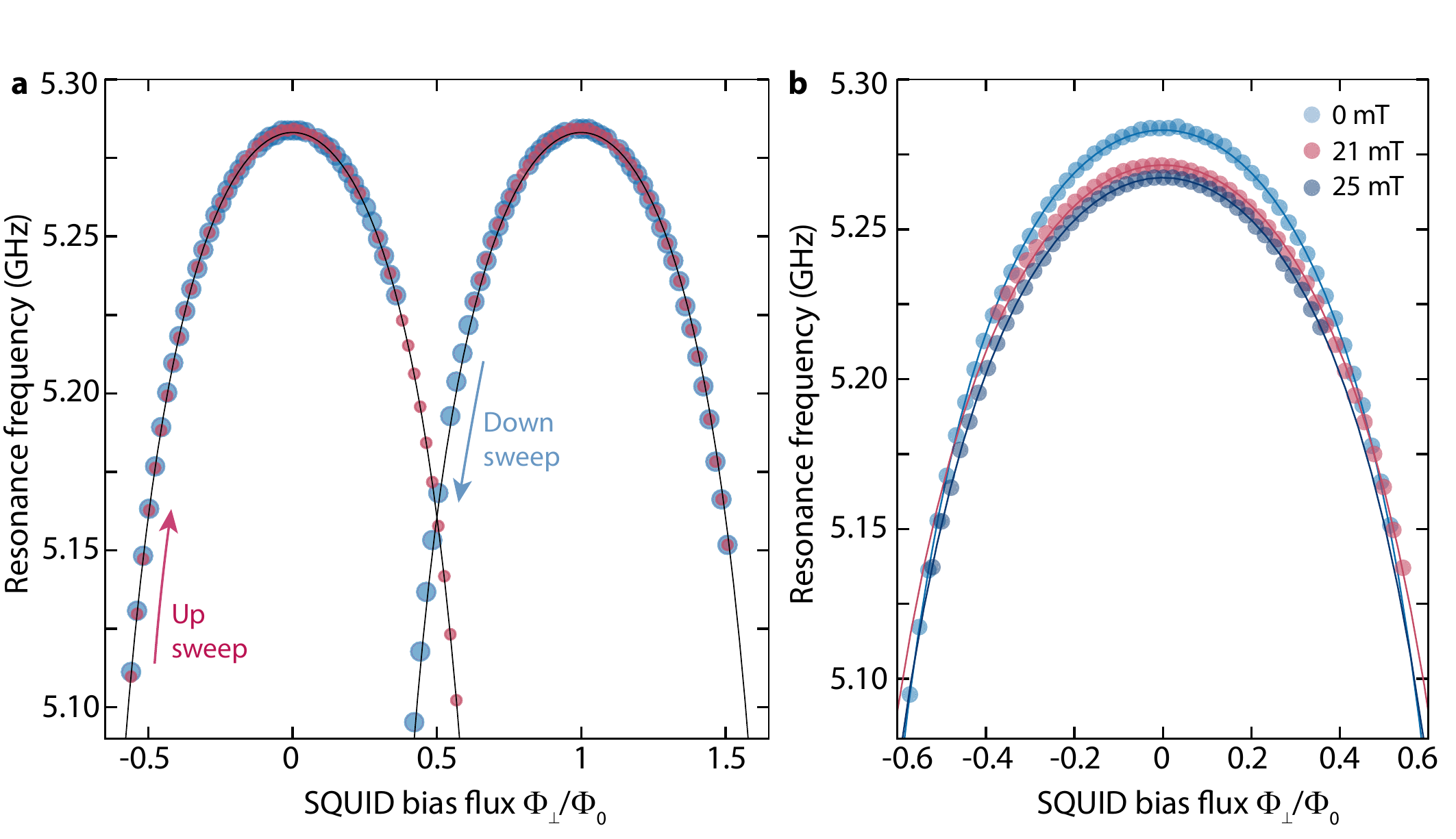}}
	\caption{\textsf{\textbf{Bias flux axis calibration and bias flux arch fitting.} In \textbf{a}, the SQUID cavity resonance frequancy vs flux bias $\Phi_\perp$ is shown for $B_\parallel = 0$. Red circles correspond to the resonance frequencies obtained during a flux upsweep, blue larger circles are data obtained during a flux downsweep. The hysteretic flux jumps around $\Phi_\perp/\Phi_0 \sim 0.5$ indicate a non-negligible loop inductance of the SQUID \cite{LevensonFalk11, Pogorzalek17}. The distance between the two shown archs corresponds to one flux quantum $\Phi_0$ and allows for a calibration of the flux axis. Lines correspond to fits using Eq.~(\ref{eqn:arch}) in combination with Eq.~(\ref{eqn:conversion}). \textbf{b}, a single arch for three different in-plane magnetic fields as labelled in the legend. With increasing $B_\parallel$, the sweetspot resonance frequency $\omega_0(\Phi_\perp = 0)$ slightly decreases and the width of the arch increases, indicating an increase in SQUID screening parameter $\beta_L$. Circles are data, lines are fits. Fit parameters are given and discussed in the text.}}
	\label{fig:Archs}
\end{figure}

The first experimental step to fit the flux-dependence of the cavity resonance frequency and to determine Josephson inductance $L_\mathrm{J}$ and screening parameter $\beta_L$ is a calibration of the bias flux axis and to find the current-to-flux conversion for the small coil generating $\Phi_\perp$, respectively.
Supplementary Fig.~\ref{fig:Archs}\textbf{a} shows as circles the experimentally obtained resonance frequencies at $B_\parallel = 0$ for a sweep of the bias flux $\Phi_\perp$.
The dataset combines the data points obtained during a bias flux upsweep and a downsweep.
This is necessary as the SQUID has a non-negligible loop inductance, which leads to a hysteretic flux response \cite{LevensonFalk11, Pogorzalek17, SRodrigues19}.
The distance between two neighboring flux archs corresponds to one flux quantum $\Phi_0$ and via this procedure the current-to-flux conversion is obtained.
Subsequently, the flux-dependence of $\omega_0$ can be fitted using Eqs.~(\ref{eqn:arch}) and (\ref{eqn:conversion}).
From the fits, we obtain the zero-bias junction critical current $I_\mathrm{c}$ and the screening parameter $\beta_L$, the corresponding fit curves are shown as lines in Supplementary Fig.~\ref{fig:Archs}\textbf{a}.

From the fit at zero in-plane field we get $I_\mathrm{c} \approx 2.6\,\mu$A and a screening parameter $\beta_L \approx 0.7$.
Using $L_\mathrm{J0} = \frac{\Phi_0}{2\pi I_\mathrm{c}}$, we get for the inductance of a single Josephson junction $L_\mathrm{J0} \approx 127\,$pH, which corresponds to a linear inductance participation ratio $\Lambda \approx 0.89$ and a total SQUID loop inductance $L_\mathrm{loop} \approx 278\,$pH.
Enabling the optomechanical coupling between the nanobeam and the SQUID cavity requires an additional in-plane magnetic transduction field, and therefore we also record the resonance frequency flux-dependence at the in-plane fields of $B_\parallel = 21\,$mT and $B_\parallel = 25\,$mT, where we operate the device for the optomechanical experiments. 
The result is shown in Supplementary Fig.~\ref{fig:Archs}\textbf{b} as circles.
From the data, we observe a small decrease of the sweetspot resonance frequency with increasing $B_\parallel$. 
In addition, we observe a slight widening of the flux arch with increasing $B_\parallel$, indicating a nonlinear increase of the kinetic contribution to the SQUID loop inductance and a consequently increased $\beta_L$.
From the fits, we get for both in-plane fields a slightly reduced critical junction current $I_\mathrm{c\parallel} \approx 2.2\,\mu$A and slightly increased screening parameters $\beta_{L, 21} \approx 0.79$ and $\beta_{L, 25} \approx 0.82$.
These value correspond to $\Lambda_\parallel \approx 0.865$, $L_\mathrm{loop, 21} \approx 371\,$pH and $L_\mathrm{loop, 25} \approx 385\,$pH.
We observe that the loop inductance seems to increase by more than both, the Josephson inductance and the linear circuit inductance due to the in-plane field.
Our suspicion is that this effect is caused by a modification of the nanobridge current-phase relation in the in-plane field, but for a final conclusion more experiments would have to be conducted.
For the optomechanical multi-photon interaction two more quantities of the SQUID cavity and their flux dependence are highly important.
The first is the flux responsivity $\mathcal{F} = \partial\omega_0/\partial\Phi_\mathrm{b}$, i.e., the change of resonance frequency with change of bias flux through the SQUID loop.
It is directly proportional to the optomechanical single-photon coupling rate $g_0$, cf. Supplementary Note~5.
The responsivity is identical to the slope of the flux tuning curve shown in Supplementary Fig.~\ref{fig:Archs}\textbf{b} and the numerically obtained results for both, experimental data and the fit curve, are shown in Supplementary Fig.~\ref{fig:Deriv}\textbf{a}.
The bias-flux operation points relevant for this paper are marked with a dotted and dashed line, respectively, and labelled as "I" and "II".
The corresponding flux responsivities are $\mathcal{F}_\mathrm{I} \approx 2\pi\cdot 300\,$MHz$/\Phi_0$ and $\mathcal{F}_\mathrm{II} \approx 2\pi\cdot520\,$MHz$/\Phi_0$, respectively, and nearly identical to each other for the two chosen in-plane fields.
\begin{figure}[h]
	\centerline {\includegraphics[trim={0.5cm 12.5cm 0.5cm 3cm},clip=True,scale=0.68]{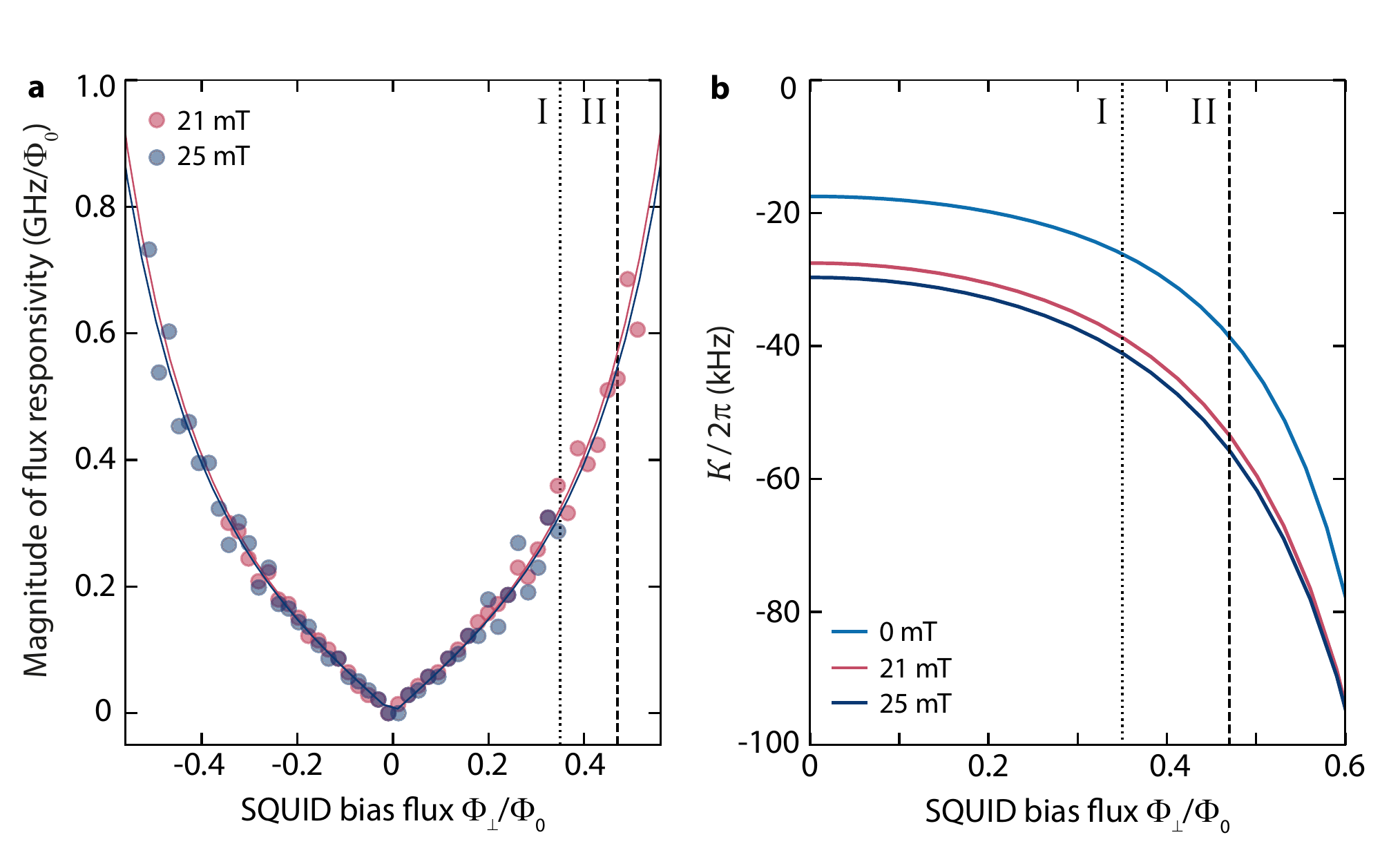}}
	\caption{\textsf{\textbf{Flux responsivity and Kerr anharmonicity at the device operation points.} In \textbf{a}, we plot the numerically obtained magnitude of the flux responsivity $\mathcal{F} = \partial\omega_0/\partial\Phi_\perp$ vs bias flux $\Phi_\perp$ for non-vanishing magnetic in-plane fields. \textbf{b} shows the Kerr anharmonicity vs bias flux for $B_\parallel = 0$, $B_\parallel = 21\,$mT and $B_\parallel = 25\,$mT. The two operation points relevant for this paper are marked by vertical dotted and dashed lines and labelled with I and II, respectively.}}
	\label{fig:Deriv}
\end{figure}
The second important quantity is the Kerr anharmonicity related to the nonlinear Josephson inductance of the SQUID.
It is given by
\begin{equation}
\mathcal{K}(\Phi_\perp) = -\frac{e^2}{2\hbar C_\mathrm{tot}}\left(\frac{L_\mathrm{S}(\Phi_\perp)}{L_\mathrm{tot}(\Phi_\perp)}\right)^3
\end{equation}
and depends in addition on the in-plane field via the in-plane dependence of the nanobridge critical current or Josephson inductance, respectively.
The result of this calculation, based on the flux arch fits of Supplementary Fig.~\ref{fig:Archs}\textbf{b} is shown in Supplementary Fig.~\ref{fig:Deriv}\textbf{b}.
The dependence of the anharmonicity on flux bias $\Phi_\perp$ shows a very similar trend for all in-plane fields with different starting values at the sweetspot $\Phi_\perp = 0$.
The completely unbiased cavity has $\mathcal{K} \approx -2\pi\cdot 17.5\,$kHz.
For an in-plane field of $B_\parallel = 21\,$mT, we obtain $\mathcal{K}_\mathrm{I, 21} \approx -2\pi\cdot 39\,$kHz and $\mathcal{K}_\mathrm{II, 21} \approx -2\pi\cdot 54\,$kHz at the operation points "I" and "II", respectively, and for $B_\parallel = 25\,$mT, we find $\mathcal{K}_\mathrm{I, 25} \approx -2\pi\cdot 41\,$kHz and $\mathcal{K}_\mathrm{II, 25} \approx -2\pi\cdot 56\,$kHz.
As the difference between the two in-plane field is small and subject to uncertainties due to uncertainties in the circuit parameters, we will work with the same approximate anharmonicities for both in-plane fields of $\mathcal{K}_\mathrm{I}\approx - 2\pi\cdot 40\,$kHz and $\mathcal{K}_\mathrm{II}\approx - 2\pi\cdot 55\,$kHz.

\section*{Supplementary Note 5: The optomechanical single-photon coupling rate}

The optomechanical single-photon coupling rate in flux-mediated optomechanics is given by \cite{SShevchuk17, SRodrigues19}
\begin{equation}
g_0 = \gamma \mathcal{F} B_\parallel l_\mathrm{m} x_\mathrm{zpf} 
\end{equation}
where $\mathcal{F}$ is the cavity frequency flux responsivity, $B_\parallel$ is the in-plane magnetic field, $l_\mathrm{m}$ is the length of the mechanical nanobeam and $\gamma$ is a scaling factor on the order of unity taking into account the mode shape of the beam.
The zero-point fluctuation amplitude of the mechanical displacement is given by
\begin{equation}
x_\mathrm{zpf} = \sqrt{\frac{\hbar}{2m\Omega_\mathrm{m}}}
\end{equation}
where $m$ is the effective mass of the beam and $\Omega_\mathrm{m}$ its resonance frequency.
For our nanobeam, cf. main paper Fig.~1\textbf{b}, we get from a scanning electron micrograph the length as approximately $l_\mathrm{m} = 18\,\mu$m.
Using the beam film thickness of $\sim 70\,$nm, its width of $\sim 500\,$nm and the density of Aluminum $\rho_\mathrm{Al} = 2700\,$kg m$^{-3}$, we get a total mass of $m = 1.7\cdot10^{-15}\,$kg.
In the experiment, we observe a mechanical resonance frequency $\Omega_\mathrm{m} \approx 2\pi\cdot5.32\,$MHz and therefore, we find a zero-point fluctuation amplitude of $x_\mathrm{zpf} \approx 30\cdot 10^{-15}\,$m $= 30\,$fm.
With the flux responsivities shown in Supplementary Fig.~\ref{fig:Deriv}\textbf{a} and assuming $\gamma \approx 1$, we calculate the corresponding single-photon coupling rates $g_0$, the result is shown in Supplementary Fig.~\ref{fig:g0}.
For the different operation points, we obtain single-photon coupling rates $g_0$ between $g_0^\mathrm{min} \approx 2\pi\cdot 1.78\,$kHz (point I) and $g_0^\mathrm{max} \approx 2\pi\cdot 3.57\,$kHz (point II).
For all presented results, we will add the corresponding coupling rates either in the figure legend or in the caption.

\begin{figure}[h]
	\centerline {\includegraphics[trim={2cm 14.0cm 2cm 6cm},clip=True,scale=0.6]{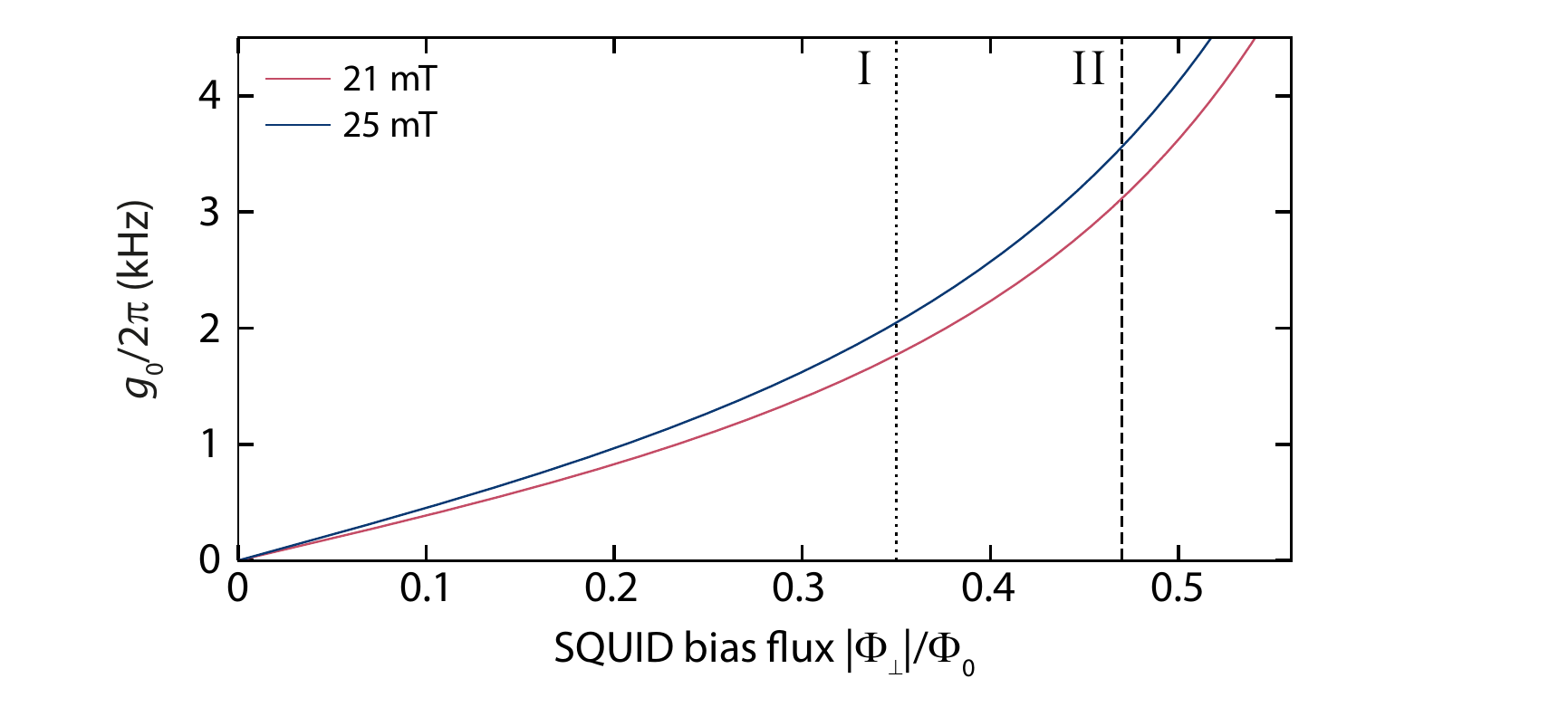}}
	\caption{\textsf{\textbf{The optomechanical single-photon coupling rate.} Using the flux responsivity $\mathcal{F}$ from the flux arch fits and the estimated mechanical zero-point-amplitude $x_\mathrm{zpf}$, we calculate the single-photon coupling rate $g_0$ for both in-plane fields as labelled in the legend. The two operation points "I" and "II" are marked with dotted and dashed lines, respectively.}}
	\label{fig:g0}
\end{figure}

\section*{Supplementary Note 6: The driven Kerr cavity}

\subsection*{Equation of motion}
We model the side-coupled SQUID cavity including the Kerr nonlinearity with the equation of motion
\begin{equation}
\dot{\alpha} = \left[i\left(\omega_0 + \mathcal{K}|\alpha|^2\right) - \frac{\kappa}{2}\right]\alpha + i\sqrt{\frac{\kappa_\mathrm{e}}{2}}S_\mathrm{in}
\end{equation}
where the intracavity field $\alpha$ is normalized such that $|\alpha|^2 = n_\mathrm{c}$ is the intracavity photon number and $|S_\mathrm{in}|^2$ corresponds to the input field photon flux.

\subsection*{Single-tone solution}

If the cavity is driven with a single tone at frequency $\omega_\mathrm{d}$, the input field is given by
\begin{equation}
S_\mathrm{in} = S_\mathrm{d}e^{i\left(\omega_\mathrm{d}t + \phi_\mathrm{d}\right)}
\end{equation}
and for the intracavity field we make the Ansatz
\begin{equation}
\alpha = \alpha_\mathrm{d}e^{i\omega_\mathrm{d}t}
\end{equation}
with real-valued $S_\mathrm{in}$ and $\alpha_\mathrm{d}$.
Any phase difference between the drive and the intracavity field is captured in the drive phase $\phi_\mathrm{d}$.
Inserting drive and intracavity field Ansatz into the equation of motion gives
\begin{equation}
\alpha_\mathrm{d}\left[\frac{\kappa}{2} + i\left(\Delta_\mathrm{d} - \mathcal{K}\alpha_\mathrm{d}^2\right)\right] = i\sqrt{\frac{\kappa_\mathrm{e}}{2}}S_\mathrm{d}e^{i\phi_\mathrm{d}}
\end{equation}
where $\Delta_\mathrm{d} = \omega_\mathrm{d} - \omega_0$ is the detuning between drive and undriven cavity resonance frequency. 
Multiplication of this equation with its complex conjugate leads to the determination polynomial for the drive intracavity photon number $n_\mathrm{d} = \alpha_\mathrm{d}^2$
\begin{equation}
\mathcal{K}^2 n_\mathrm{d}^3 - 2\mathcal{K}\Delta_\mathrm{d} n_\mathrm{d}^2 + \left(\Delta_\mathrm{d}^2 + \frac{\kappa^2}{4}\right)n_\mathrm{d} - \frac{\kappa_\mathrm{e}}{2}S_\mathrm{d}^2 = 0.
\label{eqn:Kerr_poly}
\end{equation}
In general, this polynomial has three roots for $n_\mathrm{d}$.
The real-valued solutions correspond to physical states and for certain parameters all three solutions are real-valued.
This regime corresponds to the bifurcation regime, where two of the three oscillator states are stable, one low- and one high-amplitude solution.
The middle solution is unstable and irrelevant for the experiments described here.
The phase difference between drive and intracavity oscillations can be found via
\begin{equation}
\phi_\mathrm{d} = \atan2{\left(-\frac{\kappa}{2}, \Delta_\mathrm{d} - \mathcal{K} n_\mathrm{d}\right)}.
\end{equation}

\subsection*{Nonlinear cavity response modeling}

\begin{figure}
	\centerline {\includegraphics[trim={1.cm 1.3cm 1.cm 0.5cm},clip=True,scale=0.55]{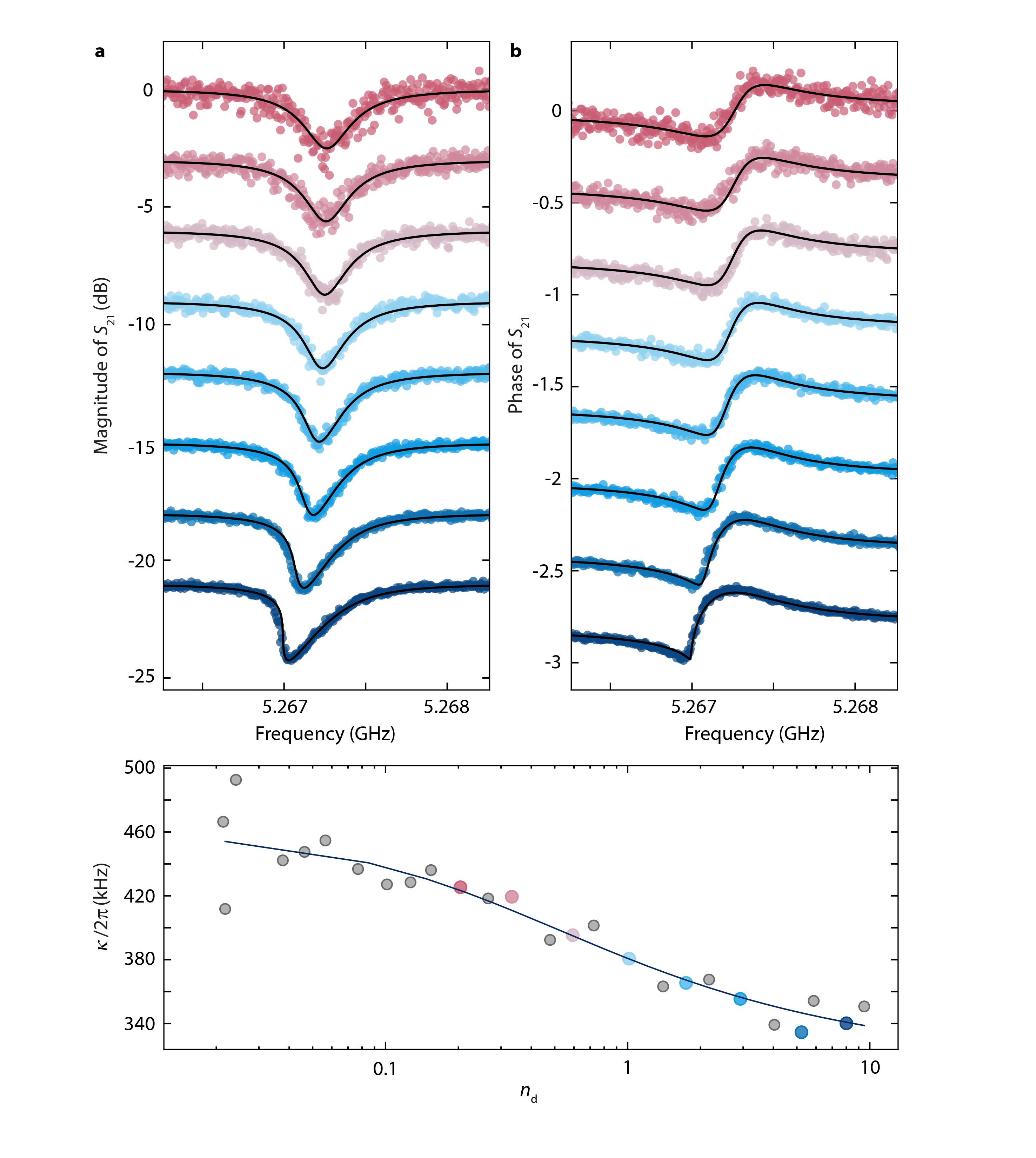}}
	\caption{\textsf{\textbf{Nonlinear cavity single-tone response.} \textbf{a} shows the magnitude and \textbf{b} the phase of the SQUID cavity response $S_{21}$ at the flux sweetspot $B_\perp = 0$ and at $B_\parallel = 25\,$mT for increasing probe power. Lowest probe power is shown in red at the top, highest power at the bottom in blue. Subsequent curves are offset by $-3\,$dB in \textbf{a} and $-0.4$ in \textbf{b} with the unshifted curves at the top. Circles are data, lines correspond to the model, for details see text. To model the experimentally obtained data accurately, we have to consider a power-dependent linewidth. The linewidths we obtain by fitting the nonlinear response curves with $\kappa$ as fit parameter are shown in $\textbf{c}$ as circles vs intracavity photons on resonance. The line in \textbf{c} shows a fit based on the two-level-system model for nonlinear dissipation in superconducting circuits. Colored circles in \textbf{c} correspond to the equally colored data in \textbf{a},\textbf{b}.}}
	\label{fig:KerrSingleTone}
\end{figure}
We measure the power dependence of the SQUID cavity response $S_{21}$ by means of a vector network analyzer at the bias flux sweetspot and at $B_\parallel = 25\,$mT.
The result is shown and discussed in Supplementary Fig.~\ref{fig:KerrSingleTone}, where in \textbf{a} the magnitude and in \textbf{b} the phase of the complex $S_{21}$ is plotted.
The data shown have been background-corrected as described above using the background fit of the first line.
To model the response, we employ the single-tone model described in the previous section and calculate the response via
\begin{eqnarray}
S_\mathrm{21} & = & 1 + i\sqrt{\frac{\kappa_\mathrm{e}}{2}}\frac{\alpha_\mathrm{d}}{S_\mathrm{d}}\nonumber\\
& = & 1+i\sqrt{\frac{\kappa_\mathrm{e}}{2}}\sqrt{\frac{\hbar\omega_\mathrm{d} n_\mathrm{d}}{P_\mathrm{d}}}e^{-i\phi_\mathrm{d}}
\end{eqnarray}
where $P_\mathrm{d}$ is the drive power on the on-chip microwave feedline.
For the model, we use the Kerr anharmonicity obtained from the independent cavity modeling $\mathcal{K} = -2\pi\cdot 30\,$kHz, the resonance frequency $\omega_0 = 2\pi \cdot 5.2672\,$GHz and the corresponding external linewidth $\kappa_\mathrm{e} = 2\pi\cdot 106.5\,$kHz as obtained from the lowest-power response of the current dataset.
As apparent from the increase in resonance absorption dip depth and the reduction of total linewidth with increasing drive power, we also have to consider nonlinear dissipation, that decreases with increasing power.
As first step in the nonlinear resonance analysis, we fit each of the nonlinear response curves using the Kerr polynomial and using a single decay rate for each power as fit parameter.
The result of this procedure for $\kappa$ is shown in Supplementary Fig.~\ref{fig:KerrSingleTone}\textbf{c}, where we plot the fit linewidth vs the photon number $n_\mathrm{d}$ on resonance.
We model this decrease of linewidth with the functional dependence for two-level systems
\begin{equation}
\kappa_\mathrm{i}(n_\mathrm{d}) = \kappa_0 + \frac{\kappa_1}{\sqrt{1+\frac{n_\mathrm{d}}{n_\mathrm{crit}}}}
\end{equation}
where $n_\mathrm{crit}$ describes the characteristic photon number for two-level saturation.
From a fit to the linewidth data we obtain $\kappa_0 = 2\pi \cdot 209\,$kHz, $\kappa_1 = 2\pi \cdot 145\,$kHz, and $n_\mathrm{crit} \approx 0.26$, cf. line in Supplementary Fig.~\ref{fig:KerrSingleTone}\textbf{c}.
As next and final step, we implement this analytical function into the Kerr polynomial and solve for the final photon number at each drive power and detuning.
To find convergence in the solution for the photon number $n_\mathrm{d}$ due to the power dependent $\kappa$ we have to iterate the polynomial solution multiple times in this approach for each frequency point, feeding back in each iteration the $\kappa(n_\mathrm{d})$ from the previous iteration.
The result is added as lines in  Supplementary Fig.~\ref{fig:KerrSingleTone}\textbf{a} and \textbf{b} and shows good agreement between experimental data and theoretical modeling. 
The only free parameter used in the end is the in-fridge attenuation between the VNA output and the sample, and we find best agreement for choosing $G_\mathrm{STK} = -89.2\,$dB.
This value is very close to the independently estimated probe attenuation of $G_\mathrm{SNR} = -89.5\,$dB at the corresponding frequency, cf. Supplementary Fig.~\ref{fig:NoiseCal} with consideration that the probe attenuation is $3\,$dB larger than the shown pump attenuation.

\subsection*{Linearized two-tone solution}

If the Kerr cavity is driven by a strong drive tone $S_\mathrm{d}$ and a weaker second tone $S_\mathrm{p}$ with frequency $\omega_\mathrm{p}$, the total input field is given by
\begin{equation}
S_\mathrm{in} = S_\mathrm{d}e^{i\left(\omega_\mathrm{d}t + \phi_\mathrm{d}\right)} + S_\mathrm{p}e^{i\omega_\mathrm{p}t}
\end{equation}
where we again choose $S_\mathrm{d}$ to be real-valued.
As Ansatz for the intracavity field, we then use
\begin{equation}
\alpha = \alpha_\mathrm{d}e^{i\omega_\mathrm{d}t} + \gamma_-e^{i\omega_\mathrm{p}t} + \gamma_+e^{i\left(2\omega_\mathrm{d} - \omega_\mathrm{p}\right)t}
\end{equation}
where $\gamma_-$ and $\gamma_+$ are complex-valued and $\gamma_+$ corresponds to the idler field of $\gamma_-$, generated by degenerate four-wave-mixing due to the Kerr nonlinearity.
Note that with this choice of Ansatz, we omit all higher order Fourier components to the total intracavity field, as in the operation regime of our device, they can be neglected to first order. 
Inserting drive and intracavity field Ansatz into the equation of motion yields 
\begin{eqnarray}
i\omega_\mathrm{d}\alpha_\mathrm{d}e^{i\Omega_\mathrm{dp}t} + i\omega_\mathrm{p}\gamma_- + i\left(2\omega_\mathrm{d} - \omega_\mathrm{p}\right)\gamma_+e^{i2\Omega_\mathrm{dp}t} & = & \left[i\omega_0 - \frac{\kappa}{2}\right]\alpha_\mathrm{d}e^{i\Omega_\mathrm{dp}t} + \left[i\omega_0 - \frac{\kappa}{2}\right]\gamma_- + \left[i\omega_0 - \frac{\kappa}{2}\right]\gamma_+e^{i2\Omega_\mathrm{dp}t}\nonumber\\
& & +~ i\mathcal{K}\left[n_\mathrm{d} + |\gamma_-|^2 + |\gamma_+|^2\right]\left(\alpha_\mathrm{d}e^{i\Omega_\mathrm{dp}t} + \gamma_- + \gamma_+e^{i2\Omega_\mathrm{dp}t}\right)\nonumber \\
& & +~ i\mathcal{K}\alpha_\mathrm{d}e^{i\Omega_\mathrm{dp}t}\left[\gamma_-^* + \gamma_+^*e^{-i2\Omega_\mathrm{dp}t}\right]\left(\alpha_\mathrm{d}e^{i\Omega_\mathrm{dp}t} + \gamma_- + \gamma_+e^{i2\Omega_\mathrm{dp}t}\right)\nonumber \\
& & +~ i\mathcal{K}\alpha_\mathrm{d}e^{-i\Omega_\mathrm{dp}t}\left[\gamma_- + \gamma_+e^{i2\Omega_\mathrm{dp}t}\right]\left(\alpha_\mathrm{d}e^{i\Omega_\mathrm{dp}t} + \gamma_- + \gamma_+e^{i2\Omega_\mathrm{dp}t}\right)\nonumber \\
& & +~i\mathcal{K}\left[\gamma_-\gamma_+^*e^{-i2\Omega_\mathrm{dp}t} + \gamma_-^*\gamma_+e^{i2\Omega_\mathrm{dp}t} \right]\left(\alpha_\mathrm{d}e^{i\Omega_\mathrm{dp}t} + \gamma_- + \gamma_+e^{i2\Omega_\mathrm{dp}t}\right)\nonumber \\
& & +~i\sqrt{\frac{\kappa_\mathrm{e}}{2}}S_\mathrm{d}e^{i\Omega_\mathrm{dp}t}e^{i\phi_d} + i\sqrt{\frac{\kappa_\mathrm{e}}{2}}S_\mathrm{p}.
\end{eqnarray}
With $n_\mathrm{d} + |\gamma_-|^2 + |\gamma_+|^2 \approx n_\mathrm{d}$ and after omitting all terms not linear in the small quantities $\gamma_-, \gamma_+$, we obtain
\begin{eqnarray}
& & \left[\frac{\kappa}{2} + i\left(\Delta_\mathrm{d} - \mathcal{K} n_\mathrm{d}\right)\right]\alpha_\mathrm{d}e^{i\Omega_\mathrm{dp}t} + \left[\frac{\kappa}{2} + i\left(\Delta_\mathrm{p} - \mathcal{K} n_\mathrm{d}\right)\right]\gamma_- + \left[\frac{\kappa}{2} + i\left(\Delta_\mathrm{p} - \mathcal{K} n_\mathrm{d} + 2\Omega_\mathrm{dp}\right)\right]\gamma_+e^{i2\Omega_\mathrm{dp}t} \nonumber \\
& = & i\mathcal{K} n_\mathrm{d}\left[\gamma_-^*e^{i2\Omega_\mathrm{dp}t} + \gamma_+^*\right] + i\mathcal{K} n_\mathrm{d}\left[\gamma_- + \gamma_+e^{i2\Omega_\mathrm{dp}t}\right] + i\sqrt{\frac{\kappa_\mathrm{e}}{2}}S_\mathrm{d}e^{i\Omega_\mathrm{dp}t}e^{i\phi_d} + i\sqrt{\frac{\kappa_\mathrm{e}}{2}}S_\mathrm{p},
\end{eqnarray}
where $\Delta_\mathrm{d} = \omega_\mathrm{d}-\omega_0$ and $\Delta_\mathrm{p} = \omega_\mathrm{p} - \omega_0$ describe the detunings of the two input field frequencies from the undriven cavity resonance.
Sorting for frequency contributions leaves us with three equations
\begin{eqnarray}
\left[\frac{\kappa}{2} + i\left(\Delta_\mathrm{d} - \mathcal{K} n_\mathrm{d}\right)\right]\alpha_\mathrm{d} & = & i\sqrt{\frac{\kappa_\mathrm{e}}{2}}S_\mathrm{d}e^{i\phi_\mathrm{d}} \\
\left[\frac{\kappa}{2} + i\left(\Delta_\mathrm{p} - 2\mathcal{K} n_\mathrm{d}\right)\right]\gamma_- - i\mathcal{K} n_\mathrm{d}\gamma_+^* & = & i\sqrt{\frac{\kappa_\mathrm{e}}{2}}S_\mathrm{p} \\
\left[\frac{\kappa}{2} + i\left(\Delta_\mathrm{p} - 2\mathcal{K} n_\mathrm{d} + 2\Omega_\mathrm{dp}\right)\right]\gamma_+ - i\mathcal{K} n_\mathrm{d}\gamma_-^* & = & 0.
\end{eqnarray}
The first of these three equations is the steady-state equation for the single-tone case and can be solved for $n_\mathrm{d}$ using the approach presented above.
To solve the other two coupled equations, we use $\Delta_\mathrm{p} = \Delta_\mathrm{d} - \Omega_\mathrm{dp}$ and define the susceptibilities
\begin{equation}
\chi_\mathrm{p} = \frac{1}{\frac{\kappa}{2} + i\left(\Delta_\mathrm{d} - 2\mathcal{K} n_\mathrm{d} - \Omega_\mathrm{dp}\right)}, ~~~~~ \chi_\mathrm{p}' = \frac{1}{\frac{\kappa}{2} + i\left(\Delta_\mathrm{d} - 2\mathcal{K} n_\mathrm{d} + \Omega_\mathrm{dp}\right)}
\end{equation}
and
\begin{equation}
\chi_\mathrm{g} = \frac{\chi_\mathrm{p}}{1-\mathcal{K}^2 n_\mathrm{d}^2 \chi_\mathrm{p}\chi_\mathrm{p}'^*}
\end{equation}
and get
\begin{eqnarray}
\gamma_- & = & i\chi_\mathrm{g}\sqrt{\frac{\kappa_\mathrm{e}}{2}}S_\mathrm{p} \label{eqn:gamma_m} \\
\gamma_+ & = & \mathcal{K} n_\mathrm{d} \chi_\mathrm{p}'\chi_\mathrm{g}^*\sqrt{\frac{\kappa_\mathrm{e}}{2}}S_\mathrm{p}\\
& = & i\mathcal{K}n_\mathrm{d}\chi_\mathrm{p}'\gamma_-^*
\label{eqn:gamma_p}
\end{eqnarray}

\subsection*{The driven Kerr-modes and their response}

To find the resonance frequency of the quasi-mode with susceptibility $\chi_\mathrm{g}$, we solve for the complex frequency $\tilde{\omega} = \omega_\mathrm{p}$, for which $\chi_\mathrm{g}^{-1} = 0$.
Therefore, the condition is
\begin{equation}
1 - \mathcal{K}^2 n_\mathrm{d}^2\chi_\mathrm{p}\chi_\mathrm{p}' = 0
\end{equation}
which is solved by
\begin{equation}
\tilde{\omega}_{1/2} = \omega_\mathrm{d} + i\frac{\kappa}{2} \pm \sqrt{\left(\Delta_\mathrm{d} - \mathcal{K}n_\mathrm{d}\right)\left(\Delta_\mathrm{d} - 3\mathcal{K}n_\mathrm{d}\right)}.
\end{equation}
where the real part corresponds to the resonance frequency and the imaginary part corresponds to half the mode linewidth.
So, as a consequence of the presence of the strong drive, the system has two resonances, which are split symmetrically with respect to the drive frequency $\omega_\mathrm{d}$.
The two Kerr-modes correspond to the cases when the cavity field $\gamma_-$ is resonant or when its idler field $\gamma_+$ is resonant and they have been discussed also in the context of nonlinear optical cavities \cite{SDrummond80, SKhandekar15} and mechanical oscillators \cite{SHuber20}.
For the experiments described here, the argument of the square root will always be positive and hence, we get as resonance frequency of signal and idler mode
\begin{equation}
\omega_\mathrm{s/i} = \omega_\mathrm{d} \pm \sqrt{\left(\Delta_\mathrm{d} - \mathcal{K}n_\mathrm{d}\right)\left(\Delta_\mathrm{d} - 3\mathcal{K}n_\mathrm{d}\right)}
\end{equation}
and both modes are having a constant linewidth of $\kappa$.
The most relevant regime for our experiment is given by $\Delta_\mathrm{d} < 0$ and $\Delta_\mathrm{d} < \mathcal{K}n_\mathrm{d}$.
Then, the signal resonance is given by
\begin{equation}
\omega_\mathrm{s} = \omega_\mathrm{d} - \sqrt{\left(\Delta_\mathrm{d} - \mathcal{K}n_\mathrm{d}\right)\left(\Delta_\mathrm{d} - 3\mathcal{K}n_\mathrm{d}\right)}
\end{equation}
and the idler resonance by
\begin{equation}
\omega_\mathrm{i} = \omega_\mathrm{d} + \sqrt{\left(\Delta_\mathrm{d} - \mathcal{K}n_\mathrm{d}\right)\left(\Delta_\mathrm{d} - 3\mathcal{K}n_\mathrm{d}\right)}.
\end{equation}
Hence, if $S_\mathrm{p}$ is a probe tone scanning the driven SQUID cavity, we expect the response to be given by
\begin{equation}
S_{21} = 1 + i\sqrt{\frac{\kappa_\mathrm{e}}{2}}\frac{\gamma_-}{S_\mathrm{p}}
\label{eqn:fitS21Kerr}
\end{equation}
and to observe two resonances symmetrically positioned around the drive, that correspond to the two Kerr-modes of the driven Kerr cavity.

\subsection*{Kerr cavity two-tone response modeling}

To observe and model the two-tone response of the driven Kerr cavity, we set a drive tone with fixed power $P_\mathrm{d}$ and fixed frequency $\omega_\mathrm{d}$ to a point of non-zero flux bias and sweep the bare resonance frequency of the cavity $\omega_0$ through the drive frequency by sweeping the bias flux $\Phi_\perp$.
For each flux bias value during the sweep, we record a response trace $S_{21}$ of the SQUID cavity using the VNA.
In Supplementary Fig.~\ref{fig:FluxSweeps} we show in comparison the result of such a measurement with and without parametric drive.
When $\omega_0 \sim \omega_\mathrm{d}$, the flux-dependence of the cavity is modified considerably with respect to the case without a drive.
The resonance frequency of the observed mode becomes nearly constant with flux and on the opposite side of the drive, a second resonance line appears.
This second mode on the right side of the drive tone with net transmission gain indicates that we are entering the quasi-mode regime and have Josephson parametric amplification in both Kerr-modes.

\begin{figure}
	\centerline {\includegraphics[trim={3.cm 7.8cm 3.cm 2.5cm},clip=True,scale=0.60]{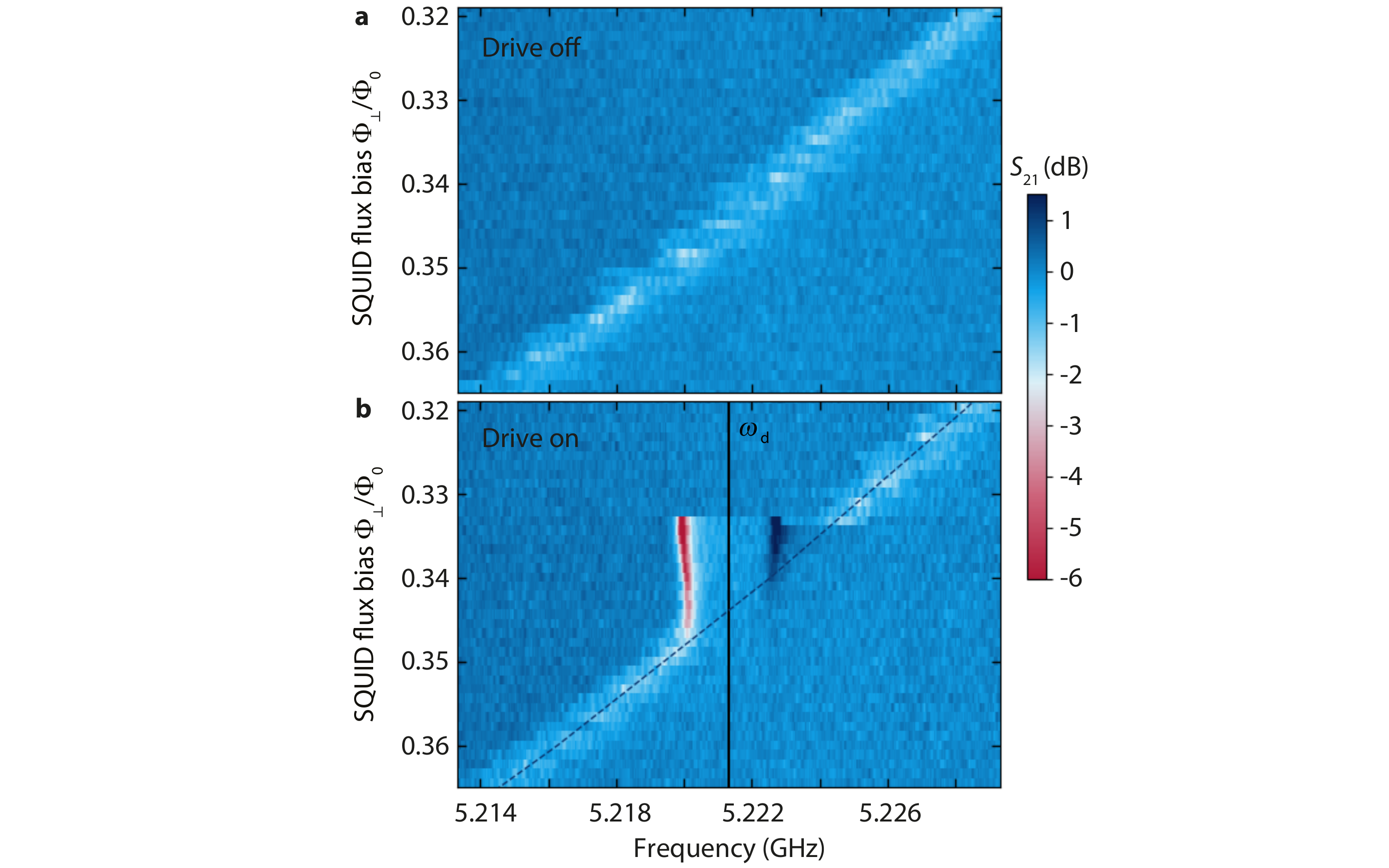}}
	\caption{\textsf{\textbf{Probe tone reponse of the SQUID cavity with and without parametric drive.} \textbf{a} shows color-coded the probe tone cavity response $S_{21}$ in a small flux and frequency window around operation point I. The cavity absorption is visible as bright and fluctuating feature. In \textbf{b} the same measurement in the same flux and frequency range is shown, but in presence of a fixed frequency drive tone at $\omega_\mathrm{d} = 2\pi\cdot 5.2213\,$GHz. The dashed line shows the theoretical resonance frequency without drive. When $\omega_\mathrm{d} \approx \omega_0$, the response of the cavity deviates significantly from the undriven response. Two modes are visible, one on the left side of the drive with increased absorption and reduced linewidth compared to the undriven case, and a second on the right side of the drive as a peak. The increased depth and reduced linewidth of the signal mode absorption dip reflects both, reduced impact of flux noise on the cavity line and Josephson parametric amplification. The Josephson amplification is also apparent in the idler mode on the right side of the drive, which shows net gain of the input signal. Due to the underlying four-wave mixing process, the resonance frequencies of the two Kerr-modes with respect to the parametric drive are equal in magnitude and opposite in sign.}}
	\label{fig:FluxSweeps}
\end{figure}

Main paper Fig.~2 shows and discusses a similar dataset in more detail including linescans and also analyzing the linewidths and the Kerr-mode resonance frequency and comparing the experimentally obtained values with theoretical calculations.
The overall behaviour of cavity and Kerr-modes is in excellent agreement with the two-tone model of a Kerr oscillator as demonstrated by the high-level agreement between theory lines and experimental data for the resonance frequencies.
The only subtlety we have to consider additionally is an effective linewidth broadening due to low-frequency flux noise out of the two-mode regime.
From the data, it is obvious that the lineshape is not just broadened but distorted also on timescales comparable to the measurement time.
Although due to this noise modulation the cavity response is not anymore described by its ideal response with an increased linewidth \cite{Brock20}, we fit it using Eq.~(\ref{eqn:fitS21}).
The apparent $\kappa'$ we obtain from this procedure is however still a good measure for the effective linewidth.
These flux-noise broadened linewidths are considerably larger than the intrinsic energy decay rate of the cavity and to take the noise-broadening and the two-level systems simultaneously into account we model the noise-free linewidth using
\begin{equation}
\kappa_\mathrm{i} = \kappa_0 + \kappa_1\left[1-\frac{n_\mathrm{d}/n_\mathrm{crit}}{\sqrt{1+n_\mathrm{d}/n_\mathrm{crit}}}\frac{1+\sqrt{1+n_\mathrm{d}/n_\mathrm{crit}}}{\left(\Delta_\mathrm{d}/\Gamma_2\right)^2 + \left(1+\sqrt{1+n_\mathrm{d}/n_\mathrm{crit}}\right)^2}\right]
\label{eqn:TotalKappa}
\end{equation}
where the first term $\kappa_0$ is the bare power and flux-independent decay rate, and the second term describes the generalized two-level system impact for detuning between drive and cavity/probe with the two-level system dephasing rate $\Gamma_2$ \cite{Capelle20}.
Finally, we phenomenologically take into account a broadening of the linewidth using
\begin{equation}
\kappa_\mathrm{i}' = \kappa_\mathrm{i}\sqrt{1 + |\mathcal{F}|^2\sigma_\Phi^2}
\label{eqn:NoiseKappa}
\end{equation}
where $\mathcal{F}$ is the flux responsivity and $\sigma_\Phi$ is the rms value of the flux fluctuations through the SQUID.
This relation seems to resemble closely what has been found numerically for tunable and frequency-fluctuating cavities \cite{Brock20}.
The line in main paper Fig.~2 is a modeling of the experimentally obtained linewidths with Eqs.~(\ref{eqn:TotalKappa}) and (\ref{eqn:NoiseKappa}).
The fit parameters are $\kappa_0 = 2\pi \cdot 200\,$kHz, $\kappa_1 = 2\pi\cdot 145 \,$kHz, $n_\mathrm{crit} = 0.26$, $\Gamma_2 = 2\pi\cdot 300\,$kHz and $ \sigma_\Phi = 5.1\cdot 10^{-3}\Phi_0$.
Note that the values used and obtained here are also in good agreement with the numbers we extracted from the modeling of the nonlinear single-tone response.
For the flux operation point II, we can use almost exactly the same parameters, except for a slightly increased $\kappa_0 = 2\pi \cdot 210\,$kHz.

\section*{Supplementary Note 7: Linearized Kerr optomechanics}

\subsection*{Classical equations of motion}

To model the optomechanical system with a Kerr nonlinearity, we use the classical equations of motion (EOM)
\begin{eqnarray}
\ddot{x} & = & -\Omega_\mathrm{m}^2 x - \Gamma_\mathrm{m}\dot{x} + \frac{\hbar G}{m}|\alpha|^2 \\
\dot{\alpha} & = & \left[i\left(\omega_0 - Gx + \mathcal{K}|\alpha|^2\right) - \frac{\kappa}{2}\right]\alpha + i\sqrt{\frac{\kappa_\mathrm{e}}{2}}S_\mathrm{in}
\end{eqnarray}
with the mechanical displacement $x$, the mechanical resonance frequency $\Omega_\mathrm{m}$, the mechanical damping rate $\Gamma_\mathrm{m}$, and the cavity pull parameter
\begin{equation}
G = -\frac{\partial\omega_0}{\partial x} = -\mathcal{F}B_\parallel l_\mathrm{m}
\end{equation}
These equations are identical to the EOMs of linear classical optomechanics \cite{SAspelmeyer14}, except for the additional Kerr term $\mathcal{K}|\alpha|^2$ in the equation for the intracavity field.

\subsection*{Single-drive solution}

For a single cavity drive field
\begin{equation}
S_\mathrm{in} = S_\mathrm{d}e^{i\left(\omega_\mathrm{d}t+\phi_\mathrm{d}\right)}
\end{equation}
we make the Ansatz
\begin{eqnarray}
x & = & x_0 \\
\alpha & = & \alpha_\mathrm{d}e^{i\omega_\mathrm{d}t}
\end{eqnarray}
and look for the steady-state solution $\ddot{x} = \dot{x} = 0$.
For the equilibrium offset displacement $x_0$, we obtain
\begin{equation}
x_0 = \frac{\hbar G}{\Omega_\mathrm{m}^2 m}n_\mathrm{d} = n_\mathrm{d}\frac{2g_0}{\Omega_\mathrm{m}}x_\mathrm{zpf}
\end{equation}
where we used
\begin{equation}
g_0 = G x_\mathrm{zpf}, ~~~ x_\mathrm{zpf} = \sqrt{\frac{\hbar}{2m\Omega_\mathrm{m}}}.
\end{equation}
For the intracavity field amplitude $\alpha_\mathrm{d}$, we find
\begin{equation}
\left[\frac{\kappa}{2} + i\left(\Delta_\mathrm{d} - \overline{\mathcal{K}}\alpha_\mathrm{d}^2\right)\right]\alpha_\mathrm{d} = i\sqrt{\frac{\kappa_\mathrm{e}}{2}}S_\mathrm{d}e^{i\phi_\mathrm{d}}.
\end{equation}
with the modified Kerr anharmonicity
\begin{equation}
\overline{\mathcal{K}} = \mathcal{K} - \frac{2g_0^2}{\Omega_\mathrm{m}}.
\end{equation}
As for our device $\mathcal{K} > 2\pi\cdot 10^4\,$Hz and $\frac{2g_0^2}{\Omega_\mathrm{m}} < 2\pi\cdot 10\,$Hz, we can assume in good approximation $\overline{\mathcal{K}} \approx \mathcal{K}$.
From here it is straightforward to calculate the intracavity drive photon number $n_\mathrm{d}$ and the phase $\phi_\mathrm{d}$ using the third order polynomial as for the bare Kerr cavity.

\subsection*{Single-drive Kerr backaction}

If we allow also for fluctuations of the displacement and the intracavity field, we get the Ansatz
\begin{eqnarray}
x & = & x_0 + \delta x(t) \\
\alpha & = & \alpha_\mathrm{d}e^{i\omega_\mathrm{d}t} + \delta\alpha(t)e^{i\omega_\mathrm{d}t}
\end{eqnarray}
and the equation of motion for the mechanical oscillator becomes
\begin{equation}
\delta\ddot{x} = -\Omega_\mathrm{m}^2 x_0 - \Omega_\mathrm{m}^2 \delta x - \Gamma_\mathrm{m} \delta\dot{x} + \frac{\hbar G}{m}\left(n_\mathrm{d} + |\delta\alpha|^2\right) + \frac{\hbar G}{m}\alpha_\mathrm{d}\left(\delta\alpha + \delta\alpha^*\right).
\end{equation}
For the intracavity field, we find
\begin{eqnarray}
i\omega_\mathrm{d}\alpha_\mathrm{d} + i\omega_\mathrm{d}\delta\alpha + \delta\dot{\alpha} & = & \left[i\omega_0 - \frac{\kappa}{2}\right]\alpha_\mathrm{d} + \left[i\omega_0 - \frac{\kappa}{2}\right]\delta\alpha \nonumber\\
& & -~iG\alpha_\mathrm{d} x_0 - iG \alpha_\mathrm{d}\delta x - iG x_0\delta\alpha - iG\delta\alpha\delta x \nonumber\\
& & +~i\mathcal{K}\left(n_\mathrm{d} + |\delta\alpha|^2\right)\left(\alpha_\mathrm{d} + \delta\alpha\right)
\nonumber\\
& & +~i\mathcal{K}\alpha_\mathrm{d}\left(\delta\alpha + \delta\alpha^*\right)\left(\alpha_\mathrm{d} + \delta\alpha\right) \nonumber \\
& & +~i\sqrt{\frac{\kappa_\mathrm{e}}{2}}S_\mathrm{d}e^{i\phi_\mathrm{d}}.
\end{eqnarray}
For the linearization, we omit now all terms not linear in the small quantities $\delta\alpha$ and $\delta x$, we apply $\overline{\mathcal{K}} \approx \mathcal{K}$ and remove the steady state solution.
The remaining equations are
\begin{eqnarray}
\delta\ddot{x} & = & -\Omega_\mathrm{m}^2 \delta x - \Gamma_\mathrm{m} \delta\dot{x} + \frac{\hbar G \alpha_\mathrm{d}}{m}\left(\delta\alpha + \delta\alpha^*\right) \\
\delta\dot{\alpha} & = & \left[-i\left(\Delta_\mathrm{d} - 2\mathcal{K}n_\mathrm{d}\right)-\frac{\kappa}{2}\right]\delta\alpha + i\mathcal{K}n_\mathrm{d}\delta\alpha^* - iG\alpha_\mathrm{d}\delta x
\end{eqnarray}
which can be Fourier-transformed to
\begin{eqnarray}
\delta x(\Omega)\left[\Omega_\mathrm{m}^2 - \Omega^2 + i\Omega\Gamma_\mathrm{m}\right] & = & \frac{\hbar G \alpha_\mathrm{d}}{m}\left[\delta\alpha(\Omega) + \delta\alpha^*(-\Omega)\right] \\
\delta\alpha(\Omega)\left[\frac{\kappa}{2} + i\left(\Delta_\mathrm{d} - 2\mathcal{K}n_\mathrm{d} + \Omega\right)\right] & = & i\mathcal{K} n_\mathrm{d}\delta\alpha^*(-\Omega) - iG\alpha_\mathrm{d}\delta x(\Omega).
\end{eqnarray}
Using the convention $\overline{\delta\alpha} = \delta\alpha^*(-\Omega)$, the observation that $\delta x(\Omega) = \delta x^*(-\Omega)$ and the definitions
\begin{equation}
\chi_\mathrm{m} = \frac{1}{\Omega_\mathrm{m}^2 - \Omega^2 + i\Omega\Gamma_\mathrm{m}}, ~~~ \chi_\mathrm{p} = \frac{1}{\frac{\kappa}{2} + i\left(\Delta_\mathrm{d} - 2\mathcal{K}n_\mathrm{d}+\Omega\right)}
\end{equation}
we write these equations as
\begin{eqnarray}
\frac{\delta x}{\chi_\mathrm{m}} & = & \frac{\hbar G \alpha_\mathrm{d}}{m}\left(\delta\alpha + \overline{\delta\alpha}\right) \\
\delta\alpha & = & i\mathcal{K}n_\mathrm{d}\chi_\mathrm{p}\overline{\delta\alpha} - iG\chi_\mathrm{p}\alpha_\mathrm{d}\delta x \\
\overline{\delta\alpha} & = & -i\mathcal{K}n_\mathrm{d}\overline{\chi}_\mathrm{p}\delta\alpha + iG\overline{\chi}_\mathrm{p}\alpha_\mathrm{d}\delta x \\
\end{eqnarray}
and solve for
\begin{eqnarray}
\delta\alpha & = & -i\chi_\mathrm{g}\alpha_\mathrm{d}G\left(1-i\mathcal{K}n_\mathrm{d}\overline{\chi}_\mathrm{p}\right)\delta x \\
\overline{\delta\alpha} & = & i\overline{\chi}_\mathrm{g}\alpha_\mathrm{d}G\left(1+i\mathcal{K}n_\mathrm{d}\chi_\mathrm{p}\right)\delta x.
\end{eqnarray}
Note that here we used the earlier definition of the two-tone Kerr susceptibility
\begin{equation}
\chi_\mathrm{g} = \frac{\chi_\mathrm{p}}{1-\mathcal{K}^2 n_\mathrm{d}^2 \chi_\mathrm{p} \overline{\chi}_\mathrm{p}}.
\end{equation}
\begin{figure}[h]
	\centerline {\includegraphics[trim={0.3cm 13.8cm 1.cm 2cm},clip=True,scale=0.55]{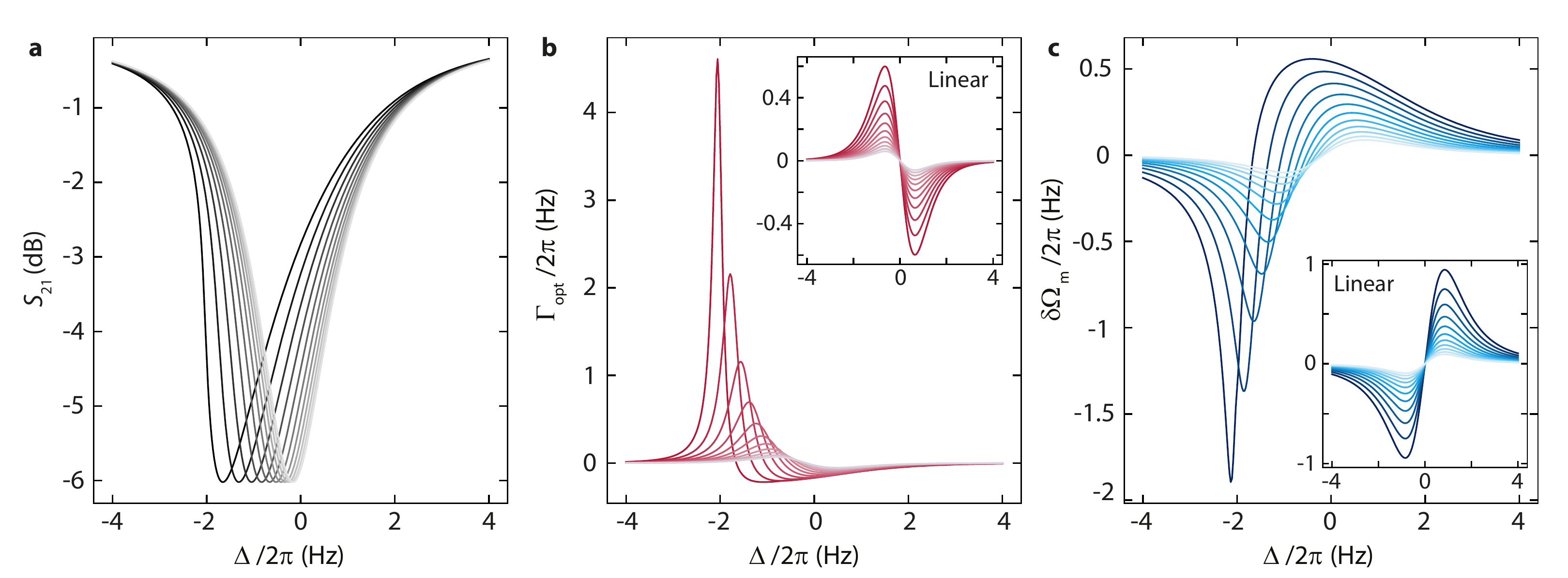}}
	\caption{\textsf{\textbf{Linearized dynamical Kerr backaction in the sideband-unresolved regime.} We calculate the cavity response and the dynamical Kerr backaction for a sideband-unresolved optomechanical system with parameters close to the device discussed in Ref.~\cite{SZoepfl20}. In \textbf{a}, we plot the magnitude of the transmission matrix element $S_{21}$ at a side-coupled cavity with a resonance frequency $\omega_0 = 2\pi\cdot 8.167\,$GHz, a total linewidth $\kappa = 2\pi\cdot 2.8\,$MHz, an external linewidth $\kappa_\mathrm{e} = 2\pi\cdot 1.4\,$MHz and an anharmonicity $\mathcal{K} = -2\pi\cdot 2.5\,$kHz for varying drive powers. We chose the drive powers such that the characteristic Duffing-like tilting of the resonance line to the left is clearly visible and keep the highest drive power below bifurcation. Using the obtained intracavity photon numbers $n_\mathrm{d}$, a mechanical resonance frequency $\Omega_\mathrm{m} = 2\pi\cdot 274.4\,$kHz, an optomechanical single-photon coupling rate $g_0 = 2\pi\cdot 57\,$Hz and our model presented in the text, we subsequently calculate the optical damping $\Gamma_\mathrm{opt}$ and the optical spring $\delta\Omega_\mathrm{m}$ induced by the drive in a Kerr cavity. The result is plotted in panels \textbf{b} and \textbf{c}, respectively, and seems to agree well with the experimental results reported in \cite{SZoepfl20}. It is furthermore interesting to compare the obtained dynamical Kerr backaction with the dynamical backaction for a completely identical system but without anharmonicity. The corresponding calculations for the linear system $\mathcal{K} = 0$ are shown as insets. The most striking and exciting difference is the eightfold enhancement of the optical damping on the "red" side of the Kerr cavity. This enhancement can also be understood by the increased slope of the intracavity field \cite{SNation08}, cf.~\textbf{a}, and the subsequently enhanced asymmetry for cavity photon up-scattering and down-scattering compared to the linear case. Different lines in each panel correspond to different drive powers and the detuning $\Delta$ labelling the $x$-axes corresponds to the detuning from the zero-power resonance frequency $\omega_0$.}}
	\label{fig:KerrZoepfl}
\end{figure}
Inserting everything into the equation of motion for the mechanical oscillator, we obtain for a (real-valued) external driving force $F_\mathrm{ex}(\Omega)$
\begin{equation}
\delta x = \chi_\mathrm{m}^\mathrm{eff} \frac{F_\mathrm{ex}}{m}	
\end{equation}
with the effective mechanical Kerr susceptibility
\begin{equation}
\frac{1}{\chi_\mathrm{m}^\mathrm{eff}} = \Omega_\mathrm{m}^2 - \Omega^2 + i\Omega \Gamma_\mathrm{m} + i2\Omega_\mathrm{m}\Sigma_\mathrm{k}
\end{equation}
where
\begin{equation}
\Sigma_\mathrm{k} = g_\alpha^2\left[\chi_\mathrm{g}\left(1-i\mathcal{K}n_\mathrm{d}\overline{\chi}_\mathrm{p}\right) - \overline{\chi}_\mathrm{g}\left(1+i\mathcal{K}n_\mathrm{d}\chi_\mathrm{p}\right)\right]
\end{equation}
describes the dynamical backaction of a single-tone driven Kerr cavity to the mechanical oscillator with the multi-photon coupling rate $g_\alpha = \alpha_\mathrm{d}g_0 =  \sqrt{n_\mathrm{d}}g_0$.
We note, that the expression for $\Sigma_\mathrm{k}$ is formally equivalent to the dynamical backaction in a linear cavity.
The first term in square brackets describes the quasi-mode susceptibility $\chi_\mathrm{g}$ for the blue motional sideband field (cooling), while the second term -- its conjugate at the opposite side of the drive tone $\overline{\chi}_\mathrm{g}$ -- is responsible for the red motional sideband field (amplification).
The additional factors in parentheses $\left(1-i\mathcal{K}n_\mathrm{d}\overline{\chi}_\mathrm{p}\right)$ and $\left(1+i\mathcal{K}n_\mathrm{d}\chi_\mathrm{p}\right)$ take into account that the blue and red motional sidebands are interfering with each other due to the Kerr-drive induced four-wave mixing.
The blue motional sideband coincides with the idler field of the red sideband and vice versa and their interference will contribute and modify the simple picture of dynamical backaction in a linear cavity.
For a high quality factor mechanical resonator and in the weak-coupling regime, we can approximate $\Omega \approx \Omega_\mathrm{m}$ and get
\begin{equation}
\frac{1}{\chi_0^\mathrm{eff}} = \frac{\Gamma_\mathrm{m}}{2} + i\left(\Omega - \Omega_\mathrm{m}\right) + \Sigma_\mathrm{k}(\Omega_\mathrm{m})
\end{equation}
with
\begin{equation}
\Sigma_\mathrm{k}(\Omega_\mathrm{m}) = g_\alpha^2\left[\chi_\mathrm{g}(\Omega_\mathrm{m})\left(1-i\mathcal{K}n_\mathrm{d}\chi_\mathrm{p}^*(-\Omega_\mathrm{m})\right) - \chi_\mathrm{g}^*(-\Omega_\mathrm{m})\left(1+i\mathcal{K}n_\mathrm{d}\chi_\mathrm{p}(\Omega_\mathrm{m})\right)\right].
\end{equation}
Note that $\chi_\mathrm{m}^\mathrm{eff}$ and $\chi_0^\mathrm{eff}$ do not have the same dimension and $2i\Omega_\mathrm{m}\chi_\mathrm{m}^\mathrm{eff} \approx \chi_0^\mathrm{eff}$ for $\Omega \approx \Omega_\mathrm{m}$.
Nevertheless we will call both a susceptibility for simplicity.
The optical spring $\delta\Omega_\mathrm{m}$ and optical damping $\Gamma_\mathrm{opt}$ are then given by
\begin{eqnarray}
\delta\Omega_\mathrm{m} & = &  -\mathrm{Im}\left[\Sigma_\mathrm{k}(\Omega_\mathrm{m})\right] \\
\Gamma_\mathrm{opt} & = & 2\mathrm{Re}\left[\Sigma_\mathrm{k}(\Omega_\mathrm{m})\right].
\end{eqnarray}
Very recently, there has been an experimental report on dynamical backaction with a SQUID Kerr cavity in the sideband-unresolved regime \cite{SZoepfl20} and we demonstrate in Supplementary Fig.~\ref{fig:KerrZoepfl} that our expressions lead to very similar results to the ones reported in Ref.~\cite{SZoepfl20} using parameters comparable with the reported ones.
\begin{figure}[h!]
	\centerline {\includegraphics[trim={0cm 0cm 0cm 0cm},clip=True,scale=0.8]{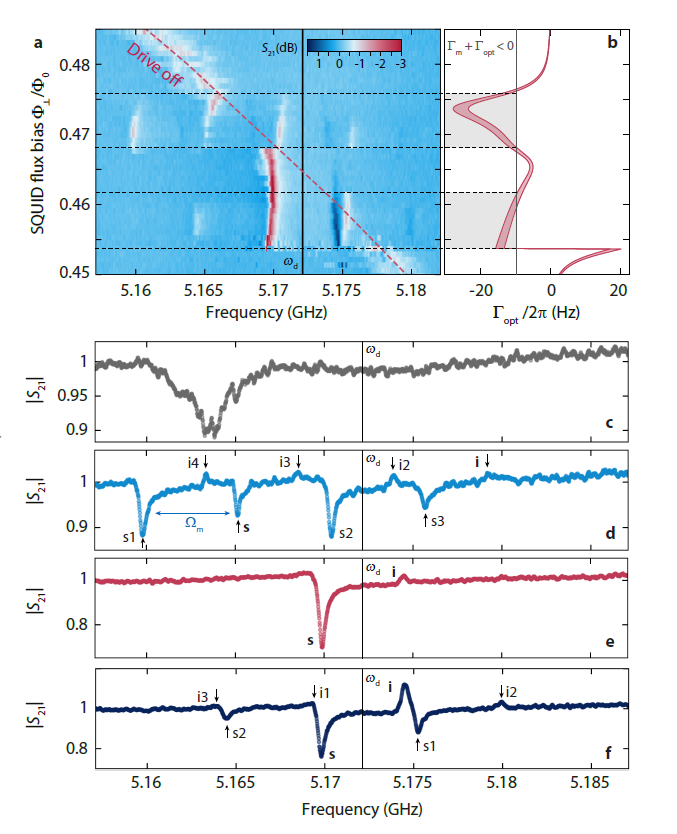}}
	\caption{\textsf{\textbf{Observing and modeling Kerr-backaction induced mechanical self-oscillations.} \textbf{a} Magnitude of the cavity response $S_{21}$ vs SQUID flux bias around operation point II in presence of a strong drive with $\omega_\mathrm{d} = 2\pi\cdot 5.1721\,$GHz. The red dashed line shows the expected resonance frequency in absence of a drive. In the response, five regimes can be discriminated, which are described in detail in the text. Black horizontal dashed lines show the boundaries of the different regimes, which are closely related to the dynamical Kerr backaction onto the mechanical resonator. \textbf{b} shows the calculated optical damping for the flux range shown in \textbf{a}. The two solid red lines are the result for $g_0 = 2\pi\cdot 3.4\,$kHz and $g_0 = 2\pi\cdot 3.7$kHz, the red-shaded area captures the range in between these values. For two particular flux ranges (which correspond to two different ranges of detunings between the bare cavity and the parametric drive and therefore to different intracavity drive photon numbers) the total damping rate of the mechanical oscillator with a intrinsic damping rate of $\Gamma_\mathrm{m} \approx 2\pi\cdot 10 \,$Hz becomes negative. These ranges are indicated by gray areas and the threshold $\Gamma_\mathrm{tot} < 0$ is indicated by a vertical black line. Here, the mechanical oscillator will become unstable and undergoes self-oscillations. These self-oscillations induce oscillations of the SQUID cavity resonance frequency due to the optomechanical interaction, which will in turn lead to the observation of multiple replicas of the cavity and idler modes in regime 2 and regime 4. \textbf{c}-\textbf{f} show individual linecuts of the cavity response shown in \textbf{a}, one for each regime from 1 to 4, where regime 1 corresponds to the highest bias flux values. In the regimes 2 and 4, shown in panels \textbf{d} and \textbf{f}, respectively, these Kerr-mode replicas are visible and labelled with s$X$ and i$X$ with $X = 1, 2, 3$. The original signal and idler Kerr- modes are labelled with \textbf{s} and \textbf{i}.}}
	\label{fig:KerrSelfOsc}
\end{figure}
Using our expressions for the Kerr backaction, we can also model with high accuracy the regimes for mechanical self-oscillation induced by the strong parametric cavity drive in the device presented here.
To do this, we perform a similar experiment to the one discussed in main paper Fig.~2, but at bias flux operation point II.
At this operation point, we have a larger $g_0 \sim 2\pi\cdot 3.56\,$kHz as well as a larger Kerr nonlinearity $\mathcal{K} \approx -2\pi\cdot 55\,$kHz.
In addition, we use a slightly higher drive power.
The result of the probe tone transmission $S_{21}$ for a bias flux sweep of the cavity resonance through the parametric drive tone is shown and discussed in Supplementary Fig.~\ref{fig:KerrSelfOsc}.
The red dashed line shows how the bare cavity resonance frequency would be moving with bias flux $\Phi_\perp$ in absence of a drive with frequency $\omega_\mathrm{d}$.
We can discriminate between five different regimes in the displayed data set.
\begin{itemize}
	\item{\textbf{Regime 1}: For the highest bias flux values, the cavity follows the undriven behaviour and its resonance frequency increases with reduced flux.
		The cavity linewidth and -shape is significantly distrorted by low-frequency flux noise.	
		Panel $\textbf{c}$ shows a linecut in this regime with a very broad and noisy single cavity absorption dip.}
	\item{\textbf{Regime 2}: At approximately $\Phi_\perp/\Phi_0 \sim 0.476$, the drive is positioned close to the blue sideband of the cavity $\omega_\mathrm{d} \approx \omega_0 + \Omega_\mathrm{m}$ and multiple resonance lines appear in $S_{21}$, four absorption modes and four gain modes are visible.
		The frequency distance between two neighboring absorption modes or two neighboring gain modes is always the mechanical frequency $\Delta\omega \approx \Omega_\mathrm{m}$.
		The appearance of a multiple-modes response is characteristic for a cavity with a strongly oscillating resonance frequency.
		In our device, the behaviour in this regime is generated and explained by the optomechanical instability and mechanical self-oscillations induced by the parametric drive being at the same time a very strong optomechanical blue-sideband pump tone.
		In panel \textbf{b}, we plot the calculated optical Kerr damping based on our equations for the dynamical backaction and on the independently determined device parameters.
		It is clearly visible that regime 2 corresponds to negative optomechanical damping, which exceeds the intrinsic mechanical linewidth $\Gamma_\mathrm{m} \approx 2\pi\cdot 10\,$Hz and therefore the mechanical oscillator is in the instability regime of self-oscillations. 
		In panel \textbf{d}, a linecut of regime 2 is shown.
		The original cavity is labelled with "\textbf{s}" and its oscillation-induced replicas with "s1", "s2", and "s3".
		In addition, we observe 4 versions of the idler Kerr-mode as small peaks, where the original mode is labelled with "\textbf{i}" and its replicas with "i1", "i2", and "i3".}
	\item{\textbf{Regime 3}: For $0.462 \lesssim \Phi_\perp/\Phi_0 \lesssim 0.468$ the observed resonances return to a single absorption dip on the left side of the drive and a single small gain mode on the right side of the pump, indicating that the cavity frequency is not oscillating anymore.
		A linecut in this regime is shown in panel \textbf{e}.}
	\item{\textbf{Regime 4}: For $0.453 \lesssim \Phi_\perp/\Phi_0 \lesssim 0.462$ the negative optical damping once again exceeds the intrinsic mechanical linewidth and a second regime of instability is entered.
		A linecut in this regime is shown in panel \textbf{f}, where three signal-modes and three idler-modes are visible and labelled as in regime 2.
		As the frequency difference between signal and idler Kerr-mode in this regime is close to the mechanical frequency, each mode almost overlaps with one replica of the corresponding mirror-mode and they form dip-peak pairs.}
	\item{\textbf{Regime 5}: For $\Phi_\perp/\Phi_0 \lesssim 0.453$ the cavity jumps to the low-amplitude oscillation branch and the impact of both, the parametric drive and the dynamical backaction on the cavity lineshape become negligible.
		The cavity continues to shift with flux just as the undriven cavity would do.
		A linecut in this regime is not explicitly shown.}
	
\end{itemize}

We note again that for the calculation of the Kerr backaction and the instability regimes as shown in panel \textbf{b}, the only free parameter was the line attenuation for the drive tone, which was adjusted to $-81.5\,$dB for a good agreement between theory and experiment, a value very close to what was obtained from the line calibration in Supplementary Note~3 for operation point II.
The other parameters used here are $\omega_0, \mathcal{K}, \mathcal{F}, \Omega_\mathrm{m}, \Gamma_\mathrm{m}, \kappa_\mathrm{e}, \kappa_0, \kappa_1, n_\mathrm{crit}, \Gamma_2, \sigma_\Phi$ and they were all obtained from independent measurements or taken from theoretical estimates in the case of $g_0$.

\subsection*{Phonon population with Kerr backaction}

To calculate the equilibrium phonon population in the mechanical oscillator with Kerr backaction, we will use the linearized equations of motion for the quantum fields $\hat{a}, \hat{a}^\dagger$ and $\hat{b}, \hat{b}^\dagger$ representing the classical intracavity fluctuations $\delta\alpha, \delta\alpha^*$ and $\beta, \beta^*$ with $\delta x =x_\mathrm{zpf}\left(\beta + \beta^*\right)$, respectively.
For the input noise, we use $\hat{\zeta}_\mathrm{m}$ for the mechanical oscillator and $\hat{\xi}_\mathrm{i}, \hat{\xi}_\mathrm{e}$ for the internal and external cavity baths, respectively.
We denote the input noise operators of the cavity at different frequencies with subscripts "$+$" and "$-$".
The equations of motion become
\begin{eqnarray}
\dot{\hat{b}} & = & i\Omega_\mathrm{m}\hat{b} - \frac{\Gamma_\mathrm{m}}{2}\hat{b} - ig_\alpha\left(\hat{a}+\hat{a}^\dagger\right) + \sqrt{\Gamma_\mathrm{m}}\hat{\zeta} \\
\dot{\hat{a}} & = & \left[-i\left(\Delta_\mathrm{d} - 2\mathcal{K}n_\mathrm{d}\right) - \frac{\kappa}{2}\right]\hat{a} + i\mathcal{K}n_\mathrm{d}\hat{a}^\dagger - ig_\alpha\left(\hat{b} + \hat{b}^\dagger\right) + \sqrt{\kappa_\mathrm{e}}\hat{\xi}_\mathrm{e} + \sqrt{\kappa_\mathrm{i}}\hat{\xi}_\mathrm{i}
\end{eqnarray}
or in Fourier space and with the equations for the creation operators too
\begin{eqnarray}
\frac{\hat{b}}{\chi_0} & = & - ig_\alpha\left(\hat{a}+\hat{a}^\dagger\right) + \sqrt{\Gamma_\mathrm{m}}\hat{\zeta} \\
\frac{\hat{b}^\dagger}{\overline{\chi}_0} & = &  ig_\alpha\left(\hat{a}+\hat{a}^\dagger\right) + \sqrt{\Gamma_\mathrm{m}}\hat{\zeta}^\dagger \\
\frac{\hat{a}}{\chi_\mathrm{p}} & = &  i\mathcal{K}n_\mathrm{d}\hat{a}^\dagger - ig_\alpha\left(\hat{b} + \hat{b}^\dagger\right) + \sqrt{\kappa_\mathrm{e}}\hat{\xi}_\mathrm{e+} + \sqrt{\kappa_\mathrm{i}}\hat{\xi}_\mathrm{i+} \\
\frac{\hat{a}^\dagger}{\overline{\chi}_\mathrm{p}} & = &  -i\mathcal{K}n_\mathrm{d}\hat{a} + ig_\alpha\left(\hat{b} + \hat{b}^\dagger\right) +\sqrt{\kappa_\mathrm{e}}\hat{\xi}_\mathrm{e-}^\dagger + \sqrt{\kappa_\mathrm{i}}\hat{\xi}_\mathrm{i-}^\dagger.
\end{eqnarray}
Note that for the external bath of the cavity, we assumed a single port, although strictly speaking the temperature on the left side of the transmission line could be different from the right side.
As first step, we shorten the expressions by using
\begin{equation}
\hat{\mathcal{N}}_\pm = \sqrt{\kappa_\mathrm{e}}\hat{\xi}_\mathrm{e\pm} + \sqrt{\kappa_\mathrm{i}}\hat{\xi}_\mathrm{i\pm}, ~~~~~\mathcal{A} = -i\mathcal{K}n_\mathrm{d}\chi_\mathrm{p}
\end{equation}
and decouple the equations for $\hat{a}$ and $\hat{a}^\dagger$.
We get
\begin{eqnarray}
\hat{a} & = & -ig_\alpha\chi_\mathrm{g}\left(1-\overline{\mathcal{A}}\right)\left(\hat{b} + \hat{b}^\dagger\right) + \chi_\mathrm{g}\left(\hat{\mathcal{N}}_+ + \overline{\mathcal{A}}\hat{N}_-^\dagger\right) \\
\hat{a}^\dagger & = & ig_\alpha\overline{\chi}_\mathrm{g}\left(1 - \mathcal{A}\right)\left(\hat{b} + \hat{b}^\dagger\right) + \overline{\chi}_\mathrm{g}\left(\hat{\mathcal{N}}_-^\dagger + \mathcal{A}\hat{N}_+\right).
\end{eqnarray}
We can go even more compact
\begin{eqnarray}
\hat{a} & = & -ig_\alpha\chi_\mathrm{k}\left(\hat{b} + \hat{b}^\dagger\right) + \chi_\mathrm{g}\hat{\mathcal{M}}_\mathrm{+} \\
\hat{a}^\dagger & = & ig_\alpha\overline{\chi}_\mathrm{k}\left(\hat{b} + \hat{b}^\dagger\right) + \overline{\chi}_\mathrm{g}\hat{\mathcal{M}}_\mathrm{-}^\dagger.
\end{eqnarray}
with
\begin{equation}
\chi_\mathrm{k} = \chi_\mathrm{g}\left(1-\overline{\mathcal{A}}\right), ~~~~~ \hat{\mathcal{M}}_+ = \hat{\mathcal{N}}_+ + \overline{\mathcal{A}}\hat{\mathcal{N}}_-^\dagger, ~~~~~ \hat{\mathcal{M}}_-^\dagger = \hat{\mathcal{N}}_-^\dagger + \mathcal{A}\hat{\mathcal{N}}_+.
\end{equation}
As next step, we substitute
\begin{eqnarray}
\hat{b} + \hat{b}^\dagger & = & -ig_\alpha\left(\chi_0 - \overline{\chi}_0\right)\left(\hat{a}+\hat{a}^\dagger\right) + \sqrt{\Gamma_\mathrm{m}}\left(\chi_0\hat{\zeta} + \overline{\chi}_0\hat{\zeta}^\dagger\right) \\
& = & -ig_\alpha\left(\chi_0 - \overline{\chi}_0\right)\left(\hat{a}+\hat{a}^\dagger\right) + \hat{\mathcal{S}}
\end{eqnarray}
and obtain
\begin{eqnarray}
\hat{a} & = & -g_\alpha^2\chi_\mathrm{k}\left(\chi_0 - \overline{\chi}_0\right)\left(\hat{a}+\hat{a}^\dagger\right) - ig_\alpha\chi_\mathrm{k}\hat{\mathcal{S}} + \chi_\mathrm{g}\hat{\mathcal{M}}_\mathrm{+} \\
\hat{a}^\dagger & = & g_\alpha^2\overline{\chi}_\mathrm{k}\left(\chi_0 - \overline{\chi}_0\right)\left(\hat{a}+\hat{a}^\dagger\right) + ig_\alpha\overline{\chi}_\mathrm{k}\hat{\mathcal{S}}^\dagger + \overline{\chi}_\mathrm{g}\hat{\mathcal{M}}_\mathrm{-}^\dagger
\end{eqnarray}
or
\begin{equation}
\hat{a} + \hat{a}^\dagger = \frac{-ig_\alpha\left(\chi_\mathrm{k}\hat{\mathcal{S}} - \overline{\chi}_\mathrm{k}\hat{\mathcal{S}}^\dagger\right) + \chi_\mathrm{g}\hat{\mathcal{M}}_\mathrm{+} + \overline{\chi}_\mathrm{g}\hat{\mathcal{M}}_\mathrm{-}^\dagger}{1+g_\alpha^2\left(\chi_0 - \overline{\chi}_0\right)\left(\chi_\mathrm{k} - \overline{\chi}_\mathrm{k}\right)}
\end{equation}
with which we can finally express $\hat{b}$ only as a function of the input noise operators
\begin{eqnarray}
\hat{b} & = &  -ig_\alpha\chi_0^\mathrm{eff} \chi_\mathrm{g}\hat{\mathcal{M}}_\mathrm{+} -ig_\alpha\chi_0^\mathrm{eff} \overline{\chi}_\mathrm{g}\hat{\mathcal{M}}_\mathrm{-}^\dagger -g_\alpha^2\chi_0^\mathrm{eff}\left(\chi_\mathrm{k}\hat{\mathcal{S}} - \overline{\chi}_\mathrm{k}\hat{\mathcal{S}}^\dagger\right) + \chi_0\sqrt{\Gamma_\mathrm{m}}\hat{\zeta}\\
& = & -ig_\alpha\chi_0^\mathrm{eff}\chi_\mathrm{k}\left( \sqrt{\kappa_\mathrm{e}}\hat{\xi}_\mathrm{e+} + \sqrt{\kappa_\mathrm{i}}\hat{\xi}_\mathrm{i+}\right) -ig_\alpha\chi_0^\mathrm{eff}\overline{\chi}_\mathrm{k}\left( \sqrt{\kappa_\mathrm{e}}\hat{\xi}_\mathrm{e-}^\dagger + \sqrt{\kappa_\mathrm{i}}\hat{\xi}_\mathrm{i-}^\dagger\right) \\
& & + ~\sqrt{\Gamma_\mathrm{m}}\chi_0\left[1-g_\alpha^2\chi_0^\mathrm{eff}\left(\chi_\mathrm{k} - \overline{\chi}_\mathrm{k}\right)\right]\hat{\zeta} - \sqrt{\Gamma_\mathrm{m}}\chi_0^\mathrm{eff}g_\alpha^2\overline{\chi}_0\left(\chi_\mathrm{k} - \overline{\chi}_\mathrm{k}\right)\hat{\zeta}^\dagger
\end{eqnarray}
where the effective mechanical susceptibiliy is given by
\begin{equation}
\chi_0^\mathrm{eff} = \frac{\chi_0}{1+g_\alpha^2\left(\chi_0 - \overline{\chi}_0\right)\left(\chi_\mathrm{k} - \overline{\chi}_\mathrm{k}\right)} \approx \frac{\chi_0}{1+g_\alpha^2\chi_0\left(\chi_\mathrm{k} - \overline{\chi}_\mathrm{k}\right)}.
\end{equation}
The last approximation is valid for a high-$Q_\mathrm{m}$ mechanical oscillator.
Using the common relations for the expectation values of the noise correlators, we can calculate the phonon power spectral density in the mechanical resonator under dynamical Kerr backaction as
\begin{eqnarray}
\langle \hat{b}^\dagger\hat{b}\rangle & = &  g_\alpha^2\left|\chi_0^\mathrm{eff}\right|^2|\chi_\mathrm{k}|^2 \kappa n_\mathrm{c} + g_\alpha^2\left|\chi_0^\mathrm{eff}\right|^2|\overline{\chi}_\mathrm{k}|^2 \kappa \left(n_\mathrm{c}+1\right)  \\
& & +~\left|\chi_0^\mathrm{eff}\right|^2 \left| 1-g_\alpha^2\overline{\chi}_0(\chi_\mathrm{k} - \overline{\chi}_\mathrm{k})\right|^2 \Gamma_\mathrm{m} n_\mathrm{m}^\mathrm{th} \\
& & +~g_\alpha^4\left|\chi_0^\mathrm{eff}\right|^2 |\overline{\chi}_0|^2\left|\chi_\mathrm{k} - \overline{\chi}_\mathrm{k}\right|^2 \Gamma_\mathrm{m} (n_\mathrm{m}^\mathrm{th}+1)
\end{eqnarray}
and the total phonon number via
\begin{equation}
n_\mathrm{m}^\mathrm{KDB} = \int_{-\infty}^\infty \langle \hat{b}^\dagger \hat{b}\rangle \frac{d\Omega}{2\pi}.
\end{equation}
This integration is performed numerically.
We also note that we assumed a constant effective cavity bath occupation
\begin{equation}
n_\mathrm{c} = \frac{\kappa_\mathrm{e}n_\mathrm{e} + \kappa_\mathrm{i}n_\mathrm{i}}{\kappa}
\end{equation}
in the relevant frequency range.
In Supplementary Fig.~\ref{fig:Phonons} we discuss the phonon occupation that results from a parametrically driven Kerr cavity for the relevant operation points I and II.
Related discussions on optomechanical cooling using a Kerr cavity can be found in Refs.~\cite{SNation08, SAsjad19, SGan19}.
\begin{figure}[h!]
	\centerline {\includegraphics[trim={1.cm 14.5cm 3.cm 2.5cm},clip=True,scale=0.58]{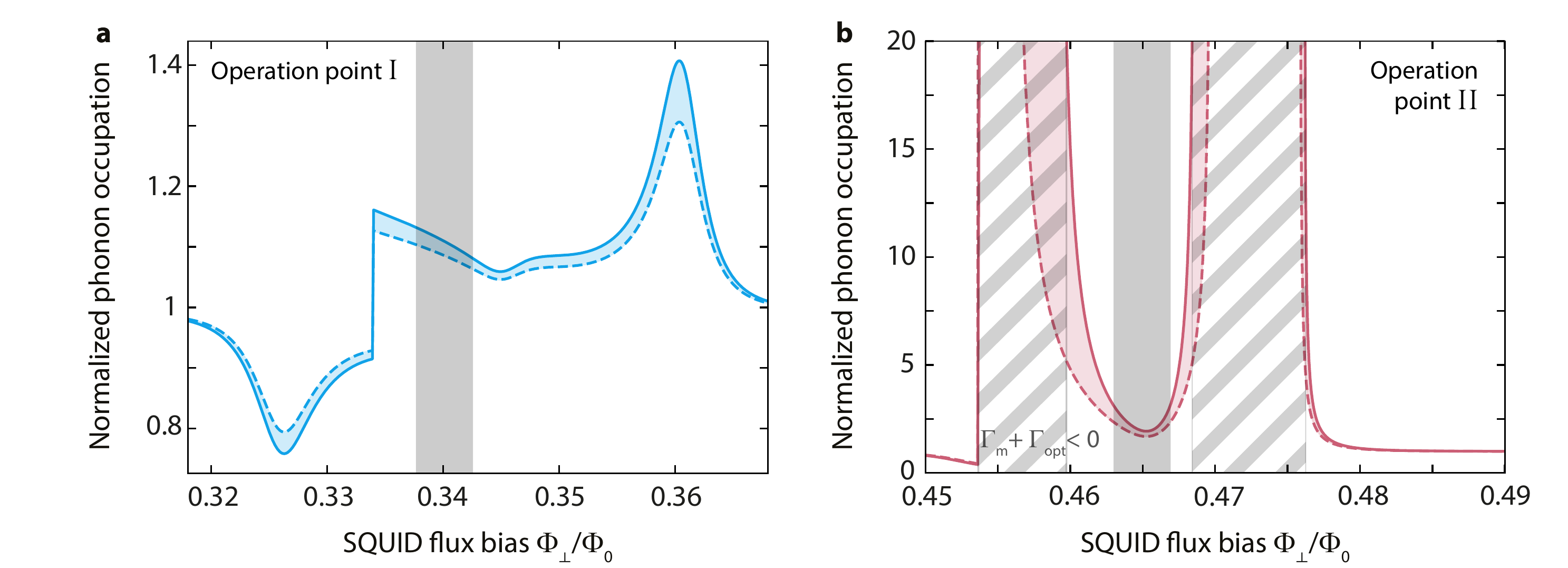}}
	\caption{\textsf{\textbf{Phonon occupation in the mechanical oscillator due to the parametric drive.} Due to the dynamical backaction induced by the parametric drive, the phonon occupation of the mechanical oscilltor will deviate from the steady-state equilibrium value. In Fig.~2 of the main paper, we show the dynamical backaction of the drive vs flux bias in the SQUID. In \textbf{a}, we show the corresponding effect on the phonon occupation, normalized to the (maximum) case without parametric drive $n_\mathrm{m}^\mathrm{th} = 80$. For $\Phi_\perp/\Phi_0 \sim 0.36$, the drive is located on the blue sideband of the cavity and amplification/heating is observed. For $\Phi_\perp/\Phi_0 \sim 0.327$, the drive is located on the red sideband and cooling of the mechanical mode is visible. In the regime of interest, where our experiments take place (gray area), the effective occuption is increased by about $10\%$. The dashed and solid blue lines correspond to $g_0 = 2\pi\cdot 1.8\,$kHz and  $g_0 = 2\pi\cdot 2\,$kHz, respectively. All other parameters are taken from independent measurements. In \textbf{b}, we show the equivalent occupation at operation point II with $g_0 = 2\pi\cdot 3.4\,$kHz and $g_0 = 2\pi\cdot 3.7\,$kHz for the dashed and solid red lines, respectively. The striped areas correspond to the instability regime $\Gamma_\mathrm{m} + \Gamma_\mathrm{opt} < 0$ in the case $g_0 = 2\pi\cdot 3.7\,$kHz. All parameters are identical to the ones used for the backaction calculation shown in Fig.~\ref{fig:KerrSelfOsc}\textbf{b}, except for the drive power which is $1\,$dB reduced as this is the regime for our cooling experiments. In the experimentally relevant flux regime, marked by solid gray shading, the phonon occupation is increased by about a factor of $\sim 2.5$ due to dynamical backaction.}}
	\label{fig:Phonons}
\end{figure}

\section*{Supplementary Note 8: Equations of motion for linearized multi-tone Kerr optomechanics}

\subsection*{Three-tone linearization}
Finally, we discuss the linearized equations of motion with three input fields, i.e.,
\begin{equation}
S_\mathrm{in} = S_\mathrm{d}e^{i\left(\omega_\mathrm{d}t + \phi_\mathrm{d}\right)} + S_\mathrm{p}e^{i\omega_\mathrm{p}t} + S_0(t)e^{i\omega_\mathrm{p}t}
\end{equation}
where $S_0(t)$ is a third, weak probe field.
We choose as Ansatz
\begin{eqnarray}
x & = & x_0 + \delta x \\
\alpha & = & \alpha_\mathrm{d}e^{i\omega_\mathrm{d}t} + \gamma_-e^{i\omega_\mathrm{p}t} + \gamma_+e^{i\left(2\omega_\mathrm{d} - \omega_\mathrm{p}\right)t} + \delta\alpha(t)e^{i\omega_\mathrm{p}t}
\end{eqnarray}
with real-valued and time-independent $\alpha_\mathrm{d}$, complex-valued and time-independent $\gamma_-, \gamma_+$ and complex-valued and time-dependent $\delta\alpha$.
For the mechanical oscillator, we get with this
\begin{eqnarray}
\delta\ddot{x} & = & -\Omega_\mathrm{m}^2x_0 - \Omega_\mathrm{m}^2\delta x - \Gamma_\mathrm{m}\delta\dot{x} \nonumber \\
& & +~\frac{\hbar G}{m}\left[\alpha_\mathrm{d}^2 + |\gamma_-|^2 + |\gamma_+|^2 + |\delta\alpha|^2 \right] \nonumber \\
& & +~\frac{\hbar G \alpha_\mathrm{d}}{m}\left[\gamma_-^* + \gamma_+^*e^{-i2\Omega_\mathrm{dp}t} + \delta\alpha^*\right]e^{i\Omega_\mathrm{dp}t} \nonumber \\
& & +~\frac{\hbar G \alpha_\mathrm{d}}{m}\left[\gamma_- + \gamma_+e^{i2\Omega_\mathrm{dp}t} + \delta\alpha\right]e^{-i\Omega_\mathrm{dp}t} \nonumber \\
& & +~\frac{\hbar G}{m}\left[ \gamma_-\gamma_+^*e^{-i2\Omega_\mathrm{dp}t} + \gamma_-^*\gamma_+e^{i2\Omega_\mathrm{dp}t}\right] \nonumber \\
& & +~\frac{\hbar G}{m}\left[ \gamma_-\delta\alpha^* + \gamma_-^*\delta\alpha\right] \nonumber \\
& & +~\frac{\hbar G}{m}\left[ \gamma_+\delta\alpha^*e^{i2\Omega_\mathrm{dp}t} + \gamma_+^*\delta\alpha e^{-i2\Omega_\mathrm{dp}t}\right].
\end{eqnarray}
In our experiments presented in the main paper we choose both $\Omega_\mathrm{dp}$ and $2\Omega_\mathrm{dp}$ to be very far-detuned from the mechanical resonance frequency $\Omega_\mathrm{m}$ and therefore we can neglect the pure driving force terms with $\pm\Omega_\mathrm{dp}$ and $\pm2\Omega_\mathrm{dp}$.
After omitting the steady-state solution we get
\begin{equation}
\delta\ddot{x} = -\Omega_\mathrm{m}^2\delta x - \Gamma_\mathrm{m}\delta\dot{x} + \frac{\hbar G \alpha_\mathrm{d}}{m}\left[\delta\alpha^*e^{i\Omega_\mathrm{dp}t} + \delta\alpha e^{-i\Omega_\mathrm{dp}t} \right] + \frac{\hbar G }{m}\left[\gamma_-\delta\alpha^* + \gamma_-^*\delta\alpha \right] + \frac{\hbar G}{m}\left[\gamma_+\delta\alpha^*e^{i2\Omega_\mathrm{dp}t} + \gamma_+^*\delta\alpha e^{-i2\Omega_\mathrm{dp}t} \right].
\end{equation}
For the intracavity field we get
\begin{eqnarray}
& & i\omega_\mathrm{d}\alpha_\mathrm{d}e^{i\Omega_\mathrm{dp}t} + i\omega_\mathrm{p}\gamma_- + i(2\omega_\mathrm{d} - \omega_\mathrm{p})\gamma_+ e^{i2\Omega_\mathrm{dp}t} + i\omega_\mathrm{p}\delta\alpha + \delta\dot{\alpha} \nonumber \\
& = & \left[ i\omega_0 - \frac{\kappa}{2}\right]\alpha_\mathrm{d}e^{i\Omega_\mathrm{dp}t} + \left[i\omega_0 - \frac{\kappa}{2}\right]\gamma_- + \left[i\omega_0 - \frac{\kappa}{2}\right]\gamma_+ e^{i2\Omega_\mathrm{dp}t} + \left[i\omega_0 - \frac{\kappa}{2}\right]\delta\alpha \nonumber \\
& & -~iGx_0\left( \alpha_\mathrm{d}e^{i\Omega_\mathrm{dp}t} + \gamma_- + \gamma_+e^{i2\Omega_\mathrm{dp}t} + \delta\alpha \right) - iG\delta x\left( \alpha_\mathrm{d}e^{i\Omega_\mathrm{dp}t} + \gamma_- + \gamma_+e^{i2\Omega_\mathrm{dp}t} + \delta\alpha \right) \nonumber \\
& & +~i\mathcal{K}\left[ \alpha_\mathrm{d}^2 + |\gamma_-|^2 + |\gamma_+|^2 + |\delta\alpha|^2 \right]\left( \alpha_\mathrm{d}e^{i\Omega_\mathrm{dp}t} + \gamma_- + \gamma_+e^{i2\Omega_\mathrm{dp}t} + \delta\alpha \right) \nonumber \\
& & +~ i\mathcal{K}\alpha_\mathrm{d}\left[ \gamma_-^* + \gamma_+^*e^{-i2\Omega_\mathrm{dp}t} + \delta\alpha^* \right]e^{i\Omega_\mathrm{dp}t}\left( \alpha_\mathrm{d}e^{i\Omega_\mathrm{dp}t} + \gamma_- + \gamma_+e^{i2\Omega_\mathrm{dp}t} + \delta\alpha \right) \nonumber \\
& & +~i\mathcal{K} \alpha_\mathrm{d}\left[ \gamma_- + \gamma_+e^{i2\Omega_\mathrm{dp}t} + \delta\alpha \right] e^{-i\Omega_\mathrm{dp}t} \left( \alpha_\mathrm{d}e^{i\Omega_\mathrm{dp}t} + \gamma_- + \gamma_+e^{i2\Omega_\mathrm{dp}t} + \delta\alpha \right) \nonumber \\
& & +~ i\mathcal{K}\left[ \gamma_-\gamma_+^* e^{-i2\Omega_\mathrm{dp}t} + \gamma_-^*\gamma_+ e^{i2\Omega_\mathrm{dp}t} \right]\left( \alpha_\mathrm{d}e^{i\Omega_\mathrm{dp}t} + \gamma_- + \gamma_+e^{i2\Omega_\mathrm{dp}t} + \delta\alpha \right) \nonumber \\
& & +~ i\mathcal{K}\left[ \gamma_-\delta\alpha^* + \gamma_-^*\delta\alpha \right]\left( \alpha_\mathrm{d}e^{i\Omega_\mathrm{dp}t} + \gamma_- + \gamma_+e^{i2\Omega_\mathrm{dp}t} + \delta\alpha \right) \nonumber \\
& & +~ i\mathcal{K}\left[ \gamma_+\delta\alpha^* e^{i2\Omega_\mathrm{dp}t} + \gamma_+^*\delta\alpha e^{-i2\Omega_\mathrm{dp}t} \right]\left( \alpha_\mathrm{d}e^{i\Omega_\mathrm{dp}t} + \gamma_- + \gamma_+e^{i2\Omega_\mathrm{dp}t} + \delta\alpha \right) \nonumber \\
& & + ~ i\sqrt{\frac{\kappa_\mathrm{e}}{2}} \left(S_\mathrm{d}e^{i\left(\Omega_\mathrm{dp}t + \phi_\mathrm{d}\right)} + S_\mathrm{p} + S_0\right).
\end{eqnarray}
Now we perform the linearization.
First with respect to the Kerr drive, i.e., we omit all terms that describe amplification induced by $\gamma_-, \gamma_+$ and include all steady-state shifts as above in $\mathcal{K}n_\mathrm{d}$. 
This corresponds to assuming $\alpha_\mathrm{d} \gg |\gamma_-|, |\gamma_+| \gg |\delta\alpha|, |Gx_0|$.
In addition, we omit terms proportional to $\delta\alpha\delta x$, i.e., do the optomechanical linearization.
The result is
\begin{eqnarray}
& & i\omega_\mathrm{d}\alpha_\mathrm{d}e^{i\Omega_\mathrm{dp}t} + i\omega_\mathrm{p}\gamma_-  + i(2\omega_\mathrm{d} - \omega_\mathrm{p})\gamma_+ e^{i2\Omega_\mathrm{dp}t} + i\omega_\mathrm{p}\delta\alpha + \delta\dot{\alpha} \nonumber \\
& = & \left[i\left(\omega_0 + \mathcal{K}n_\mathrm{d}\right) - \frac{\kappa}{2}\right]\alpha_\mathrm{d}e^{i\Omega_\mathrm{dp}t} + \left[i\left(\omega_0 + 2\mathcal{K}n_\mathrm{d}\right) - \frac{\kappa}{2}\right]\gamma_- + \left[i\left(\omega_0 + 2\mathcal{K}n_\mathrm{d}\right) - \frac{\kappa}{2}\right]\gamma_+e^{i2\Omega_\mathrm{dp}t} + \left[i\left(\omega_0 + 2\mathcal{K}n_\mathrm{d}\right) - \frac{\kappa}{2}\right]\delta\alpha \nonumber \\
& & -~iG\delta x\left( \alpha_\mathrm{d}e^{i\Omega_\mathrm{dp}t} + \gamma_- + \gamma_+e^{i2\Omega_\mathrm{dp}t} \right) + i\mathcal{K}n_\mathrm{d}\left(\gamma_-^*e^{i2\Omega_\mathrm{dp}t} + \gamma_+^* + \delta\alpha^*e^{i2\Omega_\mathrm{dp}t} \right) \nonumber \\
& & \DB{+~i2\mathcal{K}\alpha_\mathrm{d}\left[ \gamma_-^* + \gamma_+  \right]\delta\alpha e^{i\Omega_\mathrm{dp}t} + i2\mathcal{K}\alpha_\mathrm{d}\left[ \gamma_+^* + \gamma_-  \right]\delta\alpha e^{-i\Omega_\mathrm{dp}t} + i2\mathcal{K}\alpha_\mathrm{d}e^{i\Omega_\mathrm{dp}t}\gamma_-\delta\alpha^* + i2\mathcal{K}\alpha_\mathrm{d}e^{i3\Omega_\mathrm{dp}t}\gamma_+\delta\alpha^*}  \nonumber\\
& & + ~ i\sqrt{\frac{\kappa_\mathrm{e}}{2}} \left(S_\mathrm{d}e^{i\left(\Omega_\mathrm{dp}t + \phi_\mathrm{d}\right)} + S_\mathrm{p} + S_0\right).
\label{eq:finaltrain}
\end{eqnarray}
where the blue terms correspond to non-degenerate four-wave mixing contributions.
We split this equation into four equations
\begin{eqnarray}
\left[\frac{\kappa}{2} + i\left(\Delta_\mathrm{d} - \mathcal{K}n_\mathrm{d}\right) \right]\alpha_\mathrm{d} & = & i\sqrt{\frac{\kappa_\mathrm{e}}{2}}S_\mathrm{d}e^{i\phi_\mathrm{d}} \nonumber \\
\left[\frac{\kappa}{2} + i\left(\Delta_\mathrm{d} - 2\mathcal{K}n_\mathrm{d} - \Omega_\mathrm{dp}\right) \right]\gamma_- - i\mathcal{K}n_\mathrm{d}\gamma_+^* & = & i\sqrt{\frac{\kappa_\mathrm{e}}{2}}S_\mathrm{p} \nonumber \\
\left[\frac{\kappa}{2} + i\left(\Delta_\mathrm{d} - 2\mathcal{K}n_\mathrm{d} + \Omega_\mathrm{dp}\right) \right]\gamma_+ - i\mathcal{K}n_\mathrm{d}\gamma_-^* & = & 0 \nonumber \\
\delta\dot{\alpha} + \left[\frac{\kappa}{2} + i\left(\Delta_\mathrm{d} - 2\mathcal{K}n_\mathrm{d} - \Omega_\mathrm{dp}\right)\right]\delta\alpha - i\mathcal{K}n_\mathrm{d}\delta\alpha^*e^{i2\Omega_\mathrm{dp}t} + iG\delta x\left(\alpha_\mathrm{d}e^{i\Omega_\mathrm{dp}t} + \gamma_- + \gamma_+e^{i2\Omega_\mathrm{dp}t}\right) &&\nonumber \\
\DB{-~i2\mathcal{K}\alpha_\mathrm{d}\left(\left[\gamma_-^* + \gamma_+ \right]\delta\alpha e^{i\Omega_\mathrm{dp}t} + \left[\gamma_- + \gamma_+^* \right]\delta\alpha e^{-i\Omega_\mathrm{dp}t} + \gamma_-\delta\alpha^*e^{i\Omega_\mathrm{dp}t} + \gamma_+\delta\alpha^*e^{i3\Omega_\mathrm{dp}t}\right) }
& = & i\sqrt{\frac{\kappa_\mathrm{e}}{2}}S_0. \nonumber
\end{eqnarray}
The first three of these equations are the Kerr equations for a linearized two-tone driving.
Therefore, the first step of the solution is to find $n_\mathrm{d}$ using the third order polynomial as described above.
Afterwards, we solve the equations for $\gamma_-$ and $\gamma_+$.
With all these numbers at hand, we proceed to solve for the optomechanical field components.
The Fourier transforms of the remaining two optomechanical equations of motion are
\begin{eqnarray}
\frac{\delta x(\Omega)}{\chi_\mathrm{m}(\Omega)} & = & \frac{\hbar G}{m}\left[\gamma_-^*\delta\alpha(\Omega) + \gamma_-\delta\alpha^*(-\Omega)\right] + \frac{\hbar G \alpha_\mathrm{d}}{m}\left[\delta\alpha(\Omega + \Omega_\mathrm{dp}) + \delta\alpha^*(-\Omega + \Omega_\mathrm{dp})\right] \\
& & +~\frac{\hbar G}{m}\left[\gamma_+^*\delta\alpha(\Omega + 2\Omega_\mathrm{dp}) + \gamma_+\delta\alpha^*(-\Omega + 2\Omega_\mathrm{dp})\right] \\
\frac{\delta\alpha(\Omega)}{\chi_\mathrm{p}(\Omega)} & = & i\mathcal{K} n_\mathrm{d}\delta\alpha^*(-\Omega + 2\Omega_\mathrm{dp}) - iG\gamma_-\delta x(\Omega) - iG\alpha_\mathrm{d}\delta x (\Omega - \Omega_\mathrm{dp}) - iG\gamma_+\delta x(\Omega - 2\Omega_\mathrm{dp}) \nonumber\\
& & 
\DB{+~i2\mathcal{K}\alpha_\mathrm{d}\left[\gamma_-^* + \gamma_+ \right]\delta\alpha(\Omega - \Omega_\mathrm{dp}) + i2\mathcal{K}\alpha_\mathrm{d}\left[\gamma_- + \gamma_+^*\right]\delta\alpha(\Omega + \Omega_\mathrm{dp})} \nonumber \\
& & \DB{+~i2\mathcal{K}\alpha_\mathrm{d}\gamma_-\delta\alpha^*(-\Omega + \Omega_\mathrm{dp}) + i2\mathcal{K}\alpha_\mathrm{d}\gamma_+\delta\alpha^*(-\Omega + 3\Omega_\mathrm{dp}) } \nonumber \\
& & + i\sqrt{\frac{\kappa_\mathrm{e}}{2}}S_0(\Omega).
\end{eqnarray}
We introduce now the short-version
\begin{eqnarray}
f(\Omega + j\Omega_\mathrm{dp}) & = & f_j \\
f^*(-\Omega + j\Omega_\mathrm{dp}) & = & \overline{f}_j
\end{eqnarray}
for any function $f$ with $j$ being an integer, with which the equations of motion become
\begin{eqnarray}
\frac{\delta x_{0}}{\chi_\mathrm{m,0}} & = & \frac{\hbar G}{m}\left[\gamma_-^*\delta\alpha_{0} + \gamma_-\overline{\delta\alpha}_{0}\right] + \frac{\hbar G \alpha_\mathrm{d}}{m}\left[\delta\alpha_{1} + \overline{\delta\alpha}_{1}\right] +\frac{\hbar G}{m}\left[\gamma_+^*\delta\alpha_{2} + \gamma_+\overline{\delta\alpha}_{2}\right] \\
\frac{\delta\alpha_{0}}{\chi_\mathrm{p,0}} & = & \DB{i2\mathcal{K}\alpha_\mathrm{d}\gamma_+\overline{\delta\alpha}_{3}} + i\mathcal{K} n_\mathrm{d}\overline{\delta\alpha}_{2} + \DB{ i2\mathcal{K}\alpha_\mathrm{d}\gamma_-\overline{\delta\alpha}_{1} } \nonumber\\
& & 
\DB{+~i2\mathcal{K}\alpha_\mathrm{d}\left[\gamma_-^* + \gamma_+ \right]\delta\alpha_{-1} + i2\mathcal{K}\alpha_\mathrm{d}\left[\gamma_- + \gamma_+^*\right]\delta\alpha_{1}} \nonumber \\
& & - ~ iG\gamma_-\delta x_{0}- iG\alpha_\mathrm{d}\delta x_{-1} - iG\gamma_+\delta x_{-2} \nonumber \\
& & + ~ i\sqrt{\frac{\kappa_\mathrm{e}}{2}}S_{0,0}.
\end{eqnarray}
Using now
\begin{eqnarray}
\frac{\overline{\delta\alpha}_{2}}{\overline{\chi}_\mathrm{p,2}} & = & \DB{- ~ i2\mathcal{K}\alpha_\mathrm{d}\gamma_+^*\delta\alpha_{1}} - i\mathcal{K} n_\mathrm{d}\delta\alpha_{0} - \DB{ i2\mathcal{K}\alpha_\mathrm{d}\gamma_-^*\delta\alpha_{-1} } \nonumber\\
& & 
\DB{- ~ i2\mathcal{K}\alpha_\mathrm{d}\left[\gamma_- + \gamma_+^* \right]\overline{\delta\alpha}_{1} - i2\mathcal{K}\alpha_\mathrm{d}\left[\gamma_-^* + \gamma_+\right]\overline{\delta\alpha}_{3}} \nonumber \\
& & + ~ iG\gamma_-^*\delta x_{-2} + iG\alpha_\mathrm{d}\delta x_{-1} + iG\gamma_+^*\delta x_{0} \nonumber \\
& & - ~ i\sqrt{\frac{\kappa_\mathrm{e}}{2}}\overline{S}_{0,2}.
\end{eqnarray}
we can eliminate the parametric field-contribution and obtain
\begin{eqnarray}
\frac{\delta\alpha_{0}}{\chi_\mathrm{g,0}} & = & \DB{i2\mathcal{K}\alpha_\mathrm{d}\left[\gamma_+ - \overline{\mathcal{A}}_{2}\left( \gamma_-^* + \gamma_+ \right) \right]\overline{\delta\alpha}_{3} + i2\mathcal{K}\alpha_\mathrm{d}\left[\gamma_- - \overline{\mathcal{A}}_{2}\left( \gamma_- + \gamma_+^* \right) \right]\overline{\delta\alpha}_{1} } \nonumber\\
& & 
\DB{+~i2\mathcal{K}\alpha_\mathrm{d}\left[\gamma_-^* + \gamma_+ - \overline{\mathcal{A}}_{2}\gamma_-^* \right]\delta\alpha_{-1} + i2\mathcal{K}\alpha_\mathrm{d}\left[\gamma_- + \gamma_+^*  - \overline{\mathcal{A}}_{2}\gamma_+^* \right]\delta\alpha_{1}} \nonumber \\
& & - ~ iG\left[\gamma_- - \overline{\mathcal{A}}_{2}\gamma_+^* \right]\delta x_{0}- iG\alpha_\mathrm{d}\left[1  - \overline{\mathcal{A}}_{2}\right]\delta x_{-1} - iG\left[\gamma_+  - \overline{\mathcal{A}}_{2}\gamma_-^* \right]\delta x_{-2} \nonumber \\
& & + ~ i\sqrt{\frac{\kappa_\mathrm{e}}{2}}S_{0,0} -  i\overline{\mathcal{A}}_{2}\sqrt{\frac{\kappa_\mathrm{e}}{2}}\overline{S}_{0,2}
\end{eqnarray}
where we used
\begin{equation}
\mathcal{A}_j = -i\mathcal{K}n_\mathrm{d}\chi_{\mathrm{p}, j}.
\end{equation}
Instead of for $j = 0$, this equation and its parametric counterpart can easily be written down for general $j$
\begin{eqnarray}
\frac{\delta\alpha_{j}}{\chi_{\mathrm{g}, j}} & = & \DB{i2\mathcal{K}\alpha_\mathrm{d}\left[\gamma_+ - \overline{\mathcal{A}}_{2 - j}\left( \gamma_-^* + \gamma_+ \right) \right]\overline{\delta\alpha}_{3 - j} + i2\mathcal{K}\alpha_\mathrm{d}\left[\gamma_- - \overline{\mathcal{A}}_{2 - j}\left( \gamma_- + \gamma_+^* \right) \right]\overline{\delta\alpha}_{1 - j} } \nonumber\\
& & 
\DB{+~i2\mathcal{K}\alpha_\mathrm{d}\left[\gamma_-^* + \gamma_+ - \overline{\mathcal{A}}_{2 - j}\gamma_-^* \right]\delta\alpha_{j-1} + i2\mathcal{K}\alpha_\mathrm{d}\left[\gamma_- + \gamma_+^*  - \overline{\mathcal{A}}_{2 - j}\gamma_+^* \right]\delta\alpha_{j + 1}} \nonumber \\
& & - ~ iG\left[\gamma_- - \overline{\mathcal{A}}_{2 - j}\gamma_+^* \right]\delta x_{j}- iG\alpha_\mathrm{d}\left[1  - \overline{\mathcal{A}}_{2 - j}\right]\delta x_{j-1} - iG\left[\gamma_+  - \overline{\mathcal{A}}_{2 - j}\gamma_-^* \right]\delta x_{j-2} \nonumber \\
& & + ~ i\sqrt{\frac{\kappa_\mathrm{e}}{2}}S_{0,j} -  i\overline{\mathcal{A}}_{2 - j}\sqrt{\frac{\kappa_\mathrm{e}}{2}}\overline{S}_{0,2-j}
\label{eqn:3tonealpha}
\end{eqnarray}
and
\begin{eqnarray}
\frac{\overline{\delta\alpha}_{j}}{\overline{\chi}_{\mathrm{g}, j}} & = & \DB{-~i2\mathcal{K}\alpha_\mathrm{d}\left[\gamma_+^* - \mathcal{A}_{2 - j}\left( \gamma_- + \gamma_+^* \right) \right]\delta\alpha_{3 - j} - i2\mathcal{K}\alpha_\mathrm{d}\left[\gamma_-^* - \mathcal{A}_{2 - j}\left( \gamma_-^* + \gamma_+ \right) \right]\delta\alpha_{1 - j} } \nonumber\\
& & 
\DB{-~i2\mathcal{K}\alpha_\mathrm{d}\left[\gamma_- + \gamma_+^* - \mathcal{A}_{2 - j}\gamma_- \right]\overline{\delta\alpha}_{j-1} - i2\mathcal{K}\alpha_\mathrm{d}\left[\gamma_-^* + \gamma_+  - \mathcal{A}_{2 - j}\gamma_+ \right]\overline{\delta\alpha}_{j + 1}} \nonumber \\
& & + ~ iG\left[\gamma_-^* - \mathcal{A}_{2 - j}\gamma_+ \right]\delta x_{-j} + iG\alpha_\mathrm{d}\left[1  - \mathcal{A}_{2 - j}\right]\delta x_{1 - j} + iG\left[\gamma_+^*  - \mathcal{A}_{2 - j}\gamma_- \right]\delta x_{2-j} \nonumber \\
& & - ~ i\sqrt{\frac{\kappa_\mathrm{e}}{2}}\overline{S}_{0,j} + i\mathcal{A}_{2 - j}\sqrt{\frac{\kappa_\mathrm{e}}{2}}S_{0,2-j}
\end{eqnarray}
Note the particular indices in the mechanical contributions, which are due to $\overline{\delta x}_{n - j} = \delta x_{j - n}$.

\section*{Supplementary Note 9: Three-tone dynamical Kerr backaction}

To calculate the dynamical backaction induced by the doubly-driven Kerr cavity, we omit any probe drives $S_{0, j}$ for now.
Also, we only keep terms linear in $\gamma_-, \gamma_+$.
Finally, we omit $\delta x_j$ for $j \neq 0, 1, 2$, as these will not contribute to first order to the dynamical backaction.
Under these conditions, we can write down the remaining terms in the four next-iteration field components contained in Eq.~(\ref{eqn:3tonealpha}).

\begin{eqnarray}
\frac{\delta\alpha_{-1}}{\chi_{\mathrm{g}, -1}} & = &  - ~ 2G\mathcal{K}n_\mathrm{d}\left[\gamma_-\left(1 - \overline{\mathcal{A}}_{3}\right)- \gamma_+^*\overline{\mathcal{A}}_{3} \right]\left[1 - \mathcal{A}_0  \right]\overline{\chi}_{\mathrm{g}, 2}\delta x_{-1} + 2G\mathcal{K}n_\mathrm{d}\left[\gamma_- + \gamma_+^*\left( 1 - \overline{\mathcal{A}}_{3}\right) \right]\left[1 - \overline{\mathcal{A}}_2  \right]\chi_{\mathrm{g}, 0}\delta x_{-1} \nonumber \\
& & - ~ iG\left[\gamma_- - \gamma_+^*\overline{\mathcal{A}}_{3} \right]\delta x_{-1}- iG\alpha_\mathrm{d}\left[1  - \overline{\mathcal{A}}_{3}\right]\delta x_{-2} \\
\frac{\delta\alpha_{1}}{\chi_{\mathrm{g}, 1}} & = & - ~ 2G\mathcal{K} n_\mathrm{d}\left[\gamma_+\left(1 - \overline{\mathcal{A}}_{1}\right) - \gamma_-^*\overline{\mathcal{A}}_{1} \right] \left[1 - \mathcal{A}_0  \right]\overline{\chi}_{\mathrm{g}, 2}\delta x_{-1} + 2G\mathcal{K} n_\mathrm{d}\left[\gamma_+ + \gamma_-^*\left(1 - \overline{\mathcal{A}}_{1} \right) \right]\left[1 - \overline{\mathcal{A}}_2  \right]\chi_{\mathrm{g}, 0} \delta x_{-1}  \nonumber \\
& & - ~ iG\alpha_\mathrm{d}\left[1  - \overline{\mathcal{A}}_{1}\right]\delta x_{0} - iG\left[\gamma_+  - \gamma_-^*\overline{\mathcal{A}}_{1} \right]\delta x_{-1} \\
\frac{\overline{\delta\alpha}_{3}}{\overline{\chi}_{\mathrm{g}, 3}} & = & -~2G\mathcal{K} n_\mathrm{d}\left[\gamma_+^*\left( 1 - \mathcal{A}_{-1}\right) - \gamma_-\mathcal{A}_{-1} \right]\left[1 - \overline{\mathcal{A}}_2  \right]\chi_{\mathrm{g}, 0} \delta x_{-1} + 2 G \mathcal{K}n_\mathrm{d}\left[\gamma_+^* + \gamma_- \left(1 - \mathcal{A}_{-1}\right) \right] \left[1 - \mathcal{A}_0  \right]\overline{\chi}_{\mathrm{g}, 2} \delta x_{-1}  \nonumber \\
& & + ~ iG\alpha_\mathrm{d}\left[1  - \mathcal{A}_{-1}\right]\delta x_{-2} + iG\left[\gamma_+^*  - \gamma_- \mathcal{A}_{-1} \right]\delta x_{-1} \\
\frac{\overline{\delta\alpha}_{1}}{\overline{\chi}_{\mathrm{g}, 1}} & = & -~ 2 G \mathcal{K}n_\mathrm{d}\left[\gamma_-^*\left( 1 - \mathcal{A}_{1}\right) - \gamma_+ \mathcal{A}_{1} \right]\left[1 - \overline{\mathcal{A}}_2  \right]\chi_{\mathrm{g}, 0} \delta x_{-1} + 2G\mathcal{K} n_\mathrm{d}\left[\gamma_-^* + \gamma_+ \left(1 - \mathcal{A}_{1}\right) \right]\left[1 - \mathcal{A}_0  \right]\overline{\chi}_{\mathrm{g}, 2} \delta x_{-1} \nonumber \\
& & + ~ iG\left[\gamma_-^* - \gamma_+ \mathcal{A}_{1} \right]\delta x_{-1} + iG\alpha_\mathrm{d}\left[1  - \mathcal{A}_{1}\right]\delta x_{0}
\end{eqnarray}
With the unique replacements
\begin{eqnarray}
\mathcal{B}_{-1} & = & 1 - i2\mathcal{K}n_\mathrm{d}\left[(1-\overline{\mathcal{A}}_3)(1-\mathcal{A}_0)\overline{\chi}_\mathrm{g, 2} - (1-\overline{\mathcal{A}}_2)\chi_\mathrm{g, 0}  \right] \label{eq:termfirst} \\
\mathcal{D}_{-1} & = & \overline{\mathcal{A}}_3 - i2\mathcal{K}n_\mathrm{d}\left[(1-\overline{\mathcal{A}}_3)(1-\overline{\mathcal{A}}_2)\chi_\mathrm{g, 0} + \overline{\mathcal{A}}_3(1-\mathcal{A}_0)\overline{\chi}_\mathrm{g, 2}\right] \\
\mathcal{B}_{1} & = & 1 - i2\mathcal{K}n_\mathrm{d}\left[(1-\overline{\mathcal{A}}_1)(1-\mathcal{A}_0)\overline{\chi}_\mathrm{g, 2} - (1-\overline{\mathcal{A}}_2)\chi_\mathrm{g, 0}  \right] \\
\mathcal{D}_{1} & = & \overline{\mathcal{A}}_1 - i2\mathcal{K}n_\mathrm{d}\left[(1-\overline{\mathcal{A}}_1)(1-\overline{\mathcal{A}}_2)\chi_\mathrm{g, 0} + \overline{\mathcal{A}}_1(1-\mathcal{A}_0)\overline{\chi}_\mathrm{g, 2}\right] \\
\overline{\mathcal{B}}_{3} & = & 1 + i2\mathcal{K}n_\mathrm{d}\left[(1-\mathcal{A}_{-1})(1-\overline{\mathcal{A}}_2)\chi_\mathrm{g, 0} - (1-\mathcal{A}_0)\overline{\chi}_\mathrm{g, 2}  \right] \\
\overline{\mathcal{D}}_{3} & = & \mathcal{A}_{-1} + i2\mathcal{K}n_\mathrm{d}\left[(1-\mathcal{A}_{-1})(1-\mathcal{A}_0)\overline{\chi}_\mathrm{g, 2} + \mathcal{A}_{-1}(1-\overline{\mathcal{A}}_2)\chi_\mathrm{g, 0}\right] \\
\overline{\mathcal{B}}_{1} & = & 1 + i2\mathcal{K}n_\mathrm{d}\left[(1-\mathcal{A}_{1})(1-\overline{\mathcal{A}}_2)\chi_\mathrm{g, 0} - (1-\mathcal{A}_0)\overline{\chi}_\mathrm{g, 2}  \right] \\
\overline{\mathcal{D}}_{1} & = & \mathcal{A}_{1} + i2\mathcal{K}n_\mathrm{d}\left[(1-\mathcal{A}_{1})(1-\mathcal{A}_0)\overline{\chi}_\mathrm{g, 2} + \mathcal{A}_{1}(1-\overline{\mathcal{A}}_2)\chi_\mathrm{g, 0}\right] \label{eq:termlast} \\
\nonumber
\end{eqnarray}
we can write these equations shorter as
\begin{eqnarray}
\frac{\delta\alpha_{-1}}{\chi_{\mathrm{g}, -1}} & = & - ~ iG\left[\gamma_-\mathcal{B}_{-1} - \gamma_+^*\mathcal{D}_{-1} \right]\delta x_{-1}- iG\alpha_\mathrm{d}\left[1  - \overline{\mathcal{A}}_{3}\right]\delta x_{-2} \\
\frac{\delta\alpha_{1}}{\chi_{\mathrm{g}, 1}} & = & - ~ iG\alpha_\mathrm{d}\left[1  - \overline{\mathcal{A}}_{1}\right]\delta x_{0} - iG\left[\gamma_+\mathcal{B}_1  - \gamma_-^*\mathcal{D}_{1} \right]\delta x_{-1} \\
\frac{\overline{\delta\alpha}_{3}}{\overline{\chi}_{\mathrm{g}, 3}} & = &  iG\left[\gamma_+^*\overline{\mathcal{B}}_3  - \gamma_-\overline{\mathcal{D}}_3 \right]\delta x_{-1} + iG\alpha_\mathrm{d}\left[1  - \mathcal{A}_{-1}\right]\delta x_{-2} \\
\frac{\overline{\delta\alpha}_{1}}{\overline{\chi}_{\mathrm{g}, 1}} & = & iG\alpha_\mathrm{d}\left[1  - \mathcal{A}_{1}\right]\delta x_{0} + iG\left[\gamma_-^*\overline{\mathcal{B}}_1 - \gamma_+ \overline{\mathcal{D}}_{1} \right]\delta x_{-1} .
\end{eqnarray}
Note that the indices on the $\mathcal{B}$ and $\mathcal{D}$ terms are not describing a frequency shift like in the $\delta \alpha_j$, $\chi_{\mathrm{g}, j}$ and $\mathcal{A}_j$ terms.
The notation with the overline also does not refer to negative frequencies or complex conjugation here.
Instead, these rather indicate their unique definition given in Eqs.~(\ref{eq:termfirst}-\ref{eq:termlast}).
Next, we inject the field relations back into the original equation for $\delta\alpha_0$ (without the probe input) and obtain
\begin{eqnarray}
\frac{\delta\alpha_0}{\chi_\mathrm{g, 0}} & = & \DB{-2G\mathcal{K}\alpha_\mathrm{d}\overline{\chi}_{\mathrm{g}, 3}\left[\gamma_+\left(1-\overline{\mathcal{A}}_2\right) - \gamma_-^* \overline{\mathcal{A}}_2  \right]\left[\gamma_+^*\overline{\mathcal{B}}_3  - \gamma_-\overline{\mathcal{D}}_3 \right]\delta x_{-1} } \nonumber \\
& & \DB{ -2G\mathcal{K}n_\mathrm{d} \overline{\chi}_{\mathrm{g}, 3}\left[\gamma_+\left(1-\overline{\mathcal{A}}_2\right) - \gamma_-^* \overline{\mathcal{A}}_2  \right]\left[1-\mathcal{A}_{-1}  \right]  \delta x_{-2} } \nonumber \\
& & \DB{ -2G\mathcal{K}\alpha_\mathrm{d}\overline{\chi}_{\mathrm{g}, 1}\left[\gamma_-\left(1-\overline{\mathcal{A}}_2\right) - \gamma_+^* \overline{\mathcal{A}}_2  \right]\left[\gamma_-^*\overline{\mathcal{B}}_1 - \gamma_+ \overline{\mathcal{D}}_{1} \right]\delta x_{-1}   } \nonumber \\
& & \DB{ - 2G\mathcal{K}n_\mathrm{d}\overline{\chi}_{\mathrm{g}, 1}\left[\gamma_-\left(1-\overline{\mathcal{A}}_2\right) - \gamma_+^* \overline{\mathcal{A}}_2  \right] \left[1  - \mathcal{A}_{1}\right]\delta x_{0}  } \nonumber \\
& & \DB{+ 2G\mathcal{K}\alpha_\mathrm{d}\chi_\mathrm{g, -1}\left[\gamma_+ + \gamma_-^*\left(1-\overline{\mathcal{A}}_2  \right)  \right]\left[\gamma_-\mathcal{B}_{-1} - \gamma_+^*\mathcal{D}_{-1} \right]\delta x_{-1}   }\nonumber \\
& & \DB{ +2G\mathcal{K}n_\mathrm{d}\chi_\mathrm{g, -1}\left[\gamma_+ + \gamma_-^*\left(1-\overline{\mathcal{A}}_2  \right)  \right]\left[1  - \overline{\mathcal{A}}_{3}\right]\delta x_{-2}  } \nonumber \\
& & \DB{+ 2G\mathcal{K}\alpha_\mathrm{d}\chi_\mathrm{g, 1}\left[\gamma_- + \gamma_+^*\left(1-\overline{\mathcal{A}}_2  \right)  \right]\left[\gamma_+\mathcal{B}_{1} - \gamma_-^*\mathcal{D}_{1} \right]\delta x_{-1}   }\nonumber \\
& & \DB{ +2G\mathcal{K}n_\mathrm{d}\chi_\mathrm{g, 1}\left[\gamma_- + \gamma_+^*\left(1-\overline{\mathcal{A}}_2  \right)  \right]\left[1  - \overline{\mathcal{A}}_{1}\right]\delta x_{0}  }\nonumber  \\
& & -iG\left[\gamma_- - \gamma_+^*\overline{\mathcal{A}}_2  \right]\delta x_0 - iG\alpha_\mathrm{d}\left[1- \overline{\mathcal{A}}_2\right]\delta x_{-1} - iG\left[\gamma_+ - \gamma_-^*\overline{\mathcal{A}}_2\right]\delta x_{-2}.
\end{eqnarray}
We perform a final variable substitution now
\begin{eqnarray}
\mathcal{J}_-(\Omega) & = & 1 - \frac{\gamma_+^*}{\gamma_-}\overline{\mathcal{A}}_2 + \DB{ i2\mathcal{K}n_\mathrm{d}\chi_\mathrm{g, 1}\left[1 + \frac{\gamma_+^*}{\gamma_-}\left(1-\overline{\mathcal{A}}_2  \right)  \right]\left[1  - \overline{\mathcal{A}}_{1}\right] - i2\mathcal{K}n_\mathrm{d}\overline{\chi}_{\mathrm{g}, 1}\left[1-\overline{\mathcal{A}}_2 - \frac{\gamma_+^*}{\gamma_-} \overline{\mathcal{A}}_2  \right] \left[1  - \mathcal{A}_{1}\right]} \label{eqn:J_m} \\
\mathcal{J}_+(\Omega) & = & 1 - \frac{\gamma_-^*}{\gamma_+}\overline{\mathcal{A}}_2 + \DB{ i2\mathcal{K}n_\mathrm{d}\chi_\mathrm{g, -1}\left[1 + \frac{\gamma_-^*}{\gamma_+}\left(1-\overline{\mathcal{A}}_2 \right)  \right]\left[1  - \overline{\mathcal{A}}_{3}\right] - i2\mathcal{K}n_\mathrm{d}\overline{\chi}_{\mathrm{g}, 3}\left[1-\overline{\mathcal{A}}_2 - \frac{\gamma_-^*}{\gamma_+} \overline{\mathcal{A}}_2  \right] \left[1  - \mathcal{A}_{-1}\right]} \label{eqn:J_p} \\
\mathcal{J}_\alpha(\Omega) & = & 1 - \overline{\mathcal{A}}_2 + \DB{ i2\mathcal{K}\chi_\mathrm{g, 1}\left[\gamma_- + \gamma_+^*\left(1-\overline{\mathcal{A}}_2  \right)  \right]\left[\gamma_+\mathcal{B}_{1} - \gamma_-^*\mathcal{D}_{1} \right] - i2\mathcal{K}\overline{\chi}_\mathrm{g, 1}\left[\gamma_-\left(1-\overline{\mathcal{A}}_2\right) - \gamma_+^* \overline{\mathcal{A}}_2  \right]\left[\gamma_-^*\overline{\mathcal{B}}_1 - \gamma_+ \overline{\mathcal{D}}_{1} \right]} \nonumber \\
& & \DB{ + ~ i2\mathcal{K}\chi_\mathrm{g, -1} \left[\gamma_+ + \gamma_-^*\left(1-\overline{\mathcal{A}}_2  \right)  \right]\left[\gamma_-\mathcal{B}_{-1} - \gamma_+^*\mathcal{D}_{-1} \right] - i2\mathcal{K}\overline{\chi}_\mathrm{g, 3}\left[\gamma_+\left(1-\overline{\mathcal{A}}_2\right) - \gamma_-^* \overline{\mathcal{A}}_2  \right]\left[\gamma_+^*\overline{\mathcal{B}}_3  - \gamma_-\overline{\mathcal{D}}_3 \right]}
\label{eqn:J_a}
\end{eqnarray}
where the nondegenerate four-wave mixing terms are still coded in blue and we obtain
\begin{eqnarray}
\frac{\delta\alpha(\Omega)}{\chi_\mathrm{g}(\Omega)} & = & -iG\gamma_-\mathcal{J}_-(\Omega)\delta x(\Omega) - iG\alpha_\mathrm{d}\mathcal{J}_\alpha(\Omega)\delta x(\Omega - \Omega_\mathrm{dp}) - iG\gamma_+\mathcal{J}_+(\Omega)\delta x(\Omega - 2\Omega_\mathrm{dp}).
\end{eqnarray}
Inserting this result into the equation of motion for the mechanical oscillator and omitting higher order displacement terms leads to the mechanical susceptibility in the weak-coupling and high-$Q_\mathrm{m}$ regime
\begin{equation}
\chi_0^\mathrm{eff}(\Omega) = \frac{1}{\frac{\Gamma_\mathrm{m}}{2} + i\left(\Omega - \Omega_\mathrm{m}\right) + \Sigma_\mathrm{fw}(\Omega_\mathrm{m})}
\end{equation}
with the four-wave backaction
\begin{eqnarray}
\Sigma_\mathrm{fw}(\Omega_\mathrm{m}) & = & |g_-|^2\left[\chi_\mathrm{g}(\Omega_\mathrm{m})\mathcal{J}_-(\Omega_\mathrm{m}) - \chi_\mathrm{g}^*(-\Omega_\mathrm{m})\mathcal{J}_-^*(-\Omega_\mathrm{m})   \right] \nonumber \\
& & + g_\alpha^2\left[\chi_\mathrm{g}(\Omega_\mathrm{m} + \Omega_\mathrm{dp})\mathcal{J}_\alpha(\Omega_\mathrm{m} + \Omega_\mathrm{dp}) - \chi_\mathrm{g}^*(-\Omega_\mathrm{m} + \Omega_\mathrm{dp})\mathcal{J}_\alpha^*(-\Omega_\mathrm{m} + \Omega_\mathrm{dp})   \right] \nonumber \\
& & + |g_+|^2\left[\chi_\mathrm{g}(\Omega_\mathrm{m} + 2\Omega_\mathrm{dp})\mathcal{J}_+(\Omega_\mathrm{m} + 2\Omega_\mathrm{dp}) - \chi_\mathrm{g}^*(-\Omega_\mathrm{m} + 2\Omega_\mathrm{dp})\mathcal{J}_+^*(-\Omega_\mathrm{m} + 2\Omega_\mathrm{dp})   \right]
\end{eqnarray}
This multi-tone dynamical Kerr backaction has a very similar shape as a standard linear multi-tone backaction expression for several pump tones whose frequency difference is far detuned from the mechanical frequency.
The main difference, besides the modified susceptibility $\chi_\mathrm{g}$, is the $\mathcal{J}$-factors.
These $\mathcal{J}$-factors take into account that the intracavity field one mechanical resonance frequency detuned from each of the pump tones $\alpha_\mathrm{d}, \gamma_-, \gamma_+$ also has contributions from the other fields due to four-wave-mixing.
Without any Kerr nonlinearity, all $\mathcal{A}$s would be zero and all $\mathcal{J}$s would be 1.
With exclusively degenerate FWM, only the black terms in Eqs.~(\ref{eqn:J_m})-(\ref{eqn:J_a}) would survive.
These terms describe the interference between the red and blue sidebands of $\alpha_\mathrm{d}$, which are the idler fields of each other, but they also describe the interference between the red (blue) sideband of the $\gamma_-$ field with the blue (red) sideband of the $\gamma_+$ field.
Also these form two pairs of signal and idler fields.
In addition to these terms, there are the (in the equations blue-colored) non-degenerate FWM contributions.
These modify the total backaction significantly, as can be seen in the main paper Fig.~3 or in Supplementary Fig.~\ref{fig:FWOMIT_red}.
Their origin and impact can be understood in two different ways.
The first way is to consider that the cavity resonance frequency is permanently oscillating with the frequency $\Omega_\mathrm{dp}$ due to the beating of the $\alpha_\mathrm{d}$-field with the $\gamma_\pm$-fields and by taking into account the dependence of the cavity susceptibility on the total intracavity field intensity via the Kerr nonlinearity.
In this scenario, when a sideband of one of the tones is generated by mechanical motion at $\pm\Omega_\mathrm{m}$, higher-order sidebands of the scattered field will be generated by the oscillating susceptibility of the modulated cavity.
These higher-order sidebands are detuned by the cavity oscillation frequency $\Omega_\mathrm{dp}$ from the original field and they will fall on top of other first order mechanical sidebands at $\pm\Omega_\mathrm{m} \pm \Omega_\mathrm{dp}$.
The second way to understand this effect is that there are four-photon processes occurring, which involve one photon at the $\gamma_-$-frequency $\omega_\mathrm{p}$ or at the $\gamma_+$-frequency $2\omega_\mathrm{d}-\omega_\mathrm{p}$, one photon at $\omega_\mathrm{d}$ and two sideband photons at different mechanical sideband frequencies.
The result is that by these processes the intracavity fluctuation field at $\Omega$ also gets contributions from five other fluctuation frequencies, as can be clearly seen in the Eq.~(114).
Note that in the general picture, an infinite number of fields will contribute to the field at $\Omega$, but in our experimental situation, we can restrict the Fourier components to the most dominant ones.
\section*{Supplementary Note 10: Four-wave OMIT}
If we take into account input probe fields at the frequencies of the relevant field components, the relations become
\begin{eqnarray}
\frac{\delta\alpha_{-1}}{\chi_{\mathrm{g}, -1}} & = & - ~ iG\left[\gamma_-\mathcal{B}_{-1} - \gamma_+^*\mathcal{D}_{-1} \right]\delta x_{-1}- iG\alpha_\mathrm{d}\left[1  - \overline{\mathcal{A}}_{3}\right]\delta x_{-2} + i\sqrt{\frac{\kappa_\mathrm{e}}{2}}S_{0, -1} - i\overline{\mathcal{A}}_{3}\sqrt{\frac{\kappa_\mathrm{e}}{2}}\overline{S}_{0, 3} \\
\frac{\delta\alpha_{1}}{\chi_{\mathrm{g}, 1}} & = & - ~ iG\alpha_\mathrm{d}\left[1  - \overline{\mathcal{A}}_{1}\right]\delta x_{0} - iG\left[\gamma_+\mathcal{B}_1  - \gamma_-^*\mathcal{D}_{1} \right]\delta x_{-1} + i\sqrt{\frac{\kappa_\mathrm{e}}{2}}S_{0, 1} - i\overline{\mathcal{A}}_{1}\sqrt{\frac{\kappa_\mathrm{e}}{2}}\overline{S}_{0, 1} \\
\frac{\overline{\delta\alpha}_{3}}{\overline{\chi}_{\mathrm{g}, 3}} & = &  iG\left[\gamma_+^*\overline{\mathcal{B}}_3  - \gamma_-\overline{\mathcal{D}}_3 \right]\delta x_{-1} + iG\alpha_\mathrm{d}\left[1  - \mathcal{A}_{-1}\right]\delta x_{-2} - i\sqrt{\frac{\kappa_\mathrm{e}}{2}}\overline{S}_{0, 3} + i\mathcal{A}_{-1}\sqrt{\frac{\kappa_\mathrm{e}}{2}}S_{0, -1} \\
\frac{\overline{\delta\alpha}_{1}}{\overline{\chi}_{\mathrm{g}, 1}} & = & iG\alpha_\mathrm{d}\left[1  - \mathcal{A}_{1}\right]\delta x_{0} + iG\left[\gamma_-^*\overline{\mathcal{B}}_1 - \gamma_+ \overline{\mathcal{D}}_{1} \right]\delta x_{-1} - i\sqrt{\frac{\kappa_\mathrm{e}}{2}}\overline{S}_{0, 1} + i\mathcal{A}_{1}\sqrt{\frac{\kappa_\mathrm{e}}{2}}S_{0, 1} .
\end{eqnarray}
Keeping all these terms, we get for the intracavity field
\begin{eqnarray}
\frac{\delta\alpha(\Omega)}{\chi_\mathrm{g}(\Omega)} & = & -iG\gamma_-\mathcal{J}_-(\Omega)\delta x(\Omega) - iG\alpha_\mathrm{d}\mathcal{J}_\alpha(\Omega)\delta x(\Omega - \Omega_\mathrm{dp}) - iG\gamma_+\mathcal{J}_+(\Omega)\delta x(\Omega - 2\Omega_\mathrm{dp}) \nonumber \\
& & + 2\mathcal{K}\alpha_\mathrm{d}\left[\gamma_+\left(1-\overline{\mathcal{A}}_2 \right) - \gamma_-^* \overline{\mathcal{A}}_2 \right]\overline{\chi}_\mathrm{g, 3}\sqrt{\frac{\kappa_\mathrm{e}}{2}}\overline{S}_{0, 3} \nonumber \\
& & - 2\mathcal{K}\alpha_\mathrm{d}\mathcal{A}_{-1}\left[\gamma_+\left(1-\overline{\mathcal{A}}_2 \right) - \gamma_-^* \overline{\mathcal{A}}_2 \right]\overline{\chi}_\mathrm{g, 3}\sqrt{\frac{\kappa_\mathrm{e}}{2}}S_{0, -1} \nonumber \\
& & + 2\mathcal{K}\alpha_\mathrm{d}\left[\gamma_-\left(1-\overline{\mathcal{A}}_2 \right) - \gamma_+^* \overline{\mathcal{A}}_2 \right]\overline{\chi}_\mathrm{g, 1}\sqrt{\frac{\kappa_\mathrm{e}}{2}}\overline{S}_{0, 1} \nonumber \\
& & - 2\mathcal{K}\alpha_\mathrm{d}\mathcal{A}_{1}\left[\gamma_-\left(1-\overline{\mathcal{A}}_2 \right) - \gamma_+^* \overline{\mathcal{A}}_2 \right]\overline{\chi}_\mathrm{g, 1}\sqrt{\frac{\kappa_\mathrm{e}}{2}}S_{0, 1} \nonumber \\
& & - 2\mathcal{K}\alpha_\mathrm{d}\left[\gamma_+ + \gamma_-^*\left(1-\overline{\mathcal{A}}_2 \right) \right]\chi_\mathrm{g, -1}\sqrt{\frac{\kappa_\mathrm{e}}{2}}S_{0, -1} \nonumber \\
& & + 2\mathcal{K}\alpha_\mathrm{d}\overline{\mathcal{A}}_3\left[\gamma_+ + \gamma_-^*\left(1-\overline{\mathcal{A}}_2 \right) \right]\chi_\mathrm{g, -1}\sqrt{\frac{\kappa_\mathrm{e}}{2}}\overline{S}_{0, 3} \nonumber \\
& & - 2\mathcal{K}\alpha_\mathrm{d}\left[\gamma_- + \gamma_+^*\left(1-\overline{\mathcal{A}}_2 \right) \right]\chi_\mathrm{g, 1}\sqrt{\frac{\kappa_\mathrm{e}}{2}}S_{0, 1} \nonumber \\
& & + 2\mathcal{K}\alpha_\mathrm{d}\overline{\mathcal{A}}_1\left[\gamma_- + \gamma_+^*\left(1-\overline{\mathcal{A}}_2 \right) \right]\chi_\mathrm{g, 1}\sqrt{\frac{\kappa_\mathrm{e}}{2}}\overline{S}_{0, 1} \nonumber \\
& & +i\sqrt{\frac{\kappa_\mathrm{e}}{2}}\left[S_{0, 0} - \overline{\mathcal{A}}_2\overline{S}_{0, 2}\right]
\end{eqnarray}
To calculate the cavity response around the probe frequency, we will only have to keep a single term of these probe fields later, the one proportional to $S_{0, 0}$.
To express the total driving force to the mechanical oscillator though, we have to keep them all for now.
The four-wave mixing will generate fields also at frequencies that beat with the $\alpha_\mathrm{d}$ and the $\gamma_+$ field and therefore drive the mechanical oscillator.
Nevertheless, the equations can be simplified according to the experimental situation.

\subsection*{Signal resonance red-sideband pumping}
If the optomechanical pump field $\gamma_-$ is around the red sideband of the signal resonance and we probe around one mechanical frequency detuned from this pump, we have $\Omega \approx \Omega_\mathrm{m}$.
Due to the high quality factor of the mechanical oscillator, mechanical motion with $\Omega - \Omega_\mathrm{dp}$ or $\Omega - 2\Omega_\mathrm{dp}$ will be suppressed and we can neglect these terms in the equation for the field.
As equation of motion for the mechanical oscillator under these conditions, we obtain
\begin{eqnarray}
\frac{\delta x_0}{\chi_\mathrm{0, 0}^\mathrm{eff}} & = & -i\frac{\hbar G}{2m\Omega_\mathrm{m}}\left[i\gamma_-^*\chi_\mathrm{g, 0} + i\gamma_+\overline{\chi}_{g, 2}\mathcal{A}_0  \right]\sqrt{\frac{\kappa_\mathrm{e}}{2}}S_{0, 0} \nonumber \\
& & -i\frac{\hbar G \alpha_\mathrm{d}}{2m\Omega_\mathrm{m}}\left[- 2\mathcal{K}\alpha_\mathrm{d}\mathcal{A}_{0}\left[\gamma_+\left(1-\overline{\mathcal{A}}_1 \right) - \gamma_-^* \overline{\mathcal{A}}_1 \right]\chi_\mathrm{g, 1}\overline{\chi}_\mathrm{g, 2} \right]\sqrt{\frac{\kappa_\mathrm{e}}{2}}S_{0, 0} \nonumber \\
& & -i\frac{\hbar G \alpha_\mathrm{d}}{2m\Omega_\mathrm{m}}\left[- 2\mathcal{K}\alpha_\mathrm{d}\left[\gamma_+ + \gamma_-^*\left(1-\overline{\mathcal{A}}_1 \right) \right]\chi_\mathrm{g, 1}\chi_\mathrm{g, 0} \right]\sqrt{\frac{\kappa_\mathrm{e}}{2}}S_{0, 0} \nonumber \\
& & -i\frac{\hbar G \alpha_\mathrm{d}}{2m\Omega_\mathrm{m}}\left[2\mathcal{K}\alpha_\mathrm{d}\mathcal{A}_0\left[\gamma_-^* + \gamma_+\left(1-\mathcal{A}_1 \right) \right]\overline{\chi}_\mathrm{g, 1}\overline{\chi}_\mathrm{g, 2} \right]\sqrt{\frac{\kappa_\mathrm{e}}{2}}S_{0, 0} \nonumber \\
& & -i\frac{\hbar G \alpha_\mathrm{d}}{2m\Omega_\mathrm{m}}\left[2\mathcal{K}\alpha_\mathrm{d}\left[\gamma_-^*\left(1-\mathcal{A}_1 \right) - \gamma_+\mathcal{A}_1 \right]\overline{\chi}_\mathrm{g, 1}\chi_\mathrm{g, 0} \right]\sqrt{\frac{\kappa_\mathrm{e}}{2}}S_{0, 0}
\end{eqnarray}
which can also be written as
\begin{eqnarray}
\frac{\delta x_0}{\chi_\mathrm{0, 0}^\mathrm{eff}} & = & \frac{\hbar G}{2m\Omega_\mathrm{m}}\gamma_-^*\chi_\mathrm{g, 0}\left[1 + i 2\mathcal{K} n_\mathrm{d}\chi_\mathrm{g, 1}\left(1-\overline{\mathcal{A}}_1  + \frac{\gamma_+ }{\gamma_-^*} \right) -i2\mathcal{K}n_\mathrm{d}\overline{\chi}_\mathrm{g, 1}\left(1-\mathcal{A}_1 - \frac{ \gamma_+}{\gamma_-^*}\mathcal{A}_1 \right) \right]\sqrt{\frac{\kappa_\mathrm{e}}{2}}S_{0, 0} \nonumber \\
& & \frac{\hbar G}{2m\Omega_\mathrm{m}}\gamma_+\overline{\chi}_{g, 2}\mathcal{A}_0 \left[1 -i2\mathcal{K}n_\mathrm{d}\overline{\chi}_\mathrm{g, 1}\left(1-\mathcal{A}_1 + \frac{\gamma_-^*}{\gamma_+} \right) + i 2\mathcal{K}n_\mathrm{d}\chi_\mathrm{g, 1} \left(1-\overline{\mathcal{A}}_1  - \frac{\gamma_-^*}{\gamma_+} \overline{\mathcal{A}}_1 \right) \right]\sqrt{\frac{\kappa_\mathrm{e}}{2}}S_{0, 0} \nonumber \\
& = & \frac{\hbar G}{2m\Omega_\mathrm{m}}\left[\gamma_-^*\chi_\mathrm{g, 0}\mathcal{P_-} + \gamma_+\overline{\chi}_{g, 2}\mathcal{A}_0 \mathcal{P}_+\right]\sqrt{\frac{\kappa_\mathrm{e}}{2}}S_{0, 0}
\end{eqnarray}
Injecting this back into the equation for the intracavity field, we get
\begin{eqnarray}
\frac{\delta\alpha(\Omega)}{\chi_\mathrm{g}(\Omega)} & = & i\left(1 -g_-\left[g_-^*\chi_\mathrm{g}(\Omega)\mathcal{P}_-(\Omega) + g_+\chi_\mathrm{g}^*(-\Omega + 2\Omega_\mathrm{dp})\mathcal{A}(\Omega) \mathcal{P}_+(\Omega)\right]\mathcal{J}_-(\Omega)\chi_{0}^\mathrm{eff}(\Omega)\right)\sqrt{\frac{\kappa_\mathrm{e}}{2}}S_{0}(\Omega)
\label{eqn:FWOMIT_red_alpha}
\end{eqnarray}
\subsection*{Idler resonance blue-sideband pumping}
If on the other hand the pump field $\gamma_-$ is located on the blue sideband of the idler resonance and we probe around one mechanical frequency away from the corresponding $\gamma_+$ field, we have $\Omega - 2\Omega_\mathrm{dp} \approx \Omega_\mathrm{m}$.
In this case, mechanical motion with $\Omega$ and $\Omega - \Omega_\mathrm{dp}$ will be irrelevant.
Then, 
\begin{eqnarray}
\frac{\delta x_{-2}}{\chi_\mathrm{0, -2}^\mathrm{eff}} & = & -i\frac{\hbar G}{2m\Omega_\mathrm{m}}\left[i\gamma_+^*\chi_\mathrm{g, 0} + i\gamma_-\overline{\chi}_{g, 2}\mathcal{A}_0  \right]\sqrt{\frac{\kappa_\mathrm{e}}{2}}S_{0, 0} \nonumber \\
& & -i\frac{\hbar G \alpha_\mathrm{d}}{2m\Omega_\mathrm{m}}\left[- 2\mathcal{K}\alpha_\mathrm{d}\mathcal{A}_{0}\left[\gamma_-\left(1-\overline{\mathcal{A}}_3 \right) - \gamma_+^* \overline{\mathcal{A}}_3 \right]\chi_\mathrm{g, -1}\overline{\chi}_\mathrm{g, 2} \right]\sqrt{\frac{\kappa_\mathrm{e}}{2}}S_{0, 0} \nonumber \\
& & -i\frac{\hbar G \alpha_\mathrm{d}}{2m\Omega_\mathrm{m}}\left[- 2\mathcal{K}\alpha_\mathrm{d}\left[\gamma_- + \gamma_+^*\left(1-\overline{\mathcal{A}}_3 \right) \right]\chi_\mathrm{g, -1}\chi_\mathrm{g, 0} \right]\sqrt{\frac{\kappa_\mathrm{e}}{2}}S_{0, 0} \nonumber \\
& & -i\frac{\hbar G \alpha_\mathrm{d}}{2m\Omega_\mathrm{m}}\left[2\mathcal{K}\alpha_\mathrm{d}\mathcal{A}_0\left[\gamma_+^* + \gamma_-\left(1-\mathcal{A}_{-1} \right) \right]\overline{\chi}_\mathrm{g, 3}\overline{\chi}_\mathrm{g, 2} \right]\sqrt{\frac{\kappa_\mathrm{e}}{2}}S_{0, 0} \nonumber \\
& & -i\frac{\hbar G \alpha_\mathrm{d}}{2m\Omega_\mathrm{m}}\left[2\mathcal{K}\alpha_\mathrm{d}\left[\gamma_+^*\left(1-\mathcal{A}_{-1} \right) - \gamma_-\mathcal{A}_{-1} \right]\overline{\chi}_\mathrm{g, 3}\chi_\mathrm{g, 0} \right]\sqrt{\frac{\kappa_\mathrm{e}}{2}}S_{0, 0}
\end{eqnarray}
\begin{eqnarray}
\frac{\delta x_{-2}}{\chi_\mathrm{0, -2}^\mathrm{eff}} & = & \frac{\hbar G}{2m\Omega_\mathrm{m}}\gamma_+^*\chi_\mathrm{g, 0}\left[1 + i 2\mathcal{K} n_\mathrm{d}\chi_\mathrm{g, -1}\left(1-\overline{\mathcal{A}}_3  + \frac{\gamma_- }{\gamma_+^*} \right) -i2\mathcal{K}n_\mathrm{d}\overline{\chi}_\mathrm{g, 3}\left(1-\mathcal{A}_{-1} - \frac{ \gamma_-}{\gamma_+^*}\mathcal{A}_{-1} \right) \right]\sqrt{\frac{\kappa_\mathrm{e}}{2}}S_{0, 0} \nonumber \\
& & \frac{\hbar G}{2m\Omega_\mathrm{m}}\gamma_-\overline{\chi}_{g, 2}\mathcal{A}_0 \left[1 -i2\mathcal{K}n_\mathrm{d}\overline{\chi}_\mathrm{g, 3}\left(1-\mathcal{A}_{-1} + \frac{\gamma_+^*}{\gamma_-} \right) + i 2\mathcal{K}n_\mathrm{d}\chi_\mathrm{g, -1} \left(1-\overline{\mathcal{A}}_3  - \frac{\gamma_+^*}{\gamma_-} \overline{\mathcal{A}}_3 \right) \right]\sqrt{\frac{\kappa_\mathrm{e}}{2}}S_{0, 0} \nonumber \\
& = & \frac{\hbar G}{2m\Omega_\mathrm{m}}\left[\gamma_+^*\chi_\mathrm{g, 0}\mathcal{Q_+} + \gamma_-\overline{\chi}_{g, 2}\mathcal{A}_0 \mathcal{Q}_-\right]\sqrt{\frac{\kappa_\mathrm{e}}{2}}S_{0, 0}
\end{eqnarray}
Injecting this back into the equation for the intracavity field, we get
\begin{eqnarray}
\frac{\delta\alpha(\Omega)}{\chi_\mathrm{g}(\Omega)} & = & i\left(1 -g_+\left[g_+^*\chi_\mathrm{g}(\Omega)\mathcal{Q}_+(\Omega) + g_-\chi_\mathrm{g}^*(-\Omega + 2\Omega_\mathrm{dp})\mathcal{A}(\Omega) \mathcal{Q}_-(\Omega)\right]\mathcal{J}_+(\Omega)\chi_{0}^\mathrm{eff}(\Omega - 2\Omega_\mathrm{dp})\right)\sqrt{\frac{\kappa_\mathrm{e}}{2}}S_{0}(\Omega)
\label{eqn:FWOMIT_blue_alpha}
\end{eqnarray}

\subsection*{Optomechanical cavity response}

The response in both cases is given by
\begin{equation}
S_{21}(\Omega) = 1 + i\sqrt{\frac{\kappa_\mathrm{e}}{2}}\frac{\delta\alpha(\Omega)}{S_0(\Omega)}
\label{eqn:FWOMIT_S21}
\end{equation}
\section*{Supplementary Note 11: Measurement and data analysis protocol for four-wave OMIT and four-wave dynamical backaction}
\subsection*{Preparation}
\begin{itemize}
	\item {We start the experimental cycle with choosing the bias-flux operation point, either point I, and an in-plane magnetic field $B_\parallel$. We ramp the in-plane current to its corresponding value.}
	\item{A parametric drive tone is sent to the cavity with fixed frequency $\omega_\mathrm{d}$ and power $P_\mathrm{d}$ to match the chosen operation point.} 
	\item{The cavity bias flux is adjusted manually to prepare the SQUID cavity in the driven Kerr-mode state.}
	\item{The frequency of the optomechanical pump is chosen to be either on the red sideband of the signal resonance or on the blue sideband of the idler resonance. The pump is activated with fixed frequency $\omega_\mathrm{p}$ and power $P_\mathrm{p}$.}
\end{itemize}
\subsection*{The measurement}
\begin{itemize}
	\item{For the actual measurement, we start a python-based control and data acquisition script, which is programmed to wait for a terminal starting command before each data point.}
	\item{Prior to running the measurement, we input some fixed parameters to the script such as all values of the attenuators.}
	\item{We then manually adjust the probe VNA to a parameter set regarding frequency window, probe power and bandwidth in order to measure a clean OMIT response curve.}
	\item{Upon a terminal command, the script begins the acquisition and first catches all relevant information such as powers, frequencies, frequency spans, bandwidths as well as magnet DC currents from all relevant measurement equipment.}
	\item{The parameters obtained from the manually adjusted OMIT settings on the VNA are then re-used for all subsequent measurements. Based on the mechanical frequency and cavity frequency, an array of optomechanical pump frequencies is generated, which corresponds to an array of $\delta_\mathrm{p}$. At the same time a corresponding set of VNA frequency ranges is generated to track the OMIT response for all the different pump frequencies.}
	\item{The script performs a narrow-band VNA scan to measure the OMIT response and stores the data in file 1, where all subsequent narrow-band scans for varying pump frequencies are attached as well.}
	\item{The script performs a wide-band VNA scan to measure the cavity response and stores the data in file 2, where all subsequent wide-band scans for varying pump frequencies are attached as well.}
	\item{At this point the script will expect an input via the terminal, which tells whether we want to take the exact same measurement again for identical parameters or if we are going to proceed to the next pump detuning.}
	\item{After receiving our choice, the script sets the VNA to the cavity center frequency with a fixed span of $1\,$kHz and waits upon a terminal command for measuring the two VNA scans of the next point. During this waiting window, we have the possibility to counter possible bias flux drifts by manually adjusting the out-of-plane current, while permanently monitoring the cavity response at the response minimum.}
	\item{Both measures described in the latter two bullet points are critical to obtain a consistent set of data, as sometimes the bias flux and cavity starts to drift considerably on a slow timescale ($\sim$seconds). This drift can significantly distort the captured OMIT response, which cannot be measured too fast due to the small mechanical linewidth. Another risk is that the cavity leaves the driven Kerr-state without the manual feedback control loop in between measurement points.}
	\item{After the cavity is stabilized and the measurement can proceed, the script repeats the cycle from gathering all relevant parameters from all machines to taking the two VNA traces and the waiting and stabilization time.}
\end{itemize}

\begin{figure*}
	\centerline{\includegraphics[trim = {0cm, 0cm, 0cm, 0cm}, clip=True, width=0.9\textwidth]{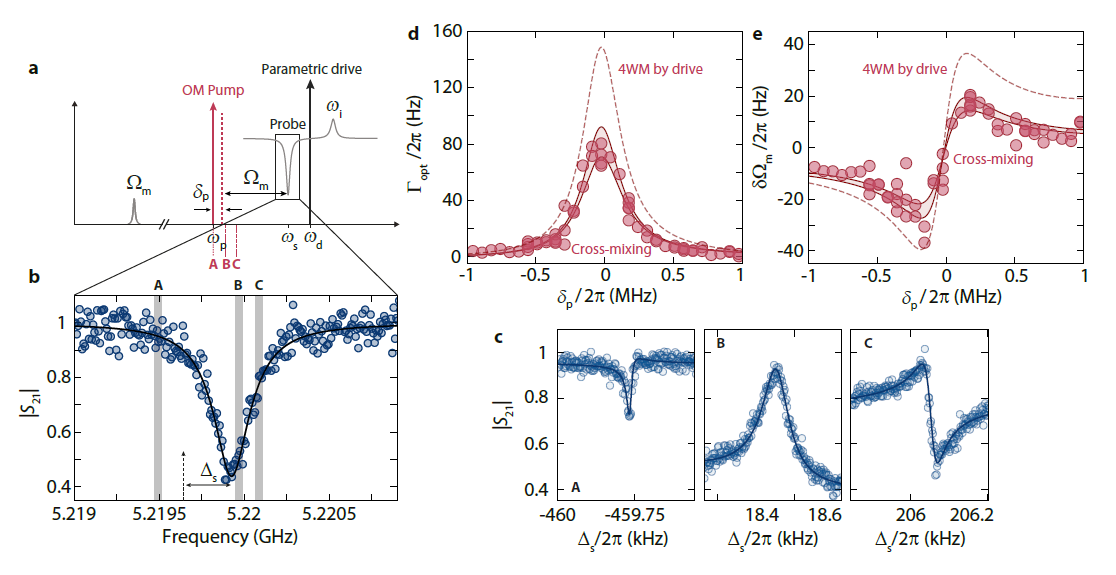}}
	\caption{\textsf{\textbf{Four-wave-OMIT and four-wave-backaction for optomechanical red-sideband pumping of the signal resonance.} Panel \textbf{a} shows schematically the experimental protocol. The SQUID cavity is prepared in the Kerr-mode state by a strong parametric drive. In addition, we apply an optomechanical pump tone one mechanical frequency red-detuned from the signal resonance $\omega_\mathrm{p} = \omega_\mathrm{s} - \Omega_\mathrm{m} + \delta_\mathrm{p}$. Finally, we use a weak probe tone around the signal resonance to detect optomechanically induced transparency. We repeat this scheme for varying detunings $\delta_\mathrm{p}$. \textbf{b} shows the Kerr-mode signal resonance transmission $S_{21}$ measured with the weak probe field in presence of the drive and pump tones. Circles are data, line is a fit. Gray vertical bars labeled with A, B, and C indicate zoom regions for the corresponding panels shown in \textbf{c} and $\Delta_\mathrm{s} = \omega - \omega_\mathrm{s}$ denotes the detuning between probe field and signal resonance. In \textbf{c}, the probe tone response in three narrow frequency windows around $\omega \approx \omega_\mathrm{p} + \Omega_\mathrm{m}$ is plotted for three different pump detunings $\delta_\mathrm{p}$, cf. panel \textbf{a} and \textbf{b}. Each probe tone response displays a narrow-band resonance, indicating the phenomenon of optomechanically induced transparency (OMIT) via excitation of the mechanical oscillator and corresponding interference between the probe field and mechanical sidebands of intracavity drive and pump fields. For each $\delta_\mathrm{p}$, we fit the OMIT response, corresponding curves are shown as lines in \textbf{c}, and extract the effective mechanical resonance frequency $\Omega_\mathrm{eff} = \Omega_\mathrm{m} + \delta\Omega_\mathrm{m}$ and the effective mechanical linewidth $\Gamma_\mathrm{eff} = \Gamma_\mathrm{m} + \Gamma_\mathrm{opt}$. The contributions $\delta\Omega_\mathrm{m}$ and $\Gamma_\mathrm{opt}$, induced by dynamical backaction of the total intracavity fields, are plotted in panels \textbf{d} and \textbf{e} as circles vs detuning of the pump tone from the red sideband of the signal resonance. The result of analytical calculations is shown as two solid lines with shaded area, where the range described by the lines captures uncertainties in the device parameters, cf. text. The dashed line shows the result of equivalent calculations without taking into account cross-mixing terms.}}
	\label{fig:FWOMIT_red}
\end{figure*}

\subsection*{Data analysis}
\begin{itemize}
	\item{Data analysis starts with a fit of the wide-band signal resonance response $S_{21}$ using Eq.~(\ref{eqn:fitS21}). From this fit, we obtain effective parameters for $\kappa'$, $\kappa_\mathrm{e}$ and $\omega_\mathrm{s}$ and a fit of the complex background.}
	\item{Using the background fit parameters, we calculate the complex background in the narrow-band frequency window of the corresponding OMIT response scan and divide it off the measured signal.}
	\item{We convert the frequency axis of the OMIT response to frequencies relative to $\gamma_-$ in the red-sideband case and relative to $\gamma_+$ in the blue-sideband case and fit the background corrected and frequency-shifted OMIT response using Eq.~(\ref{eqn:fitS21}) as well. As we are only interested in the resonance frequency and the effective linewidth of the OMIT resonance at this point, this procedure is as reliable but significantly simpler than using the full four-wave OMIT expression. As fit parameters we obtain $\Omega_\mathrm{m}^\mathrm{eff}$ and $\Gamma_\mathrm{eff}$.}
	\item{We substract the corresponding bare values $\Omega_\mathrm{m}$ and $\Gamma_\mathrm{m}$ to obtain the dynamical backaction contributions $\delta\Omega_\mathrm{m}$ and $\Gamma_\mathrm{opt}$. We note that to make them match with theory for the red-sideband case and the blue-sideband case simultaneously with a single set of otherwise identical parameters, the bare values slightly differ between the red and blue case. For the mechanical resonance frequency, this blue-red-difference is about $8\,$Hz and for the linewidth it is about $1\,$Hz. As the bare mechanical frequency and linewidth depend strongly on temperature (cf. Supplementary Note 14), a small temperature difference of the mechanical oscillator during the two measurements might be the origin of these differences.}
	\item{For the range of theoretical values (shaded area between the two solid lines in main paper Fig.~3 and Supplementary Fig.~\ref{fig:FWOMIT_red}), we consider uncertainties in the parameters going into the theoretical calculations. These include variations of the optomechanical single-photon coupling rate $g_0 = 1.85\pm0.05\,$kHz, of the driven cavity linewidth $\kappa' = 349\pm20\,$kHz and of the bare cavity resonance frequency $\omega_0 = 5.2236\pm0.1\,$MHz.}
\end{itemize}

\section*{Supplementary Note 12: Data for OMIT and dynamical backaction on the red sideband of the signal resonance}
In this section, we present data on four-wave-OMIT and dynamical four-wave backaction for an optomechanical pump field $\gamma_-$ on the red sideband of the signal resonance with $\omega_\mathrm{p} =\omega_\mathrm{s} - \Omega_\mathrm{m} + \delta_\mathrm{p}$, where $\delta_\mathrm{p}$ is the detuning between the pump tone and the red sideband.
The experimental setting is schematically shown in Supplementary Fig.~\ref{fig:FWOMIT_red}\textbf{a}
In this configuration, we follow the usual OMIT protocol in the experiment, i.e., we pump at a frequency $\omega_\mathrm{p}$ around the signal resonance red sideband and probe the cavity response $S_{21}$ in a narrow frequency window around $\omega \approx \omega_\mathrm{p} + \Omega_\mathrm{m}$.
In addition, we measure the transmission in a wider range to capture the complete cavity absorption.
We repeat this scheme for different detunings $\delta_\mathrm{p}$, where the range of $\delta_\mathrm{p}$ is chosen to cover more than 2 signal resonance linewidths around the red sideband.
One exemplary wide-band cavity transmission is displayed in Supplementary Fig.~\ref{fig:FWOMIT_red}\textbf{b}, in \textbf{c} three corresponding narrow-band measurements are shown for three different $\delta_\mathrm{p}$, clearly showing the characteristic OMIT window, representing the mechanical oscillator.
From a fit to the OMIT response, shown as lines, we extract the mechanical oscillator resonance frequency and the effective mechanical linewidth.
After subtracting the bare values, we obtain the dynamical backaction contributions $\delta\Omega_\mathrm{m}$ and $\Gamma_\mathrm{opt}$, which are plotted in \textbf{d} and \textbf{e} as circles. 

\section*{Supplementary Note 13: Multi-tone Kerr optomechanics with noise input}
Working with quantum formalism for the equations of motion with noise input we obtain for the mechanical oscillator
\begin{eqnarray}
\frac{\hat{b}_0}{\chi_{0, 0}} & = & -i\left(g_-^*\hat{a}_0 + g_-\hat{a}_0^\dagger \right) - ig_\alpha\left(\hat{a}_1 + \hat{a}_1^\dagger \right) - i\left( g_+^* \hat{a}_2 + g_+ \hat{a}_2^\dagger \right) + \sqrt{\Gamma_\mathrm{m}}\hat{\zeta}_0 \nonumber \\
\frac{\hat{b}_0^\dagger}{\overline{\chi}_{0, 0}} & = & i\left(g_-^*\hat{a}_0 + g_-\hat{a}_0^\dagger \right) + ig_\alpha\left(\hat{a}_1 + \hat{a}_1^\dagger \right) + i\left( g_+^* \hat{a}_2 + g_+ \hat{a}_2^\dagger \right) + \sqrt{\Gamma_\mathrm{m}}\hat{\zeta}_0^\dagger
\end{eqnarray}
and for the intracavity fluctuation fields
\begin{eqnarray}
\frac{\hat{a}_0}{\chi_\mathrm{p, 0}} & = & i\mathcal{K}n_\mathrm{d}\hat{a}_2^\dagger - ig_-\left( \hat{b}_0 + \hat{b}_0^\dagger  \right) - i g_\alpha\left(\hat{b}_{-1} + \hat{b}_{1}^\dagger \right) - ig_+\left(\hat{b}_{-2} + \hat{b}_{2}^\dagger  \right) \nonumber \\
& & \DB{+ ~ i2\mathcal{K}\alpha_\mathrm{d}\left[\gamma_-^* + \gamma_+ \right]\hat{a}_{-1} + i2\mathcal{K}\alpha_\mathrm{d}\left[\gamma_- + \gamma_+^* \right]\hat{a}_{1} + i2\mathcal{K}\alpha_\mathrm{d}\gamma_- \hat{a}_1^\dagger + i2\mathcal{K}\alpha_\mathrm{d}\gamma_+ \hat{a}_3^\dagger } \nonumber \\
& & + \sqrt{\kappa_\mathrm{e}}\hat{\xi}_\mathrm{e0+} + \sqrt{\kappa_\mathrm{i}}\hat{\xi}_\mathrm{i0+} \\
\frac{\hat{a}_2^\dagger}{\overline{\chi}_\mathrm{p, 2}} & = & -i\mathcal{K}n_\mathrm{d}\hat{a}_0 + ig_-^*\left( \hat{b}_{-2} + \hat{b}_{2}^\dagger  \right) + i g_\alpha\left(\hat{b}_{-1} + \hat{b}_{-1}^\dagger \right) + ig_+^*\left(\hat{b}_{0} + \hat{b}_{0}^\dagger  \right) \nonumber \\
& & \DB{- ~ i2\mathcal{K}\alpha_\mathrm{d}\left[\gamma_- + \gamma_+^* \right]\hat{a}_{1}^\dagger - i2\mathcal{K}\alpha_\mathrm{d}\left[\gamma_-^* + \gamma_+ \right]\hat{a}_{3}^\dagger - i2\mathcal{K}\alpha_\mathrm{d}\gamma_-^* \hat{a}_{-1} - i2\mathcal{K}\alpha_\mathrm{d}\gamma_+^* \hat{a}_1 } \nonumber \\
& & + \sqrt{\kappa_\mathrm{e}}\hat{\xi}_\mathrm{e2-}^\dagger + \sqrt{\kappa_\mathrm{i}}\hat{\xi}_\mathrm{i2-}^\dagger
\end{eqnarray}
The latter two equations can be combined into
\begin{eqnarray}
\frac{\hat{a}_0}{\chi_\mathrm{g, 0}} & = & \DB{ i2\mathcal{K}\alpha_\mathrm{d}\left[\gamma_-\left(1 - \overline{\mathcal{A}}_2 \right) -\gamma_+^*\overline{\mathcal{A}}_2 \right]\hat{a}_1^\dagger + i2\mathcal{K}\alpha_\mathrm{d}\left[\gamma_+\left(1 - \overline{\mathcal{A}}_2 \right) -\gamma_-^*\overline{\mathcal{A}}_2 \right] \hat{a}_3^\dagger } \nonumber \\
& & \DB{ + ~ i2\mathcal{K}\alpha_\mathrm{d}\left[\gamma_+ + \gamma_-^*\left(1 - \overline{\mathcal{A}}_2 \right)  \right]\hat{a}_{-1} + i2\mathcal{K}\alpha_\mathrm{d}\left[\gamma_- + \gamma_+^*\left(1 - \overline{\mathcal{A}}_2\right) \right]\hat{a}_{1} } \nonumber \\
& & - i\left(g_- - g_+^*\overline{\mathcal{A}}_2 \right)\left(\hat{b}_0 + \hat{b}_0^\dagger \right) \nonumber \\
& & - i g_\alpha\left(1-\overline{\mathcal{A}}_2 \right)\left(\hat{b}_{-1} + \hat{b}_{1}^\dagger \right) \nonumber \\
& & - i\left(g_+ - g_-^*\overline{\mathcal{A}}_2 \right)\left(\hat{b}_{-2} + \hat{b}_{2}^\dagger\right)   \nonumber \\
& & + \hat{\mathcal{N}}_{0+} + \overline{\mathcal{A}}_2\hat{\mathcal{N}}_{2-}^\dagger
\end{eqnarray}
with
\begin{equation} \hat{\mathcal{N}}_{0+} =  \sqrt{\kappa_\mathrm{e}}\hat{\xi}_\mathrm{e0+} + \sqrt{\kappa_\mathrm{i}}\hat{\xi}_\mathrm{i0+} ~~~~~ \hat{\mathcal{N}}_{2-}^\dagger = \sqrt{\kappa_\mathrm{e}}\hat{\xi}_\mathrm{e2-}^\dagger +  \sqrt{\kappa_\mathrm{i}}\hat{\xi}_\mathrm{i2-}^\dagger. 
\end{equation}
Just as for the classical equations, we need now the next iteration field components
\begin{eqnarray}
\frac{\hat{a}_{-1}}{\chi_\mathrm{g, -1}} & = & - i\left[g_-\mathcal{B}_{-1} - g_+^*\mathcal{D}_{-1} \right]\left(\hat{b}_{-1} + \hat{b}_1^\dagger \right) - i g_\alpha\left(1-\overline{\mathcal{A}}_3 \right)\left(\hat{b}_{-2} + \hat{b}_{2}^\dagger \right) + \hat{\mathcal{N}}_{-1+} + \overline{\mathcal{A}}_3\hat{\mathcal{N}}_{3-}^\dagger \\
\frac{\hat{a}_{1}}{\chi_\mathrm{g, 1}} & = & - i g_\alpha\left(1-\overline{\mathcal{A}}_1 \right)\left(\hat{b}_{0} + \hat{b}_{0}^\dagger \right) - i\left[g_+\mathcal{B}_{1} - g_-^*\mathcal{D}_{1} \right]\left(\hat{b}_{-1} + \hat{b}_1^\dagger \right) + \hat{\mathcal{N}}_{1+} + \overline{\mathcal{A}}_1\hat{\mathcal{N}}_{1-}^\dagger \\
\frac{\hat{a}_{1}^\dagger}{\overline{\chi}_\mathrm{g, 1}} & = & i g_\alpha\left(1-\mathcal{A}_1 \right)\left(\hat{b}_{0} + \hat{b}_{0}^\dagger \right) + i\left[g_-^*\overline{\mathcal{B}}_{1} - g_+\overline{\mathcal{D}}_{1} \right]\left(\hat{b}_{-1} + \hat{b}_1^\dagger \right) + \hat{\mathcal{N}}_{1-}^\dagger + \mathcal{A}_1\hat{\mathcal{N}}_{1+} \\
\frac{\hat{a}_{3}^\dagger}{\overline{\chi}_\mathrm{g, 3}} & = & i\left[g_+^*\overline{\mathcal{B}}_{3} - g_-\overline{\mathcal{D}}_{3} \right]\left(\hat{b}_{-1} + \hat{b}_1^\dagger \right) + i g_\alpha\left(1-\mathcal{A}_{-1} \right)\left(\hat{b}_{-2} + \hat{b}_{2}^\dagger \right) + \hat{\mathcal{N}}_{3-}^\dagger + \mathcal{A}_{-1}\hat{\mathcal{N}}_{-1+}
\end{eqnarray}
which lead to an expression for $\hat{a}_0$ given by
\begin{eqnarray}
\frac{\hat{a}_0}{\chi_\mathrm{g, 0}} & = & -ig_-\mathcal{J}_-\left(\hat{b}_0 + \hat{b}_0^\dagger \right) -ig_\alpha\mathcal{J}_\alpha\left(\hat{b}_{-1} + \hat{b}_{1}^\dagger \right)-ig_+\mathcal{J}_+\left(\hat{b}_{-2} + \hat{b}_{2}^\dagger \right) \nonumber \\
& & + i2\mathcal{K}\alpha_\mathrm{d}\left( \left[\gamma_-\left(1 - \overline{\mathcal{A}}_2 \right) -\gamma_+^*\overline{\mathcal{A}}_2 \right]\mathcal{A}_1\overline{\chi}_\mathrm{g, 1} + \left[\gamma_- + \gamma_+^*\left(1 - \overline{\mathcal{A}}_2\right) \right]\chi_\mathrm{g, 1} \right)\hat{\mathcal{N}}_{1+} \nonumber \\
& & + i2\mathcal{K}\alpha_\mathrm{d}\left( \left[\gamma_-\left(1 - \overline{\mathcal{A}}_2 \right) -\gamma_+^*\overline{\mathcal{A}}_2 \right]\overline{\chi}_\mathrm{g, 1} + \left[\gamma_- + \gamma_+^*\left(1 - \overline{\mathcal{A}}_2\right) \right]\overline{\mathcal{A}}_1\chi_\mathrm{g, 1} \right)\hat{\mathcal{N}}_{1-}^\dagger \nonumber \\
& & + i2\mathcal{K}\alpha_\mathrm{d}\left( \left[\gamma_+\left(1 - \overline{\mathcal{A}}_2 \right) -\gamma_-^*\overline{\mathcal{A}}_2 \right]\overline{\chi}_\mathrm{g, 3} + \left[\gamma_+ + \gamma_-^*\left(1 - \overline{\mathcal{A}}_2\right) \right]\overline{\mathcal{A}}_3\chi_\mathrm{g, -1} \right)\hat{\mathcal{N}}_{3-}^\dagger \nonumber \\
& & + i2\mathcal{K}\alpha_\mathrm{d}\left( \left[\gamma_+\left(1 - \overline{\mathcal{A}}_2 \right) -\gamma_-^*\overline{\mathcal{A}}_2 \right]\mathcal{A}_{-1}\overline{\chi}_\mathrm{g, 3} + \left[\gamma_+ + \gamma_-^*\left(1 - \overline{\mathcal{A}}_2\right) \right]\chi_\mathrm{g, -1} \right)\hat{\mathcal{N}}_{-1+} \nonumber \\
& & + \hat{\mathcal{N}}_{0+} + \overline{\mathcal{A}}_2\hat{\mathcal{N}}_{2-}^\dagger \nonumber \\
& = & -ig_-\mathcal{J}_-\left(\hat{b}_0 + \hat{b}_0^\dagger \right) -ig_\alpha\mathcal{J}_\alpha\left(\hat{b}_{-1} + \hat{b}_{1}^\dagger \right)-ig_+\mathcal{J}_+\left(\hat{b}_{-2} + \hat{b}_{2}^\dagger \right) \nonumber \\
& & + \mathcal{Y}_{1+}\hat{\mathcal{N}}_{1+}   + \overline{\mathcal{Y}}_{1-}\hat{\mathcal{N}}_{1-}^\dagger + \overline{\mathcal{Y}}_{3-}\hat{\mathcal{N}}_{3-}^\dagger + \mathcal{Y}_{-1+}\hat{\mathcal{N}}_{-1+} + \hat{\mathcal{N}}_{0+} + \overline{\mathcal{A}}_2\hat{\mathcal{N}}_{2-}^\dagger 
\end{eqnarray}
or, in its shortest version
\begin{eqnarray}
\frac{\hat{a}_0}{\chi_\mathrm{g, 0}} & = & -ig_-\mathcal{J}_-(\Omega)\left(\hat{b}_0 + \hat{b}_0^\dagger \right) -ig_\alpha\mathcal{J}_\alpha(\Omega)\left(\hat{b}_{-1} + \hat{b}_{1}^\dagger \right)-ig_+\mathcal{J}_+(\Omega)\left(\hat{b}_{-2} + \hat{b}_{2}^\dagger \right) + \hat{\mathcal{M}}_{0+}
\end{eqnarray}
and the corresponding equation for $\hat{a}^\dagger$
\begin{eqnarray}
\frac{\hat{a}_0^\dagger}{\overline{\chi}_\mathrm{g, 0}} & = &ig_-^*\mathcal{J}_-^*(-\Omega)\left(\hat{b}_0 + \hat{b}_0^\dagger \right) +ig_\alpha\mathcal{J}_\alpha^*(-\Omega)\left(\hat{b}_{1} + \hat{b}_{-1}^\dagger \right)+ig_+\mathcal{J}_+^*(-\Omega)\left(\hat{b}_{2} + \hat{b}_{-2}^\dagger \right) + \hat{\mathcal{M}}_{0-}^\dagger.
\end{eqnarray}
\subsection*{Signal resonance red sideband pumping}
Next we use
\begin{equation}
\hat{b}_0 + \hat{b}_0^\dagger = -i\left(\chi_{0, 0} - \overline{\chi}_{0, 0}  \right)\left(g_-^*\hat{a}_0 + g_-\hat{a}_0^\dagger \right) -ig_\alpha\left(\chi_{0, 0} - \overline{\chi}_{0, 0}  \right)\left(\hat{a}_{1} + \hat{a}_{1}^\dagger \right) -i\left(\chi_{0, 0} - \overline{\chi}_{0, 0}  \right)\left(g_+^*\hat{a}_{2} + g_+\hat{a}_2^\dagger \right) + \hat{S}
\end{equation}
and keep only first order terms to obtain
\begin{eqnarray}
\hat{b}_0 + \hat{b}_0^\dagger & = & -|g_-|^2\left(\chi_{0, 0} - \overline{\chi}_{0, 0}  \right)\left[\mathcal{J}_-(\Omega)\chi_\mathrm{g, 0} - \mathcal{J}_-^*(-\Omega)\overline{\chi}_\mathrm{g, 0} \right]\left(  \hat{b}_0 + \hat{b}_0^\dagger \right)\nonumber \\
& & -g_\alpha^2\left(\chi_{0, 0} - \overline{\chi}_{0, 0}  \right)\left[\mathcal{J}_\alpha(\Omega + \Omega_\mathrm{dp})\chi_\mathrm{g, 1} - \mathcal{J}_\alpha^*(-\Omega + \Omega_\mathrm{dp})\overline{\chi}_\mathrm{g, 1} \right]\left( \hat{b}_0 + \hat{b}_0^\dagger \right)\nonumber \\
& & -|g_+|^2\left(\chi_{0, 0} - \overline{\chi}_{0, 0}  \right)\left[\mathcal{J}_+(\Omega + 2\Omega_\mathrm{dp})\chi_\mathrm{g, 2} - \mathcal{J}_+^*(-\Omega + 2\Omega_\mathrm{dp})\overline{\chi}_\mathrm{g, 2} \right]\left( \hat{b}_0 + \hat{b}_0^\dagger \right)\nonumber \\
& & -i\left(\chi_{0, 0} - \overline{\chi}_{0, 0}  \right)\left[g_-^*\chi_\mathrm{g, 0}\hat{\mathcal{M}}_{0+} + g_-\overline{\chi}_\mathrm{g, 0}\hat{\mathcal{M}}_{0-}^\dagger + g_\alpha\chi_\mathrm{g, 1}\hat{\mathcal{M}}_{1+} + g_\alpha\overline{\chi}_\mathrm{g, 1}\hat{\mathcal{M}}_{1-}^\dagger + g_+^*\chi_\mathrm{g, 2}\hat{\mathcal{M}}_{2+} + g_+\overline{\chi}_\mathrm{g, 2}\hat{\mathcal{M}}_{2-}^\dagger  \right] \nonumber \\
& & + \hat{S}
\end{eqnarray}
We can find our earlier obtained four-wave dynamical backaction in this relation and write
\begin{eqnarray}
\hat{b}_0 + \hat{b}_0^\dagger & = & -i\frac{\chi_{0, 0} - \overline{\chi}_{0, 0}  }{1+\left( \chi_{0, 0} - \overline{\chi}_{0, 0}\right)\Sigma_\mathrm{fw}(\Omega_\mathrm{m})}\left[g_-^*\chi_\mathrm{g, 0}\hat{\mathcal{M}}_{0+} + g_-\overline{\chi}_\mathrm{g, 0}\hat{\mathcal{M}}_{0-}^\dagger + g_\alpha\chi_\mathrm{g, 1}\hat{\mathcal{M}}_{1+} + g_\alpha\overline{\chi}_\mathrm{g, 1}\hat{\mathcal{M}}_{1-}^\dagger + g_+^*\chi_\mathrm{g, 2}\hat{\mathcal{M}}_{2+} + g_+\overline{\chi}_\mathrm{g, 2}\hat{\mathcal{M}}_{2-}^\dagger  \right] \nonumber \\
& & + \frac{\hat{S}}{1+\left( \chi_{0, 0} - \overline{\chi}_{0, 0}\right)\Sigma_\mathrm{fw}(\Omega_\mathrm{m})}
\end{eqnarray}
For a pump on the red sideband of the signal resonance, a high mechanical quality factor and the detection frequency to be $\Omega \approx \Omega_\mathrm{m}$, we can simplify the relations, i.e., keep only dominant terms and obtain
\begin{eqnarray}
\hat{a}_0 & = & -ig_-\mathcal{J}_-(\Omega)\chi_\mathrm{g,0}\hat{b}_0 + \chi_\mathrm{g, 0}\hat{\mathcal{M}}_{0+} \label{eqn:SimpleFields_1} \\
\hat{a}_0^\dagger & = & i g_-^*\mathcal{J}_-^*(-\Omega)\overline{\chi}_\mathrm{g, 0}\hat{b}_0 + \overline{\chi}_\mathrm{g, 0}\hat{\mathcal{M}}_{0-}^\dagger \\
\hat{a}_1 & = & -ig_\alpha \mathcal{J}_\alpha(\Omega + \Omega_\mathrm{dp})\chi_\mathrm{g,1} \hat{b}_0 + \chi_\mathrm{g, 1}\hat{\mathcal{M}}_{1+} \\
\hat{a}_{1}^\dagger & = & ig_\alpha \mathcal{J}_\alpha^*(-\Omega + \Omega_\mathrm{dp})\overline{\chi}_\mathrm{g,1}\hat{b}_0 + \overline{\chi}_\mathrm{g, 1}\hat{\mathcal{M}}_{1-}^\dagger \\
\hat{a}_2 & = & -ig_+ \mathcal{J}_+(\Omega + 2\Omega_\mathrm{dp})\chi_\mathrm{g,2} \hat{b}_0 + \chi_\mathrm{g, 2}\hat{\mathcal{M}}_{2+} \\
\hat{a}_{2}^\dagger & = & ig_+^* \mathcal{J}_+^*(-\Omega + 2\Omega_\mathrm{dp})\overline{\chi}_\mathrm{g,2}\hat{b}_0 + \overline{\chi}_\mathrm{g, 2}\hat{\mathcal{M}}_{2-}^\dagger. \label{eqn:SimpleFields_6}
\end{eqnarray}
For the detection frequency range, we therefore get
\begin{eqnarray}
\hat{a}_0 & = & -g_-\mathcal{J}_-(\Omega)\chi_\mathrm{g, 0}\chi_{0,0}^\mathrm{eff}\left[g_-^*\chi_\mathrm{g, 0}\hat{\mathcal{M}}_{0+} + g_-\overline{\chi}_\mathrm{g, 0}\hat{\mathcal{M}}_{0-}^\dagger + g_\alpha\chi_\mathrm{g, 1}\hat{\mathcal{M}}_{1+} + g_\alpha\overline{\chi}_\mathrm{g, 1}\hat{\mathcal{M}}_{1-}^\dagger + g_+^*\chi_\mathrm{g, 2}\hat{\mathcal{M}}_{2+} + g_+\overline{\chi}_\mathrm{g, 2}\hat{\mathcal{M}}_{2-}^\dagger  \right] \nonumber \\
& & -ig_-\mathcal{J}_-(\Omega)\chi_\mathrm{g, 0}\chi_{0,0}^\mathrm{eff}\sqrt{\Gamma_\mathrm{m}}\hat{\zeta} + \chi_\mathrm{g,0}\hat{\mathcal{M}}_{0+}
\end{eqnarray}
where we applied $\overline{\chi}_{0,0} \approx 0$ for $\Omega \approx \Omega_\mathrm{m}$.
Note that the cavity noise is well described for all frequencies in this approximation, but the upconverted mechanical noise is limited to one of the many sidebands.
We can resolve and sort now for input noise frequency components, where we only keep cavity input noise terms around the signal and the idler resonances.
The result is
\begin{eqnarray}
\hat{a}_0 & = & -g_-\mathcal{J}_-(\Omega)\chi_\mathrm{g, 0}\chi_{0,0}^\mathrm{eff}\left[g_-^*\chi_\mathrm{g, 0}\hat{\mathcal{M}}_{0+} + g_-\overline{\chi}_\mathrm{g, 0}\hat{\mathcal{M}}_{0-}^\dagger + g_\alpha\chi_\mathrm{g, 1}\hat{\mathcal{M}}_{1+} + g_\alpha\overline{\chi}_\mathrm{g, 1}\hat{\mathcal{M}}_{1-}^\dagger + g_+\chi_\mathrm{g, 2}\hat{\mathcal{M}}_{2+} + g_+^*\overline{\chi}_\mathrm{g, 2}\hat{\mathcal{M}}_{2-}^\dagger  \right] \nonumber \\
& & -ig_-\mathcal{J}_-(\Omega)\chi_\mathrm{g, 0}\chi_{0,0}^\mathrm{eff}\sqrt{\Gamma_\mathrm{m}}\hat{\zeta} + \chi_\mathrm{g,0}\hat{\mathcal{M}}_{0+} \nonumber \\
& \approx &  -ig_-\mathcal{J}_-(\Omega)\chi_\mathrm{g, 0}\chi_{0,0}^\mathrm{eff}\left[\sqrt{\Gamma_\mathrm{m}}\hat{\zeta} - i\left(g_-^*\chi_\mathrm{g, 0}\mathcal{P}_- + g_+\mathcal{A}_0\mathcal{P}_+\overline{\chi}_\mathrm{g, 2} \right)\hat{\mathcal{N}}_{0+} - i\left(g_-^*\overline{\mathcal{A}}_2\mathcal{P}_-\chi_\mathrm{g, 0} + g_+\mathcal{P}_+\overline{\chi}_\mathrm{g, 2} \right)\hat{\mathcal{N}}_{2-}^\dagger \right] \nonumber \\
& & + \chi_\mathrm{g,0}\hat{\mathcal{N}}_{0+} + \chi_\mathrm{g,0}\overline{\mathcal{A}}_2\hat{\mathcal{N}}_{2-}^\dagger 
\end{eqnarray}
with
\begin{eqnarray}
\mathcal{P}_- & = & \left[1 + i 2\mathcal{K} n_\mathrm{d}\chi_\mathrm{g, 1}\left(1-\overline{\mathcal{A}}_1  + \frac{g_+ }{g_-^*} \right) -i2\mathcal{K}n_\mathrm{d}\overline{\chi}_\mathrm{g, 1}\left(1-\mathcal{A}_1 - \frac{ g_+}{g_-^*}\mathcal{A}_1 \right) \right]        \\
\mathcal{P}_+ & = & \left[1 - i 2\mathcal{K} n_\mathrm{d}\overline{\chi}_\mathrm{g, 1}\left(1-\mathcal{A}_1  + \frac{g_-^* }{g_+} \right) +i2\mathcal{K}n_\mathrm{d}\chi_\mathrm{g, 1}\left(1-\overline{\mathcal{A}}_1 - \frac{ g_-^*}{g_+}\overline{\mathcal{A}}_1 \right) \right]
\end{eqnarray}
For the cavity output field on one side of the feedline, we get
\begin{eqnarray}
\hat{a}_\mathrm{out} & = & \hat{\xi}_\mathrm{e0+}^\mathrm{left} - \sqrt{\frac{\kappa_\mathrm{e}}{2}}\hat{a}_0 \nonumber \\
& = & -ig_-\mathcal{J}_-(\Omega)\chi_\mathrm{g, 0}\chi_{0,0}^\mathrm{eff}\sqrt{\frac{\kappa_\mathrm{e}}{2}}\sqrt{\Gamma_\mathrm{m}}\hat{\zeta} \nonumber \\
& & + \chi_\mathrm{g, 0}\sqrt{\frac{\kappa_\mathrm{e}}{2}}\left[1 - g_-\mathcal{J}_-(\Omega)\chi_{0,0}^\mathrm{eff}\left(g_-^*\chi_\mathrm{g, 0}\mathcal{P}_- + g_+\mathcal{A}_0\mathcal{P}_+\overline{\chi}_\mathrm{g, 2} \right)     \right]\sqrt{\kappa_\mathrm{i}}\hat{\xi}_\mathrm{i0+} \nonumber \\
& & + \chi_\mathrm{g, 0}\sqrt{\frac{\kappa_\mathrm{e}}{2}}\left[1 - g_-\mathcal{J}_-(\Omega)\chi_{0,0}^\mathrm{eff}\left(g_-^*\chi_\mathrm{g, 0}\mathcal{P}_- + g_+\mathcal{A}_0\mathcal{P}_+\overline{\chi}_\mathrm{g, 2} \right)     \right]\sqrt{\frac{\kappa_\mathrm{e}}{2}}\hat{\xi}_\mathrm{e0+}^\mathrm{right} \nonumber \\
& & + \chi_\mathrm{g, 0}\sqrt{\frac{\kappa_\mathrm{e}}{2}}\left[\overline{\mathcal{A}}_2 - g_-\mathcal{J}_-(\Omega)\chi_{0,0}^\mathrm{eff}\left(g_-^*\chi_\mathrm{g, 0}\overline{\mathcal{A}}_2\mathcal{P}_- + g_+\mathcal{P}_+\overline{\chi}_\mathrm{g, 2} \right)     \right]\sqrt{\kappa_\mathrm{i}}\hat{\xi}_\mathrm{i2-}^\dagger \nonumber \\
& & + \chi_\mathrm{g, 0}\sqrt{\frac{\kappa_\mathrm{e}}{2}}\left[\overline{\mathcal{A}}_2 - g_-\mathcal{J}_-(\Omega)\chi_{0,0}^\mathrm{eff}\left(g_-^*\chi_\mathrm{g, 0}\overline{\mathcal{A}}_2\mathcal{P}_- + g_+\mathcal{P}_+\overline{\chi}_\mathrm{g, 2} \right)     \right]\sqrt{\kappa_\mathrm{e}}\hat{\xi}_\mathrm{e2-}^\dagger \nonumber \\
& & + \left(1 - \chi_\mathrm{g, 0}\sqrt{\frac{\kappa_\mathrm{e}}{2}}\left[1 - g_-\mathcal{J}_-(\Omega)\chi_{0,0}^\mathrm{eff}\left(g_-^*\chi_\mathrm{g, 0}\mathcal{P}_- + g_+\mathcal{A}_0\mathcal{P}_+\overline{\chi}_\mathrm{g, 2} \right)     \right]\sqrt{\frac{\kappa_\mathrm{e}}{2}}\right)\hat{\xi}_\mathrm{e0+}^\mathrm{left}
\label{eqn:noise_output_red}
\end{eqnarray}
where we split the relevant external input noise into the contributions from the left and the right side of the feedline.
This can be used directly to calculate the symmetrized output field power spectral density in units of phonons as
\begin{equation}
\frac{S(\omega)}{\hbar\omega} = n_\mathrm{add} + \frac{1}{2}\langle \hat{a}_\mathrm{out}^\dagger \hat{a}_\mathrm{out} + \hat{a}_\mathrm{out} \hat{a}_\mathrm{out}^\dagger\rangle.
\label{eqn:Sout_red}
\end{equation}
The total number of noise photons added by our detection chain is found to be $n_\mathrm{add}\approx 13$ from a thermal calibration of the residual occupation of the mechanical oscillator, cf. Supplementary Notes 14 and 15.
To calculate the corresponding phonon occupation, we use the relations (\ref{eqn:SimpleFields_1}) - (\ref{eqn:SimpleFields_6}) and keep only cavity noise input terms for $\hat{\mathcal{N}}_{0+}$ and $\hat{\mathcal{N}}_{2-}^\dagger$
\begin{eqnarray}
\frac{\hat{b}_0}{\chi_\mathrm{0, 0}^\mathrm{eff}} & = & -ig_-^*\mathcal{P}_-\chi_\mathrm{g, 0}\left[\hat{\mathcal{N}}_{0+} + \overline{\mathcal{A}}_2 \hat{\mathcal{N}}_{2-}^\dagger  \right] -ig_+\mathcal{P}_+\overline{\chi}_\mathrm{g, 2}\left[\hat{\mathcal{N}}_{2-}^\dagger + \mathcal{A}_0 \hat{\mathcal{N}}_{0+} \right] + \sqrt{\Gamma_\mathrm{m}}\hat{\zeta}
\end{eqnarray}
which gives the mechanical power spectral density
\begin{equation}
\langle \hat{b}_0^\dagger\hat{b}_0 \rangle = |\chi_{0, 0}^\mathrm{eff}|^2 |g_-^*\mathcal{P}_-\chi_\mathrm{g, 0} + g_+\mathcal{P}_+\mathcal{A}_0\overline{\chi}_\mathrm{g, 2}|^2 \kappa n_\mathrm{c}^\mathrm{th} + |\chi_{0, 0}^\mathrm{eff}|^2 |g_-^*\mathcal{P}_-\overline{\mathcal{A}}_2\chi_\mathrm{g, 0} + g_+\mathcal{P}_+\overline{\chi}_\mathrm{g, 2}|^2\kappa\left(n_\mathrm{c}^\mathrm{th} + 1\right) + |\chi_{0, 0}^\mathrm{eff}|^2\Gamma_\mathrm{m}n_\mathrm{m}^\mathrm{th}.
\label{eqn:PhononPSD_red}
\end{equation}
The integration of this relation over all frequencies then results in the effective phonon occupation in presence of the optomechanical coupling.

\subsection*{Idler resonance blue sideband pumping}

In this case, we use
\begin{equation}
\hat{b}_{-2} + \hat{b}_2^\dagger = -i\left(\chi_{0, -2} - \overline{\chi}_{0, 2}  \right)\left(g_-^*\hat{a}_{-2} + g_-\hat{a}_{2}^\dagger \right) -ig_\alpha\left(\chi_{0, -2} - \overline{\chi}_{0, -2}  \right)\left(\hat{a}_{-1} + \hat{a}_{3}^\dagger \right) -i\left(\chi_{0, -2} - \overline{\chi}_{0, 2}  \right)\left(g_+^*\hat{a}_{0} + g_+\hat{a}_4^\dagger \right) + \hat{S}
\end{equation}
and keep only first order terms to obtain
\begin{eqnarray}
\hat{b}_{-2} + \hat{b}_2^\dagger & = & -|g_-|^2\left(\chi_{0, -2} - \overline{\chi}_{0, 2}  \right)\left[\mathcal{J}_-(\Omega-2\Omega_\mathrm{dp})\chi_\mathrm{g, -2} - \mathcal{J}_-^*(-\Omega + 2\Omega_\mathrm{dp})\overline{\chi}_\mathrm{g, 2} \right]\left(  \hat{b}_{-2} + \hat{b}_2^\dagger \right)\nonumber \\
& & -g_\alpha^2\left(\chi_{0, -2} - \overline{\chi}_{0, 2}  \right)\left[\mathcal{J}_\alpha(\Omega - \Omega_\mathrm{dp})\chi_\mathrm{g, -1} - \mathcal{J}_\alpha^*(-\Omega + 3\Omega_\mathrm{dp})\overline{\chi}_\mathrm{g, 3} \right]\left( \hat{b}_{-2} + \hat{b}_2^\dagger \right)\nonumber \\
& & -|g_+|^2\left(\chi_{0, -2} - \overline{\chi}_{0, 2}  \right)\left[\mathcal{J}_+(\Omega)\chi_\mathrm{g, 0} - \mathcal{J}_+^*(-\Omega + 4\Omega_\mathrm{dp})\overline{\chi}_\mathrm{g, 4} \right]\left( \hat{b}_{-2} + \hat{b}_2^\dagger \right)\nonumber \\
& & -i\left(\chi_{0, -2} - \overline{\chi}_{0, 2}  \right)\left[g_-^*\chi_\mathrm{g, -2}\hat{\mathcal{M}}_{-2+} + g_-\overline{\chi}_\mathrm{g, 2}\hat{\mathcal{M}}_{2-}^\dagger + g_\alpha\chi_\mathrm{g, -1}\hat{\mathcal{M}}_{-1+} + g_\alpha\overline{\chi}_\mathrm{g, 3}\hat{\mathcal{M}}_{3-}^\dagger + g_+^*\chi_\mathrm{g, 0}\hat{\mathcal{M}}_{0+} + g_+\overline{\chi}_\mathrm{g, 4}\hat{\mathcal{M}}_{4-}^\dagger  \right] \nonumber \\
& & + \hat{S}
\end{eqnarray}
We can find our earlier obtained four-wave dynamical backaction in this relation and write
\begin{eqnarray}
\hat{b}_{-2} + \hat{b}_2^\dagger & = & -i\frac{\chi_{0, -2} - \overline{\chi}_{0, 2}  }{1+\left( \chi_{0, -2} - \overline{\chi}_{0, 2}\right)\Sigma_\mathrm{fw}(\Omega_\mathrm{m})}\bigg[g_-^*\chi_\mathrm{g, -2}\hat{\mathcal{M}}_{-2+} + g_-\overline{\chi}_\mathrm{g, 2}\hat{\mathcal{M}}_{2-}^\dagger + g_\alpha\chi_\mathrm{g, -1}\hat{\mathcal{M}}_{-1+} \nonumber \\
& & ~~~~~~~~~~~~~~~~~~~~~~~~~~~~~~~~~~~~~~~~~ + g_\alpha\overline{\chi}_\mathrm{g, 3}\hat{\mathcal{M}}_{3-}^\dagger + g_+^*\chi_\mathrm{g, 0}\hat{\mathcal{M}}_{0+} + g_+\overline{\chi}_\mathrm{g, 4}\hat{\mathcal{M}}_{4-}^\dagger  \bigg] \nonumber \\
& & + \frac{\hat{S}}{1+\left( \chi_{0, -2} - \overline{\chi}_{0, 2}\right)\Sigma_\mathrm{fw}(\Omega_\mathrm{m})}
\end{eqnarray}
For a pump on the blue sideband of the idler resonance, a high mechanical quality factor and the detection frequency to be $\Omega \approx \Omega_\mathrm{m} - 2\Omega_\mathrm{dp}$, we can simplify the cavity field operator relations, i.e., keep only dominant terms and obtain
\begin{eqnarray}
\hat{a}_{-2} & = & -ig_-\mathcal{J}_-(\Omega - 2\Omega_\mathrm{dp})\chi_\mathrm{g,-2}\hat{b}_{-2} + \chi_\mathrm{g, -2}\hat{\mathcal{M}}_{-2+} \label{eqn:SimpleFieldsBlue_1} \\
\hat{a}_2^\dagger & = & i g_-^*\mathcal{J}_-^*(-\Omega + 2\Omega_\mathrm{dp})\overline{\chi}_\mathrm{g, 2}\hat{b}_{-2} + \overline{\chi}_\mathrm{g, 2}\hat{\mathcal{M}}_{2-}^\dagger \\
\hat{a}_{-1} & = & -ig_\alpha \mathcal{J}_\alpha(\Omega - \Omega_\mathrm{dp})\chi_\mathrm{g, -1} \hat{b}_{-2} + \chi_\mathrm{g, -1}\hat{\mathcal{M}}_{-1+} \\
\hat{a}_{3}^\dagger & = & ig_\alpha \mathcal{J}_\alpha^*(-\Omega + 3\Omega_\mathrm{dp})\overline{\chi}_\mathrm{g,3}\hat{b}_{-2} + \overline{\chi}_\mathrm{g, 3}\hat{\mathcal{M}}_{3-}^\dagger \\
\hat{a}_0 & = & -ig_+ \mathcal{J}_+(\Omega)\chi_\mathrm{g,0} \hat{b}_{-2} + \chi_\mathrm{g, 0}\hat{\mathcal{M}}_{0+} \\
\hat{a}_{4}^\dagger & = & ig_+^* \mathcal{J}_+^*(-\Omega + 4\Omega_\mathrm{dp})\overline{\chi}_\mathrm{g,4}\hat{b}_{-2} + \overline{\chi}_\mathrm{g, 4}\hat{\mathcal{M}}_{4-}^\dagger. \label{eqn:SimpleFieldsBlue_6}
\end{eqnarray}
For the detection frequency range, we therefore get
\begin{eqnarray}
\hat{a}_0 & = & -g_+\mathcal{J}_+(\Omega)\chi_\mathrm{g, 0}\chi_{0,-2}^\mathrm{eff}\left[g_-^*\chi_\mathrm{g, -2}\hat{\mathcal{M}}_{-2+} + g_-\overline{\chi}_\mathrm{g, 2}\hat{\mathcal{M}}_{2-}^\dagger + g_\alpha\chi_\mathrm{g, -1}\hat{\mathcal{M}}_{-1+} + g_\alpha\overline{\chi}_\mathrm{g, 3}\hat{\mathcal{M}}_{3-}^\dagger + g_+^*\chi_\mathrm{g, 0}\hat{\mathcal{M}}_{0+} + g_+\overline{\chi}_\mathrm{g, 4}\hat{\mathcal{M}}_{4-}^\dagger  \right] \nonumber \\
& & -ig_+\mathcal{J}_+(\Omega)\chi_\mathrm{g, 0}\chi_{0,-2}^\mathrm{eff}\sqrt{\Gamma_\mathrm{m}}\hat{\zeta} + \chi_\mathrm{g,0}\hat{\mathcal{M}}_{0+}
\end{eqnarray}
where we applied $\overline{\chi}_{0,2} \approx 0$ for $\Omega \approx \Omega_\mathrm{m} - 2\Omega_\mathrm{dp}$.
We can resolve and sort now for input noise frequency components again, where we only keep cavity input noise terms around the signal and the idler resonances.
The result is
\begin{eqnarray}
\hat{a}_0 & \approx &  -ig_+\mathcal{J}_+(\Omega)\chi_\mathrm{g, 0}\chi_{0,-2}^\mathrm{eff}\left[\sqrt{\Gamma_\mathrm{m}}\hat{\zeta} - i\left(g_+^*\mathcal{Q}_+ \chi_\mathrm{g, 0} + g_-\mathcal{A}_0\mathcal{Q}_-\overline{\chi}_\mathrm{g, 2} \right)\hat{\mathcal{N}}_{0+} - i\left(g_+^*\overline{\mathcal{A}}_2\mathcal{Q}_+\chi_\mathrm{g, 0} + g_-\mathcal{Q}_-\overline{\chi}_\mathrm{g, 2} \right)\hat{\mathcal{N}}_{2-}^\dagger \right] \nonumber \\
& & + \chi_\mathrm{g,0}\hat{\mathcal{N}}_{0+} + \chi_\mathrm{g,0}\overline{\mathcal{A}}_2\hat{\mathcal{N}}_{2-}^\dagger 
\end{eqnarray}
with
\begin{eqnarray}
\mathcal{Q}_- & = & \left[1 + i 2\mathcal{K} n_\mathrm{d}\chi_\mathrm{g, -1}\left(1-\overline{\mathcal{A}}_3  - \frac{g_+^* }{g_-}\overline{\mathcal{A}}_3 \right) -i2\mathcal{K}n_\mathrm{d}\overline{\chi}_\mathrm{g, 3}\left(1-\mathcal{A}_{-1} - \frac{ g_+^*}{g_-} \right) \right]        \\
\mathcal{Q}_+ & = & \left[1 - i 2\mathcal{K} n_\mathrm{d}\overline{\chi}_\mathrm{g, 3}\left(1-\mathcal{A}_{-1}  - \frac{g_- }{g_+^*}\mathcal{A}_{-1} \right) +i2\mathcal{K}n_\mathrm{d}\chi_\mathrm{g, -1}\left(1-\overline{\mathcal{A}}_3 - \frac{ g_-}{g_+^*}\overline{\mathcal{A}}_3 \right) \right]
\end{eqnarray}
For the cavity output field on one side of the feedline, we get
\begin{eqnarray}
\hat{a}_\mathrm{out} & = & \hat{\xi}_\mathrm{e0+}^\mathrm{left} - \sqrt{\frac{\kappa_\mathrm{e}}{2}}\hat{a}_0 \nonumber \\
& = & ig_+\mathcal{J}_+(\Omega)\chi_\mathrm{g, 0}\chi_{0,-2}^\mathrm{eff}\sqrt{\frac{\kappa_\mathrm{e}}{2}}\sqrt{\Gamma_\mathrm{m}}\hat{\zeta} \nonumber \\
& & - \chi_\mathrm{g, 0}\sqrt{\frac{\kappa_\mathrm{e}}{2}}\left[1 - g_+\mathcal{J}_+(\Omega)\chi_{0,-2}^\mathrm{eff}\left(g_+^* \mathcal{Q}_+ \chi_\mathrm{g, 0} + g_-\mathcal{A}_0\mathcal{Q}_-\overline{\chi}_\mathrm{g, 2} \right)     \right]\sqrt{\kappa_\mathrm{i}}\hat{\xi}_\mathrm{i0+} \nonumber \\
& & - \chi_\mathrm{g, 0}\sqrt{\frac{\kappa_\mathrm{e}}{2}}\left[1 - g_+\mathcal{J}_+(\Omega)\chi_{0,-2}^\mathrm{eff}\left(g_+^* \mathcal{Q}_+ \chi_\mathrm{g, 0} + g_-\mathcal{A}_0\mathcal{Q}_-\overline{\chi}_\mathrm{g, 2} \right)     \right]\sqrt{\frac{\kappa_\mathrm{e}}{2}}\hat{\xi}_\mathrm{e0+}^\mathrm{right} \nonumber \\
& & - \chi_\mathrm{g, 0}\sqrt{\frac{\kappa_\mathrm{e}}{2}}\left[\overline{\mathcal{A}}_2 - g_+\mathcal{J}_+(\Omega)\chi_{0,-2}^\mathrm{eff}\left(g_+^*\overline{\mathcal{A}}_2\mathcal{Q}_+ \chi_\mathrm{g, 0} + g_-\mathcal{Q}_-\overline{\chi}_\mathrm{g, 2} \right)     \right]\sqrt{\kappa_\mathrm{i}}\hat{\xi}_\mathrm{i2-}^\dagger \nonumber \\
& & - \chi_\mathrm{g, 0}\sqrt{\frac{\kappa_\mathrm{e}}{2}}\left[\overline{\mathcal{A}}_2 - g_+\mathcal{J}_+(\Omega)\chi_{0,-2}^\mathrm{eff}\left(g_+^*\overline{\mathcal{A}}_2\mathcal{Q}_+ \chi_\mathrm{g, 0} + g_-\mathcal{Q}_-\overline{\chi}_\mathrm{g, 2} \right)        \right]\sqrt{\kappa_\mathrm{e}}\hat{\xi}_\mathrm{e2-}^\dagger \nonumber \\
& & + \left(1 - \chi_\mathrm{g, 0}\sqrt{\frac{\kappa_\mathrm{e}}{2}}\left[1 - g_+\mathcal{J}_+(\Omega)\chi_{0,-2}^\mathrm{eff}\left(g_+^* \mathcal{Q}_+ \chi_\mathrm{g, 0} + g_-\mathcal{A}_0\mathcal{Q}_-\overline{\chi}_\mathrm{g, 2} \right)    \right]\sqrt{\frac{\kappa_\mathrm{e}}{2}}\right)\hat{\xi}_\mathrm{e0+}^\mathrm{left}
\label{eqn:noise_output_blue}
\end{eqnarray}
This can be used just as in the red sideband case to calculate the output field power spectral density in units of phonons.
To calculate the corresponding phonon occupation, we use relations (\ref{eqn:SimpleFieldsBlue_1}) - (\ref{eqn:SimpleFieldsBlue_6}) and keep only cavity noise input terms for $\hat{\mathcal{N}}_{0+}$ and $\hat{\mathcal{N}}_{2-}^\dagger$
\begin{eqnarray}
\frac{\hat{b}_{-2}}{\chi_\mathrm{0, -2}^\mathrm{eff}} & = & -ig_+^*\mathcal{Q}_+\chi_\mathrm{g, 0}\left[\hat{\mathcal{N}}_{0+} + \overline{\mathcal{A}}_2 \hat{\mathcal{N}}_{2-}^\dagger  \right] -ig_-\mathcal{Q}_-\overline{\chi}_\mathrm{g, 2}\left[\hat{\mathcal{N}}_{2-}^\dagger + \mathcal{A}_0 \hat{\mathcal{N}}_{0+} \right] + \sqrt{\Gamma_\mathrm{m}}\hat{\zeta}
\end{eqnarray}
which gives the mechanical power spectral density
\begin{equation}
\langle \hat{b}_{-2}^\dagger\hat{b}_{-2} \rangle = |\chi_{0, -2}^\mathrm{eff}|^2 |g_+^*\mathcal{Q}_+\chi_\mathrm{g, 0} + g_-\mathcal{Q}_-\mathcal{A}_0\overline{\chi}_\mathrm{g, 2}|^2 \kappa n_\mathrm{c}^\mathrm{th} + |\chi_{0, -2}^\mathrm{eff}|^2 |g_+^*\mathcal{Q}_+\overline{\mathcal{A}}_2\chi_\mathrm{g, 0} + g_-\mathcal{Q}_-\overline{\chi}_\mathrm{g, 2}|^2 \kappa \left(n_\mathrm{c}^\mathrm{th} + 1\right) + |\chi_{0, -2}^\mathrm{eff}|^2\Gamma_\mathrm{m}n_\mathrm{m}^\mathrm{th}.
\label{eqn:PhononPSD_blue}
\end{equation}
The integration of this relation over all frequencies then results in the effective phonon occupation in presence of the optomechanical coupling.

\section*{Supplementary Note 14: Measurement and analysis protocols for thermal noise detection and four-wave cooling}
\subsection*{Preparation}
\begin{itemize}
	\item{We start the experimental cycle with choosing a bias flux operation point, either point I or point II, and an in-plane magnetic field $B_\parallel$. We ramp the in-plane magnet current to its corresponding value.}
	\item{A parametric drive tone is sent to the cavity with fixed frequency $\omega_\mathrm{d}$ and power $P_\mathrm{d}$ to match the chosen operation point.}
	\item{The cavity flux bias is adjusted manually to prepare the SQUID cavity in the driven Kerr mode state.}
	\item{The frequency of the optomechanical pump is chosen to be either on the red sideband of the signal resonance or on the blue sideband of the idler resonance. The pump is activated with frequency $\omega_\mathrm{p}$ and power $P_\mathrm{p}$.}
\end{itemize}
\subsection*{Thermal calibration measurement}
\begin{itemize}
	\item{The optomechanical pump is positioned on the red sideband of the signal resonance and the fridge is sitting at base temperature.}
	\item{For the actual measurement, we start a python-based control and data acquisition script.}
	\item{Prior to running the measurement, we input some fixed parameters to the script such as the values of all external attenuators in the input lines and the number of room-temperature amplifiers on the output line.}
	\item{We manually adjust the probe VNA to the parameter set regarding frequency window, probe power and bandwidth in order to measure a clean OMIT response. We hereby choose a red-sideband pump power, that is large enough to yield a clear OMIT response, but low enough to keep the effective cooperativity in the regime $\sim 1$.}
	\item{Upon a terminal input command, the script begins data acquisition and first catches all relevant parameters such as powers, frequencies, frequency spans, bandwidths as well as magnet DC currents from all participating electronic devices.}
	\item{Afterwards, the script takes three datasets.}
	\item{First, it performs a scan of the narrow-band OMIT window using the VNA and the manually adjusted settings.}
	\item{Secondly, the center frequency and frequency span are sent to the spectrum analyzer to measure a power spectrum in the OMIT frequency range. During this spectrum analyzer measurement, the VNA frequency is set to a frequency $4\,$kHz detuned from the detection window of the spectrum analyzer to avoid any interference and the cavity response at a single frequency point is monitored permanently during the spectrum acquisition. This measure enables to control the cavity response by out-of-plane current feedback for the case the cavity is drifting during the spectrum measurement.}
	\item{Lastly, a wide-band VNA scan of the complete cavity response is taken.}
	\item{Each of the three data traces is stored in a separate file.}
	\item{For most temperatures, we repeat the measurement once.}
	\item{We adjust the fridge temperature to its new set value and after a temperature settling time of $\sim$10 minutes, we begin the cycle from the beginning at the new base temperature.}
	\item{Parametric drive and optomechanical pump powers $P_\mathrm{d}$ and $P_\mathrm{p}$, respectively, were adjusted during the temperature sweep to keep driven cavity state and effective cooperativity nearly constant.}
\end{itemize}

\subsection*{Four-wave-cooling measurement}
\begin{itemize}
	\item{For the four-wave-cooling measurement we start a python-based control and data acquisition script, which is programmed to wait for an input terminal starting command before each data point.}
	\item{Prior to running the measurement, we input some fixed parameters to the script such as the values of all external attenuators in the input lines and the number of room-temperature amplifiers on the output lines.}
	\item{We set the parametric drive power and the optomechanical pump powers to the desired values $P_\mathrm{d}$ and $P_\mathrm{p}$, respectively, and manually adjust the probe VNA to the parameter set regarding frequency window, probe power and bandwidth in order to measure a clean OMIT response.}
	\item{Upon a terminal input command, the script begins data acquisition and first catches all relevant parameters such as powers, frequencies, frequency spans, bandwidths as well as magnet DC currents from all participating electronic devices.}
	\item{At this point, the script waits for the final command to measure three traces. Once we observe a stable response in the OMIT window on the VNA screen, the script is continued.}
	\item{The first data trace acquired by the script is a narrow-band VNA scan of the OMIT response using the manually adjusted settings.}
	\item{Secondly, the center frequency and frequency span of the VNA trace are sent to the spectrum analyzer and an output power spectrum in the OMIT frequency range is acquired. During this spectrum measurement, the probe VNA frequency is set to a value several spectrum analyzer frequency spans detuned from the detection window of the spectrum analyzer window to avoid any interference between the VNA signal and the power spectrum. The VNA is continuously scanning a single frequency point of the cavity response to enable monitoring and feedback control of the bias flux for the case the cavity response is drifting due to flux drifts.}
	\item{Lastly, a wide-band VNA trace of the complete signal resonance $S_{21}$ is acquired.}
	\item{The three data traces are stored in individual files, where also the subsequent equivalent traces for the next settings are appended.}
	\item{The VNA setting are reset to the OMIT window and manual VNA control is enabled by the measurement script.}
	\item{At this point, we can choose by a terminal input between a repetition of the same measurement or a continuation to the next settings.}
	\item{In case of continuation to the next settings, the optomechanical pump power $P_\mathrm{p}$ is manually set to its new value and the VNA probe settings are adjusted for a the next measurement iteration. The most important parameter is the frequency range for the OMIT and the spectrum analyzer window, which has to be increased with increasing dynamical backaction due to the considerable increase of the mechanical linewidth from the bare value of $\sim15\,$Hz to the largest effective linewidth of $\sim 1.5\,$kHz for the highest achieved effective cooperativity.}
	\item{Depending on the optomechanical pump power, we also adjust the parametric drive power $P_\mathrm{d}$ in some cases to keep the driven cavity response nearly constant. We suspect this measure is necessary as for large optomechanical pump powers, the experiment is at the edge of the linearized regime with respect to $\gamma_-, \gamma_+$ and parametric drive depletion is occurring.}
	\item{Once the new parameters are adjusted, the data acquisition cycle starts from the beginning.}
\end{itemize}
\subsection*{Thermal calibration data analysis}
\begin{figure*}
	\centerline{\includegraphics[trim = {0cm, 0cm, 0cm, 0cm}, clip=True, width=0.8\textwidth]{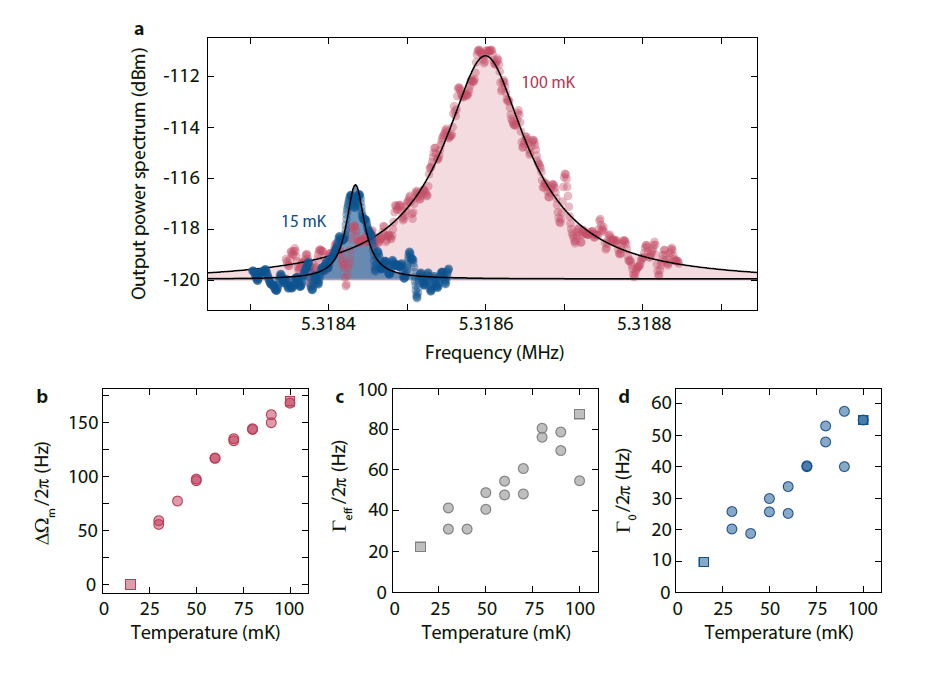}}
	\caption{\textsf{\textbf{Temperature dependence of the mechanical oscillator.} \textbf{a} shows the power spectra of the signal resonance output field during optomechanical red-sideband pumping at two different refrigerator base temperatures $T_\mathrm{b}^\mathrm{min} = 15\,$mK and $T_\mathrm{b}^\mathrm{max} = 100\,$mK. The measurement scheme is detailed in Supplementary Note 14. Points show data, lines and shaded areas show fits to the full four-wave model, where all system parameters have been obtained from a VNA measurement of the cavity and the OMIT response, except for the thermal phonon occupation $n_\mathrm{m}^\mathrm{th}(T_\mathrm{b})$ and the total output gain conversion, defining the absolute power scale of the spectra. Frequency-axis is given with respect to the red sideband optomechanical pump frequency $\omega_\mathrm{p}$. \textbf{b} shows the resonance frequency shift $\Delta\Omega_\mathrm{m} = \Omega_\mathrm{m}(T) - \Omega_\mathrm{m}(15\,\textrm{mK})$ of the mechanical oscillator with dilution refrigerator base temperature. Panels \textbf{c} and \textbf{d} show the effective and intrinsic mechanical linewidth $\Gamma_\mathrm{eff}(T)$ and $\Gamma_\mathrm{m}(T)$ vs base temperature $T_\mathrm{b}$. The effective linewidth is broadened by dynamical backaction and obtained from a fit to the OMIT response, the intrinsic linewidth is obtained from a fit to the OMIT response using the full four-wave model. The parameters obtained from this temperature dependence are used to calculate the residual thermal phonon occupation vs fridge temperature, the result is shown in main paper Fig.~4, where the values for each temperature have been averaged. Thermal calibration measurements were done at operation point I with an in-plane field of $B_\parallel = 21\,$mT. Square points in \textbf{b}-\textbf{d} indicate the results for the two datasets shown in $\textbf{a}$.}}
	\label{fig:thermalcal}
\end{figure*}

\begin{itemize}
	\item{The data analysis of the thermal calibration begins with fitting the cavity response $S_{21}$ for each temperature using Eq.~(\ref{eqn:fitS21}). This fit provides us with a value for the linewidth $\kappa'$ and a fit function for the complex background.}
	\item{We divide off the complex background from both, the cavity response and the OMIT response in their corresponding frequency ranges. In addition we correct for a small phase rotation which is intrinsic to the Kerr cavity susceptibility and therefore not captured by the complex background.}
	\item{Next, we use the full Kerr-mode model function Eq.~(\ref{eqn:fitS21Kerr}) to fit the background-corrected cavity response once again. As fixed parameter for this second fit, we use $\kappa'$ from the first fit, the parametric drive power $P_\mathrm{d}$ and the in-line attenuation. We also use the independently obtained $\kappa_\mathrm{e} = 2\pi\cdot 120 \pm 20\,$kHz of the undriven cavity and allow for small variations, necessary to match the observed resonances. As fit parameters in this second fit, we get the detuning between drive and undriven cavity $\Delta_\mathrm{d}$ and by using the Kerr polynomial (\ref{eqn:Kerr_poly}) we get the corresponding intracavity drive photon number $n_\mathrm{d}$. }
	\item{Next, we fit the background-corrected and rotated OMIT VNA response with Eq.~(\ref{eqn:fitS21}) and obtain $\Gamma_\mathrm{eff}$ and $Omega_\textrm{m}$ as fit parameter.}
	\item{We fit the OMIT VNA response with the full four-wave OMIT model Eqs.~(\ref{eqn:FWOMIT_red_alpha}) and (\ref{eqn:FWOMIT_S21}). We input as fixed parameters $\mathcal{K}$ and $g_0$ from the flux arch, the parametric drive power $P_\mathrm{d}$, the optomechanical pump power $P_\mathrm{p}$ and the in-line attenuation. Also, we use the fitted $\kappa_\mathrm{e}$ from the previous full model cavity fit. As starting values for the remaining model parameters, we use $\Delta_\mathrm{d}$, $n_\mathrm{d}$ and $\kappa'$ from the previous fits of the cavity. In the routine, we allow for a change of the bare cavity resonance frequency, i.e., of $\Delta_\mathrm{d}$ due to possible flux fluctuations. Based on the characteristic polynomial Eq.~(\ref{eqn:Kerr_poly}), we then dynamically adjust $n_\mathrm{d}$ to the modified $\Delta_\mathrm{d}$. Additionally, we allow for changes in $\kappa'$ of up to $\pm100\,$kHz, with a lower limit of $\kappa_\mathrm{min}' = 2\pi\cdot 320\,$kHz. As in the experiment between the VNA cavity scan and the VNA OMIT scan several minutes pass, during which the thermal noise spectrum is recorded, these allowed changes in fit parameters reflect possible drifts of the bare cavity resonance due to flux fluctuations in this time span. We calculate the intracavity $\gamma_-$ and $\gamma_+$ fields based on Eqs.~(\ref{eqn:gamma_m}) and (\ref{eqn:gamma_p}). Ultimately, we obtain from this OMIT fit with the full model a value for the bare mechanical linewidth $\Gamma_\mathrm{m}$ and the mechanical resonance frequency $\Omega_\mathrm{m}$. As last fit parameter, we allow for a small correction of the OMIT resonance circle, which is necessary due to the uncertainty of the background extraction during the initial cavity fitting routine at the cavity resonance frequency. This is taken into account by allowing for a multiplication of the OMIT response by a small complex scaling factor $\left(1+x\right)e^{i\beta}$ with $x, \beta \ll 1$.}
	\item{Finally, we fit the measured thermal noise spectrum using the full model Eqs.~(\ref{eqn:noise_output_red}) and (\ref{eqn:Sout_red}). We input as fixed parameters here all relevant quantities as obtained from the previous full model OMIT fit. The only remaining fit parameters at this point are the total detection output gain, converting the PSD in numbers of quanta to an absolute power and the uncalibrated equilibrium occupation of the mechanical oscillator $n_\mathrm{m}'$. The last used parameter $n_\mathrm{add}$ is adjusted to match the temperature dependence of the uncalibrated $n_\mathrm{m}'$ in the linear regime to the Bose distribution, which corresponds to a calibration of $n_\mathrm{m}'$ to $n_\mathrm{m}^\mathrm{th}$. As a result we obtain a number for the added photons $n_\mathrm{add} \approx 14$ and the residual thermal phonon occupation shown in main paper Fig.~4. In Supplementary Note 15, we present some additional data on the temperature dependence of $\Omega_\mathrm{m}$ and $\Gamma_\mathrm{m}$ as obtained from this procedure.}
	\item{Note: Due to slow SQUID cavity resonance frequency fluctuations and drifts, the measurement time of the spectrum acquisition was limited and we took these data with a reasonable compromise between number of data points, frequency span and bandwidth. For the lowest cooperativities however, when the mechanical linewdith is close to the intrinsic linewidth, the resolution bandwidth of the spectrum analyzer and the mechanical linewidth are comparable in size. In this case the effect of the resolution bandwidth is to smoothen and broaden the real mechanical power spectral density Lorentzian. To consider this effect in the fit curve for the power spectral density, we apply a moving-average-filter with the corresponding width in frequency space to the theory curve within the fitting routine itself. In the thermal calibration experiment, the resolution bandwidth of the spectrum analyzer was set to $5\,$Hz and therefore considerably smaller than the effective mechanical linewidths, cf. Supplementary Fig.~\ref{fig:thermalcal}.} 
	\item{Error bars in thermal occupation: During the fitting procedure we simultaneously calculate the corresponding error bars for each point. These translate the impact of deviations of the cavity from its operation point. From this we estimate the difference in thermal occupation by fitting the thermal noise spectrum with the cavity parameters $\kappa'$, $\Delta_\textrm{d}$ and $n_\textrm{d}$. This difference is plotted as the error in y-axis of the inset of in Fig.~4\textbf{c} of the main paper.}
\end{itemize}

\subsection*{Four-wave cooling data analysis}
\begin{itemize}
	\item{The data analysis of the four-wave cooling experiment begins with fitting the cavity response $S_{21}$ for each power setting of $P_\mathrm{d}$ and $P_\mathrm{p}$ using Eq.~(\ref{eqn:fitS21}). This fit provides us with a value for the linewidth $\kappa'$ and a fit function for the complex background. }
	\item{We divide off the complex background from both, the cavity response and the OMIT response in their corresponding frequency ranges. In addition, we correct for a small phase rotation which is intrinsic to the driven Kerr cavity susceptibility, and therefore not captured by the complex background.}
	\item{Next, we use the full Kerr-mode model function Eq.~(\ref{eqn:fitS21Kerr}) to fit the cavity response once again. As fixed parameter for this second fit, we use $\kappa'$ from the first fit, the parametric drive power $P_\mathrm{d}$ and the in-line attenuation. We also allow for small variations of the bare external decay rate of each operation point $\kappa^\textrm{I,II}_\mathrm{e}$ and allow for $\kappa_\mathrm{e} = 2\pi\cdot (\kappa^\textrm{I,II}_\mathrm{e} \pm 10)\,$kHz, which is necessary to match the observed resonances for all powers. As fit parameters in this second fit, we get the detuning between parametric drive and undriven cavity $\Delta_\mathrm{d}$. Additionally, by using the Kerr polynomial Eq.~(\ref{eqn:Kerr_poly}) within the fit routine we get the corresponding intracavity drive photon number $n_\mathrm{d}$.}
	\item{We fit the background-corrected and rotated OMIT VNA response with Eq.~(\ref{eqn:fitS21}) and obtain $\Gamma_\mathrm{eff}$ and $\Omega_\mathrm{m}$ as fit parameters.}
	\item{We fit the OMIT VNA response with the full four-wave OMIT model Eqs.~(\ref{eqn:FWOMIT_red_alpha}) and (\ref{eqn:FWOMIT_S21}) for red-signal-sideband pumping or Eqs.~(\ref{eqn:FWOMIT_blue_alpha}) and (\ref{eqn:FWOMIT_S21}) for the blue-idler-sideband case, respectively. We input as fixed parameters $\mathcal{K}$ and $g_0$ as determined from their flux dependence, the parametric drive power $P_\mathrm{d}$, the optomechanical pump power $P_\mathrm{p}$ and the input line attenuation. Also, we use the $\kappa_\mathrm{e}$ as determined from the full model cavity fit. As starting values for the remaining model parameters, we use $\Delta_\mathrm{d}$, $n_\mathrm{d}$ and $\kappa'$ from the previous fits of the cavity. In the routine, we allow for adjustments of the bare cavity resonance frequency, i.e., of $\Delta_\mathrm{d}$ up to $\pm2\,$MHz due to possible flux fluctuations, cf. operation range in main paper Fig.~2. Based on the characteristic polynomial Eq.~(\ref{eqn:Kerr_poly}), the intracavity drive photon number $n_\mathrm{d}$ is adjusted correspondingly within the fit routine. Additionally, we allow for adjustments of the total linewidth $\kappa'$ of up to $\pm100\,$kHz, but with a lower limit $\kappa_\mathrm{min, I}' = 2\pi\cdot 320\,$kHz at operation point I and $\kappa_\mathrm{min, II}' = 2\pi\cdot 350\,$kHz at point II. Several minutes pass in the experiment between the VNA scan of the OMIT response and the VNA scan of the signal resonance, during which the thermal noise spectrum is recorded. We allow for adjustments of some of fit parameters between the two scans, which reflects possible drifts of the bare cavity resonance due to flux drifts in this time span. We note that in principle also $\mathcal{K}$ and $g_0$ might experience small drifts due to the effective change in bias flux. As the flux-drift related variations of these two parameters are small within the operation range however, cf. main paper Fig.~2 and Supplementary Note 4, we work with constant average values for the analysis here. Based on Eqs.~(\ref{eqn:gamma_m}) and (\ref{eqn:gamma_p}), we calculate also the intracavity fields $\gamma_-$ and $\gamma_+$. As last fit parameter, we allow for a small correction of the OMIT resonance circle, which is necessary due to the uncertainty of the background extraction during the initial cavity fitting routine at the cavity resonance frequency. This is taken into account by allowing for a multiplication of the OMIT response by a small complex scaling factor $\left(1+x\right)e^{i\beta}$ with $x, \beta \ll 1$.}
	\item{Finally, we fit the measured thermal noise spectrum using the full model Eqs.~(\ref{eqn:noise_output_red}) and (\ref{eqn:Sout_red}) for the red-signal-sideband case and Eqs.~(\ref{eqn:noise_output_blue}) and (\ref{eqn:Sout_red}) for the blue-idler-sideband case, respectively. Once again, in this fitting procedure we allow for fluctuations of the cavity decay rate $\delta\kappa'=\pm 2\pi\cdot60\,$kHz and $\delta\Delta_\mathrm{d} = \pm 2\pi\cdot 0.8\,$MHz which are limited to small deviations around the average values obtained from the cavity and OMIT fitting routines. Note that the OMIT measurement and the cavity scan were taken prior and posterior (respectively) to the thermal noise detection and therefore we account for possible deviations of the cavity state due to the fluctuations in the system. Finally, the only remaining fit parameters are the total detection output gain and the equilibrium occupation of the mechanical oscillator $n_\mathrm{m}^\mathrm{th}$. Note that this last fit parameter was allowed to vary between 70-90 phonons for blue sideband driving and 60-90 phonons for red-sideband driving. Without these restriction, the fit often fails and the corresponding thermal phonon numbers are oscillating unsystematically between 50 and 130 phonons. The corresponding values without restrictions are considered in the error bars though, see below. The number of added photons $n_\mathrm{add} \approx 14$ was determined via the thermal calibration procedure.}
	\item{Note: Due to slow SQUID cavity resonance frequency fluctuations and drifts, the measurement time of the spectrum acquisition was limited and we took these data with a reasonable compromise between number of data points, frequency span and bandwidth. For the lowest cooperativities however, when the mechanical linewdith is close to the intrinsic linewidth, the resolution bandwidth of the spectrum analyzer and the mechanical linewidth are comparable in size. In this case the effect of the resolution bandwidth is to smoothen and broaden the mechanical power spectral density Lorentzian. To consider this effect in the fit curve for the power spectral density, we apply a moving-average-filter with the corresponding width in frequency space to the theory curve within the fitting routine itself. An additional broadening effect of the spectrum might arise due to slow mechanical frequency fluctuations induced by a variation of the optical spring during bias flux drifts, cf. main paper Fig.~2.} 
	\item{Based on the full ensemble of system parameters obtained by this multi-step analysis and fit procedure, we finally infer the resulting cooled phonon number $n_\mathrm{m}$ of the mechanical oscillator by integrating Eq.~(\ref{eqn:PhononPSD_red}) for the red-signal-sideband case and Eq.~(\ref{eqn:PhononPSD_blue}) for the blue-idler-sideband case, respectively. The results are plotted in main paper Fig.~4 and in Supplementary Fig.~\ref{fig:RSB_cooling}.}
	\item{Error bars for cooled number of phonons: During the fitting procedure we simultaneously calculate the corresponding error bars of each point. These translate the impact of deviations of the cavity from its operation point and of a different thermal occupation in the extraction of cooled photon number. For this we estimate the difference in the cooled phonons by fitting the thermal noise spectrum with the cavity parameters, which were extracted from the OMIT fit, in this case without any restriction to the thermal phonon number. This difference is plotted as the error in the y-axis of Fig.~4 and in Supplementary Fig.~\ref{fig:RSB_cooling}. Furthermore, we calculate the error in the extraction of the effective mechanical linewidth by computing the difference of $\Gamma_\textrm{eff}$ obtained from the full model noise fit and the one obtained from the fit of the OMIT VNA response with Eq.~(\ref{eqn:fitS21}). In addition, we consider an uncertainty in $\Gamma_\textrm{m}$ of $\pm1\,$Hz The sum of these errors is plotted as the error bar in the $\Gamma_\mathrm{eff}/\Gamma_\textrm{m}$ direction.}
\end{itemize}

\begin{figure*}
	\centerline{\includegraphics[trim = {0cm, 0cm, 0cm, 0cm}, clip=True, width=0.9\textwidth]{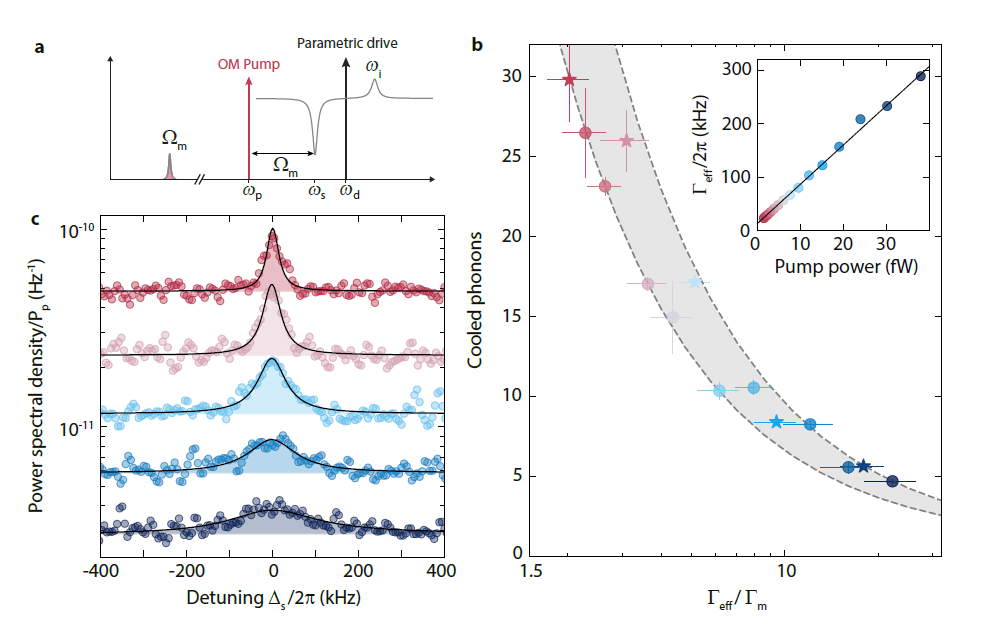}}
	\caption{\textsf{\textbf{Red-sideband four-wave cooling of the mechanical oscillator.} \textbf{a} Schematic representation of the experiment. A parametric drive is used to activate the Kerr quasi-mode state and an optomechanical pump is sent to the red sideband of the signal resonance $\omega_\mathrm{p} \approx \omega_\mathrm{s} - \Omega_\mathrm{m}$. From $S_{21}$ measurements of signal resonance and OMIT, as well as a signal mode output power spectrum, we determine the cooled mechanical phonon occupation. The result for increasing sideband pump power $P_\mathrm{p}$ and effective mechanical linewidth, respectively, is shown in panel \textbf{b}. Inset shows that $\Gamma_\mathrm{eff} \propto P_\mathrm{p}$. Circles and stars are data, the lowest achieved occupation is $n_\mathrm{m}\sim 4$, limited by cavity bifurcation instability. Dashed lines and shaded areas display the theoretical value range of $n_\mathrm{m}$, taking into account $60 < n_\mathrm{m}^\mathrm{th} < 90$. Data were taken at operation point I and at $B_\parallel = 21\,$mT. In addition, the error bars represent possible deviations in the extraction of the plotted values based on their difference from the ones calculated based on the parameters extracted from the corresponding OMIT fit. For more details, see Supplementary Note 14. Panel \textbf{c} shows selected power spectra data for the points in \textbf{b}, which are plotted as stars. }}
	\label{fig:RSB_cooling}
\end{figure*}
\section*{Supplementary Note 15: Thermal calibration of the residual mechanical phonon occupation}
We perform the thermal calibration measurement and data analysis as described in Supplementary Note~14.
In Supplementary Fig.~\ref{fig:thermalcal} we present some of the results obtained from this experiment.
In particular, we show the detected thermal noise spectrum in units of quanta for the lowest and highest temperatures $T_\mathrm{b}^\mathrm{min} = 15\,$mK and $T_\mathrm{b}^\mathrm{max} = 100\,$mK, including the fit from the full model, and we show the obtained temperature dependence of the mechanical parameters $\Omega_\mathrm{m}$, $\Gamma_\mathrm{eff}$ and $\Gamma_\mathrm{m}$.
From the data it is clear, that both, the mechanical frequency and the intrinsic mechanical linewidth increase significantly with temperature and have a strong dependence on temperature at the lowest temperatures.
Either intrinsic or pump-power-induced small variations of the chip temperature might therefore be a source for the observation that for good agreement of our datasets with the theory, we have to consider variations between the datasets of some $10\,$Hz in the mechanical resonance frequency and of a few Hz for the mechanical linewidth in the range $10\,$Hz$<\frac{\Gamma_\mathrm{m}}{2\pi} < 15\,$Hz.
\section*{Supplementary Note 16: Four-wave-cooling with a pump on the red sideband of the signal resonance}
We perform a four-wave cooling experiment with an optomechanical pump positioned on the red sideband of the signal resonance, cf. Supplementary Fig.~\ref{fig:RSB_cooling}.
Data acquisition and data analysis are described in Supplementary Note 14.

\section*{Supplementary References}

\end{document}